\documentclass[longauth]{aa} 
\usepackage{natbib}
\usepackage{graphicx}
\usepackage{txfonts}
\usepackage[colorlinks,allcolors=blue]{hyperref}
\usepackage{scalerel}
\usepackage{booktabs}
\usepackage{multirow}
\usepackage{placeins}

\newcommand{\orcit}[1]{\protect\href{https://orcid.org/#1}{\protect\includegraphics[width=8pt]{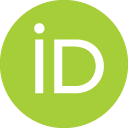}}}

\newcommand{\Gaia}{\textit{Gaia}\xspace}
\newcommand{\rvflag}{\texttt{flag\_rv}\xspace}

\newcommand{\mags}{{\rm mag}}
\newcommand{\kms}{{\rm km\,s}^{-1}}
\newcommand{\days}{{\rm days}}

\newcommand{\Teff}{T_{\rm eff}}

\newcommand{\gbp}{G_{\rm BP}}
\newcommand{\grp}{G_{\rm RP}}
\newcommand{\grvs}{G_{\rm RVS}}
\newcommand{\rv}{V_R}
\newcommand{\meanrv}{\overline{\rv}}
\newcommand{\medianrv}{\langle\rv\rangle}
\newcommand{\zprv}{\rv^0}
\newcommand{\pubrv}{\rv^{\rm DR3}}
\newcommand{\erv}{\varepsilon_{\rv}}
\newcommand{\meanerv}{\overline{\varepsilon}_{\rv}}

\newcommand{\pg}{P_{G}}
\newcommand{\pbp}{P_{G_{\rm BP}}}
\newcommand{\prp}{P_{G_{\rm RP}}}
\newcommand{\prv}{P_{\rv}}
\newcommand{\pph}{P_{\rm ph}}

\newcommand{\durationrv}{\Delta t_{\rv}}
\newcommand{\ncycrv}{n_{\rv}^{\rm cyc}}

\newcommand{\ampg}{A_G}
\newcommand{\amprv}{A_{\rv}}

\newcommand{\nobsrv}{N_{\rv}}
\newcommand{\nobsrvraw}{N_{\rv}^{\rm raw}}
\newcommand{\noutrv}{N_{\rv}^{\rm out}}
\newcommand{\snrv}{{\rm SN}_{\rv}}

\newcommand{\wbprp}{W_{\scaleto{\rm BP,RP}{4.5pt}}}

\makeatletter
\renewcommand*\maketitle{%
  \thispagestyle{firstpage}
\begingroup
    \if@wideboxfn
    \setlength\bibindent{1.4\parindent}
    \else
    \setlength\bibindent{\parindent}
    \fi
    \renewcommand*\thefootnote{\@fnsymbol\c@footnote}%
    \renewcommand\@makefntext[1]{%
    \ifaa@longfn\hsize\textwidth\fi
    \noindent
    \hb@xt@\bibindent{\hss\@makefnmark\enspace}##1}
  \ifaa@twocolumn
  \begingroup
    \begin{aa@strip}
          \aa@maketitle
    \end{aa@strip}
    \@thanks	  	
  \endgroup
  \else
    \begingroup
      \let\thanks\footnote
      \aa@maketitle
    \endgroup
  \fi
\endgroup
  \setcounter{footnote}{0}%
}
\makeatother
\begin{document} 

  \title{\Gaia Focused Product Release: Radial velocity time series of long-period variables}
  \titlerunning{\Gaia FPR LPV radial velocities}
  
  \authorrunning{Gaia Collaboration}
\author{
{\it Gaia} Collaboration
\and M.        ~Trabucchi                     \orcit{0000-0002-1429-2388} \inst{\ref{inst:0001},\ref{inst:0002}}
\and N.        ~Mowlavi                       \orcit{0000-0003-1578-6993} \inst{\ref{inst:0002}}
\and T.        ~Lebzelter                     \orcit{0000-0002-0702-7551} \inst{\ref{inst:0004}}
\and I.        ~Lecoeur-Taibi                 \orcit{0000-0003-0029-8575} \inst{\ref{inst:0005}}
\and M.        ~Audard                        \orcit{0000-0003-4721-034X} \inst{\ref{inst:0002},\ref{inst:0005}}
\and L.        ~Eyer                          \orcit{0000-0002-0182-8040} \inst{\ref{inst:0002}}
\and P.        ~Garc\'{i}a-Lario              \orcit{0000-0003-4039-8212} \inst{\ref{inst:0009}}
\and P.        ~Gavras                        \orcit{0000-0002-4383-4836} \inst{\ref{inst:0010}}
\and B.        ~Holl                          \orcit{0000-0001-6220-3266} \inst{\ref{inst:0002},\ref{inst:0005}}
\and G.        ~Jevardat de Fombelle          \orcit{0000-0001-6166-8221} \inst{\ref{inst:0002}}
\and K.        ~Nienartowicz                  \orcit{0000-0001-5415-0547} \inst{\ref{inst:0014},\ref{inst:0005}}
\and L.        ~Rimoldini                     \orcit{0000-0002-0306-585X} \inst{\ref{inst:0005}}
\and P.        ~Sartoretti                    \orcit{0000-0002-6574-7565} \inst{\ref{inst:0017}}
\and R.        ~Blomme                        \orcit{0000-0002-2526-346X} \inst{\ref{inst:0018}}
\and Y.        ~Fr\'{e}mat                    \orcit{0000-0002-4645-6017} \inst{\ref{inst:0018}}
\and O.        ~Marchal                       \orcit{ 0000-0001-7461-892} \inst{\ref{inst:0020}}
\and Y.        ~Damerdji                      \orcit{0000-0002-3107-4024} \inst{\ref{inst:0021},\ref{inst:0022}}
\and A.G.A.    ~Brown                         \orcit{0000-0002-7419-9679} \inst{\ref{inst:0023}}
\and A.        ~Guerrier                                                  \inst{\ref{inst:0024}}
\and P.        ~Panuzzo                       \orcit{0000-0002-0016-8271} \inst{\ref{inst:0017}}
\and D.        ~Katz                          \orcit{0000-0001-7986-3164} \inst{\ref{inst:0017}}
\and G.M.      ~Seabroke                      \orcit{0000-0003-4072-9536} \inst{\ref{inst:0027}}
\and K.        ~Benson                                                    \inst{\ref{inst:0027}}
\and R.        ~Haigron                                                   \inst{\ref{inst:0017}}
\and M.        ~Smith                                                     \inst{\ref{inst:0027}}
\and A.        ~Lobel                         \orcit{0000-0001-5030-019X} \inst{\ref{inst:0018}}
\and A.        ~Vallenari                     \orcit{0000-0003-0014-519X} \inst{\ref{inst:0032}}
\and T.        ~Prusti                        \orcit{0000-0003-3120-7867} \inst{\ref{inst:0033}}
\and J.H.J.    ~de Bruijne                    \orcit{0000-0001-6459-8599} \inst{\ref{inst:0033}}
\and F.        ~Arenou                        \orcit{0000-0003-2837-3899} \inst{\ref{inst:0017}}
\and C.        ~Babusiaux                     \orcit{0000-0002-7631-348X} \inst{\ref{inst:0036}}
\and A.        ~Barbier                       \orcit{0009-0004-0983-931X} \inst{\ref{inst:0024}}
\and M.        ~Biermann                      \orcit{0000-0002-5791-9056} \inst{\ref{inst:0038}}
\and O.L.      ~Creevey                       \orcit{0000-0003-1853-6631} \inst{\ref{inst:0039}}
\and C.        ~Ducourant                     \orcit{0000-0003-4843-8979} \inst{\ref{inst:0040}}
\and D.W.      ~Evans                         \orcit{0000-0002-6685-5998} \inst{\ref{inst:0041}}
\and R.        ~Guerra                        \orcit{0000-0002-9850-8982} \inst{\ref{inst:0009}}
\and A.        ~Hutton                                                    \inst{\ref{inst:0043}}
\and C.        ~Jordi                         \orcit{0000-0001-5495-9602} \inst{\ref{inst:0044},\ref{inst:0045},\ref{inst:0046}}
\and S.A.      ~Klioner                       \orcit{0000-0003-4682-7831} \inst{\ref{inst:0047}}
\and U.        ~Lammers                       \orcit{0000-0001-8309-3801} \inst{\ref{inst:0009}}
\and L.        ~Lindegren                     \orcit{0000-0002-5443-3026} \inst{\ref{inst:0049}}
\and X.        ~Luri                          \orcit{0000-0001-5428-9397} \inst{\ref{inst:0044},\ref{inst:0045},\ref{inst:0046}}
\and F.        ~Mignard                                                   \inst{\ref{inst:0039}}
\and S.        ~Randich                       \orcit{0000-0003-2438-0899} \inst{\ref{inst:0054}}
\and R.        ~Smiljanic                     \orcit{0000-0003-0942-7855} \inst{\ref{inst:0055}}
\and P.        ~Tanga                         \orcit{0000-0002-2718-997X} \inst{\ref{inst:0039}}
\and N.A.      ~Walton                        \orcit{0000-0003-3983-8778} \inst{\ref{inst:0041}}
\and C.A.L.    ~Bailer-Jones                                              \inst{\ref{inst:0058}}
\and U.        ~Bastian                       \orcit{0000-0002-8667-1715} \inst{\ref{inst:0038}}
\and M.        ~Cropper                       \orcit{0000-0003-4571-9468} \inst{\ref{inst:0027}}
\and R.        ~Drimmel                       \orcit{0000-0002-1777-5502} \inst{\ref{inst:0061}}
\and M.G.      ~Lattanzi                      \orcit{0000-0003-0429-7748} \inst{\ref{inst:0061},\ref{inst:0063}}
\and C.        ~Soubiran                      \orcit{0000-0003-3304-8134} \inst{\ref{inst:0040}}
\and F.        ~van Leeuwen                   \orcit{0000-0003-1781-4441} \inst{\ref{inst:0041}}
\and J.        ~Bakker                                                    \inst{\ref{inst:0009}}
\and J.        ~Casta\~{n}eda                 \orcit{0000-0001-7820-946X} \inst{\ref{inst:0067},\ref{inst:0044},\ref{inst:0046}}
\and F.        ~De Angeli                     \orcit{0000-0003-1879-0488} \inst{\ref{inst:0041}}
\and C.        ~Fabricius                     \orcit{0000-0003-2639-1372} \inst{\ref{inst:0046},\ref{inst:0044},\ref{inst:0045}}
\and M.        ~Fouesneau                     \orcit{0000-0001-9256-5516} \inst{\ref{inst:0058}}
\and L.        ~Galluccio                     \orcit{0000-0002-8541-0476} \inst{\ref{inst:0039}}
\and E.        ~Masana                        \orcit{0000-0002-4819-329X} \inst{\ref{inst:0046},\ref{inst:0044},\ref{inst:0045}}
\and R.        ~Messineo                                                  \inst{\ref{inst:0079}}
\and C.        ~Nicolas                                                   \inst{\ref{inst:0024}}
\and F.        ~Pailler                       \orcit{0000-0002-4834-481X} \inst{\ref{inst:0024}}
\and F.        ~Riclet                                                    \inst{\ref{inst:0024}}
\and W.        ~Roux                          \orcit{0000-0002-7816-1950} \inst{\ref{inst:0024}}
\and R.        ~Sordo                         \orcit{0000-0003-4979-0659} \inst{\ref{inst:0032}}
\and F.        ~Th\'{e}venin                  \orcit{0000-0002-5032-2476} \inst{\ref{inst:0039}}
\and G.        ~Gracia-Abril                                              \inst{\ref{inst:0086},\ref{inst:0038}}
\and J.        ~Portell                       \orcit{0000-0002-8886-8925} \inst{\ref{inst:0044},\ref{inst:0045},\ref{inst:0046}}
\and D.        ~Teyssier                      \orcit{0000-0002-6261-5292} \inst{\ref{inst:0091}}
\and M.        ~Altmann                       \orcit{0000-0002-0530-0913} \inst{\ref{inst:0038},\ref{inst:0093}}
\and J.        ~Berthier                      \orcit{0000-0003-1846-6485} \inst{\ref{inst:0094}}
\and P.W.      ~Burgess                       \orcit{0009-0002-6668-4559} \inst{\ref{inst:0041}}
\and D.        ~Busonero                      \orcit{0000-0002-3903-7076} \inst{\ref{inst:0061}}
\and G.        ~Busso                         \orcit{0000-0003-0937-9849} \inst{\ref{inst:0041}}
\and H.        ~C\'{a}novas                   \orcit{0000-0001-7668-8022} \inst{\ref{inst:0091}}
\and B.        ~Carry                         \orcit{0000-0001-5242-3089} \inst{\ref{inst:0039}}
\and N.        ~Cheek                                                     \inst{\ref{inst:0100}}
\and G.        ~Clementini                    \orcit{0000-0001-9206-9723} \inst{\ref{inst:0101}}
\and M.        ~Davidson                      \orcit{0000-0001-9271-4411} \inst{\ref{inst:0102}}
\and P.        ~de Teodoro                                                \inst{\ref{inst:0009}}
\and L.        ~Delchambre                    \orcit{0000-0003-2559-408X} \inst{\ref{inst:0021}}
\and A.        ~Dell'Oro                      \orcit{0000-0003-1561-9685} \inst{\ref{inst:0054}}
\and E.        ~Fraile Garcia                 \orcit{0000-0001-7742-9663} \inst{\ref{inst:0010}}
\and D.        ~Garabato                      \orcit{0000-0002-7133-6623} \inst{\ref{inst:0107}}
\and N.        ~Garralda Torres                                           \inst{\ref{inst:0108}}
\and N.C.      ~Hambly                        \orcit{0000-0002-9901-9064} \inst{\ref{inst:0102}}
\and D.L.      ~Harrison                      \orcit{0000-0001-8687-6588} \inst{\ref{inst:0041},\ref{inst:0111}}
\and D.        ~Hatzidimitriou                \orcit{0000-0002-5415-0464} \inst{\ref{inst:0112}}
\and J.        ~Hern\'{a}ndez                 \orcit{0000-0002-0361-4994} \inst{\ref{inst:0009}}
\and S.T.      ~Hodgkin                       \orcit{0000-0002-5470-3962} \inst{\ref{inst:0041}}
\and S.        ~Jamal                         \orcit{0000-0002-3929-6668} \inst{\ref{inst:0058}}
\and S.        ~Jordan                        \orcit{0000-0001-6316-6831} \inst{\ref{inst:0038}}
\and A.        ~Krone-Martins                 \orcit{0000-0002-2308-6623} \inst{\ref{inst:0117},\ref{inst:0118}}
\and A.C.      ~Lanzafame                     \orcit{0000-0002-2697-3607} \inst{\ref{inst:0119},\ref{inst:0120}}
\and W.        ~L\"{ o}ffler                                              \inst{\ref{inst:0038}}
\and A.        ~Lorca                         \orcit{0000-0002-7985-250X} \inst{\ref{inst:0043}}
\and P.M.      ~Marrese                       \orcit{0000-0002-8162-3810} \inst{\ref{inst:0123},\ref{inst:0124}}
\and A.        ~Moitinho                      \orcit{0000-0003-0822-5995} \inst{\ref{inst:0118}}
\and K.        ~Muinonen                      \orcit{0000-0001-8058-2642} \inst{\ref{inst:0126},\ref{inst:0127}}
\and M.        ~Nu\~{n}ez Campos                                          \inst{\ref{inst:0043}}
\and I.        ~Oreshina-Slezak                                           \inst{\ref{inst:0039}}
\and P.        ~Osborne                       \orcit{0000-0003-4482-3538} \inst{\ref{inst:0041}}
\and E.        ~Pancino                       \orcit{0000-0003-0788-5879} \inst{\ref{inst:0054},\ref{inst:0124}}
\and T.        ~Pauwels                                                   \inst{\ref{inst:0018}}
\and A.        ~Recio-Blanco                  \orcit{0000-0002-6550-7377} \inst{\ref{inst:0039}}
\and M.        ~Riello                        \orcit{0000-0002-3134-0935} \inst{\ref{inst:0041}}
\and A.C.      ~Robin                         \orcit{0000-0001-8654-9499} \inst{\ref{inst:0136}}
\and T.        ~Roegiers                      \orcit{0000-0002-1231-4440} \inst{\ref{inst:0137}}
\and L.M.      ~Sarro                         \orcit{0000-0002-5622-5191} \inst{\ref{inst:0138}}
\and M.        ~Schultheis                    \orcit{0000-0002-6590-1657} \inst{\ref{inst:0039}}
\and C.        ~Siopis                        \orcit{0000-0002-6267-2924} \inst{\ref{inst:0140}}
\and A.        ~Sozzetti                      \orcit{0000-0002-7504-365X} \inst{\ref{inst:0061}}
\and E.        ~Utrilla                                                   \inst{\ref{inst:0043}}
\and M.        ~van Leeuwen                   \orcit{0000-0001-9698-2392} \inst{\ref{inst:0041}}
\and K.        ~Weingrill                     \orcit{0000-0002-8163-2493} \inst{\ref{inst:0144}}
\and U.        ~Abbas                         \orcit{0000-0002-5076-766X} \inst{\ref{inst:0061}}
\and P.        ~\'{A}brah\'{a}m               \orcit{0000-0001-6015-646X} \inst{\ref{inst:0146},\ref{inst:0147}}
\and A.        ~Abreu Aramburu                \orcit{0000-0003-3959-0856} \inst{\ref{inst:0108}}
\and C.        ~Aerts                         \orcit{0000-0003-1822-7126} \inst{\ref{inst:0149},\ref{inst:0150},\ref{inst:0058}}
\and G.        ~Altavilla                     \orcit{0000-0002-9934-1352} \inst{\ref{inst:0123},\ref{inst:0124}}
\and M.A.      ~\'{A}lvarez                   \orcit{0000-0002-6786-2620} \inst{\ref{inst:0107}}
\and J.        ~Alves                         \orcit{0000-0002-4355-0921} \inst{\ref{inst:0004}}
\and F.        ~Anders                                                    \inst{\ref{inst:0044},\ref{inst:0045},\ref{inst:0046}}
\and R.I.      ~Anderson                      \orcit{0000-0001-8089-4419} \inst{\ref{inst:0159}}
\and T.        ~Antoja                        \orcit{0000-0003-2595-5148} \inst{\ref{inst:0044},\ref{inst:0045},\ref{inst:0046}}
\and D.        ~Baines                        \orcit{0000-0002-6923-3756} \inst{\ref{inst:0163}}
\and S.G.      ~Baker                         \orcit{0000-0002-6436-1257} \inst{\ref{inst:0027}}
\and Z.        ~Balog                         \orcit{0000-0003-1748-2926} \inst{\ref{inst:0038},\ref{inst:0058}}
\and C.        ~Barache                                                   \inst{\ref{inst:0093}}
\and D.        ~Barbato                                                   \inst{\ref{inst:0002},\ref{inst:0061}}
\and M.        ~Barros                        \orcit{0000-0002-9728-9618} \inst{\ref{inst:0170}}
\and M.A.      ~Barstow                       \orcit{0000-0002-7116-3259} \inst{\ref{inst:0171}}
\and S.        ~Bartolom\'{e}                 \orcit{0000-0002-6290-6030} \inst{\ref{inst:0046},\ref{inst:0044},\ref{inst:0045}}
\and D.        ~Bashi                         \orcit{0000-0002-9035-2645} \inst{\ref{inst:0175},\ref{inst:0176}}
\and N.        ~Bauchet                       \orcit{0000-0002-2307-8973} \inst{\ref{inst:0017}}
\and N.        ~Baudeau                                                   \inst{\ref{inst:0178}}
\and U.        ~Becciani                      \orcit{0000-0002-4389-8688} \inst{\ref{inst:0119}}
\and L.R.      ~Bedin                                                     \inst{\ref{inst:0032}}
\and I.        ~Bellas-Velidis                                            \inst{\ref{inst:0181}}
\and M.        ~Bellazzini                    \orcit{0000-0001-8200-810X} \inst{\ref{inst:0101}}
\and W.        ~Beordo                        \orcit{0000-0002-5094-1306} \inst{\ref{inst:0061},\ref{inst:0063}}
\and A.        ~Berihuete                     \orcit{0000-0002-8589-4423} \inst{\ref{inst:0185}}
\and M.        ~Bernet                        \orcit{0000-0001-7503-1010} \inst{\ref{inst:0044},\ref{inst:0045},\ref{inst:0046}}
\and C.        ~Bertolotto                                                \inst{\ref{inst:0079}}
\and S.        ~Bertone                       \orcit{0000-0001-9885-8440} \inst{\ref{inst:0061}}
\and L.        ~Bianchi                       \orcit{0000-0002-7999-4372} \inst{\ref{inst:0191}}
\and A.        ~Binnenfeld                    \orcit{0000-0002-9319-3838} \inst{\ref{inst:0192}}
\and A.        ~Blazere                                                   \inst{\ref{inst:0193}}
\and T.        ~Boch                          \orcit{0000-0001-5818-2781} \inst{\ref{inst:0020}}
\and A.        ~Bombrun                                                   \inst{\ref{inst:0195}}
\and S.        ~Bouquillon                                                \inst{\ref{inst:0093},\ref{inst:0197}}
\and A.        ~Bragaglia                     \orcit{0000-0002-0338-7883} \inst{\ref{inst:0101}}
\and J.        ~Braine                        \orcit{0000-0003-1740-1284} \inst{\ref{inst:0040}}
\and L.        ~Bramante                                                  \inst{\ref{inst:0079}}
\and E.        ~Breedt                        \orcit{0000-0001-6180-3438} \inst{\ref{inst:0041}}
\and A.        ~Bressan                       \orcit{0000-0002-7922-8440} \inst{\ref{inst:0202}}
\and N.        ~Brouillet                     \orcit{0000-0002-3274-7024} \inst{\ref{inst:0040}}
\and E.        ~Brugaletta                    \orcit{0000-0003-2598-6737} \inst{\ref{inst:0119}}
\and B.        ~Bucciarelli                   \orcit{0000-0002-5303-0268} \inst{\ref{inst:0061},\ref{inst:0063}}
\and A.G.      ~Butkevich                     \orcit{0000-0002-4098-3588} \inst{\ref{inst:0061}}
\and R.        ~Buzzi                         \orcit{0000-0001-9389-5701} \inst{\ref{inst:0061}}
\and E.        ~Caffau                        \orcit{0000-0001-6011-6134} \inst{\ref{inst:0017}}
\and R.        ~Cancelliere                   \orcit{0000-0002-9120-3799} \inst{\ref{inst:0210}}
\and S.        ~Cannizzo                                                  \inst{\ref{inst:0211}}
\and R.        ~Carballo                      \orcit{0000-0001-7412-2498} \inst{\ref{inst:0213}}
\and T.        ~Carlucci                                                  \inst{\ref{inst:0093}}
\and M.I.      ~Carnerero                     \orcit{0000-0001-5843-5515} \inst{\ref{inst:0061}}
\and J.M.      ~Carrasco                      \orcit{0000-0002-3029-5853} \inst{\ref{inst:0046},\ref{inst:0044},\ref{inst:0045}}
\and J.        ~Carretero                     \orcit{0000-0002-3130-0204} \inst{\ref{inst:0219},\ref{inst:0220}}
\and S.        ~Carton                                                    \inst{\ref{inst:0211}}
\and L.        ~Casamiquela                   \orcit{0000-0001-5238-8674} \inst{\ref{inst:0040},\ref{inst:0017}}
\and M.        ~Castellani                    \orcit{0000-0002-7650-7428} \inst{\ref{inst:0123}}
\and A.        ~Castro-Ginard                 \orcit{0000-0002-9419-3725} \inst{\ref{inst:0023}}
\and V.        ~Cesare                        \orcit{0000-0003-1119-4237} \inst{\ref{inst:0119}}
\and P.        ~Charlot                       \orcit{0000-0002-9142-716X} \inst{\ref{inst:0040}}
\and L.        ~Chemin                        \orcit{0000-0002-3834-7937} \inst{\ref{inst:0228}}
\and V.        ~Chiaramida                                                \inst{\ref{inst:0079}}
\and A.        ~Chiavassa                     \orcit{0000-0003-3891-7554} \inst{\ref{inst:0039}}
\and N.        ~Chornay                       \orcit{0000-0002-8767-3907} \inst{\ref{inst:0041},\ref{inst:0005}}
\and R.        ~Collins                       \orcit{0000-0001-8437-1703} \inst{\ref{inst:0102}}
\and G.        ~Contursi                      \orcit{0000-0001-5370-1511} \inst{\ref{inst:0039}}
\and W.J.      ~Cooper                        \orcit{0000-0003-3501-8967} \inst{\ref{inst:0235},\ref{inst:0061}}
\and T.        ~Cornez                                                    \inst{\ref{inst:0211}}
\and M.        ~Crosta                        \orcit{0000-0003-4369-3786} \inst{\ref{inst:0061},\ref{inst:0239}}
\and C.        ~Crowley                       \orcit{0000-0002-9391-9360} \inst{\ref{inst:0195}}
\and C.        ~Dafonte                       \orcit{0000-0003-4693-7555} \inst{\ref{inst:0107}}
\and M.        ~David                         \orcit{0000-0002-4172-3112} \inst{\ref{inst:0242}}
\and P.        ~de Laverny                    \orcit{0000-0002-2817-4104} \inst{\ref{inst:0039}}
\and F.        ~De Luise                      \orcit{0000-0002-6570-8208} \inst{\ref{inst:0245}}
\and R.        ~De March                      \orcit{0000-0003-0567-842X} \inst{\ref{inst:0079}}
\and J.        ~De Ridder                     \orcit{0000-0001-6726-2863} \inst{\ref{inst:0149}}
\and R.        ~de Souza                      \orcit{0009-0007-7669-0254} \inst{\ref{inst:0248}}
\and A.        ~de Torres                                                 \inst{\ref{inst:0195}}
\and E.F.      ~del Peloso                                                \inst{\ref{inst:0038}}
\and M.        ~Delbo                         \orcit{0000-0002-8963-2404} \inst{\ref{inst:0039}}
\and A.        ~Delgado                                                   \inst{\ref{inst:0010}}
\and T.E.      ~Dharmawardena                 \orcit{0000-0002-9583-5216} \inst{\ref{inst:0058}}
\and S.        ~Diakite                                                   \inst{\ref{inst:0254}}
\and C.        ~Diener                                                    \inst{\ref{inst:0041}}
\and E.        ~Distefano                     \orcit{0000-0002-2448-2513} \inst{\ref{inst:0119}}
\and C.        ~Dolding                                                   \inst{\ref{inst:0027}}
\and K.        ~Dsilva                        \orcit{0000-0002-1476-9772} \inst{\ref{inst:0140}}
\and J.        ~Dur\'{a}n                                                 \inst{\ref{inst:0010}}
\and H.        ~Enke                          \orcit{0000-0002-2366-8316} \inst{\ref{inst:0144}}
\and P.        ~Esquej                        \orcit{0000-0001-8195-628X} \inst{\ref{inst:0010}}
\and C.        ~Fabre                                                     \inst{\ref{inst:0193}}
\and M.        ~Fabrizio                      \orcit{0000-0001-5829-111X} \inst{\ref{inst:0123},\ref{inst:0124}}
\and S.        ~Faigler                       \orcit{0000-0002-8368-5724} \inst{\ref{inst:0175}}
\and M.        ~Fatovi\'{c}                   \orcit{0000-0003-1911-4326} \inst{\ref{inst:0266}}
\and G.        ~Fedorets                      \orcit{0000-0002-8418-4809} \inst{\ref{inst:0126},\ref{inst:0268}}
\and J.        ~Fern\'{a}ndez-Hern\'{a}ndez                               \inst{\ref{inst:0010}}
\and P.        ~Fernique                      \orcit{0000-0002-3304-2923} \inst{\ref{inst:0020}}
\and F.        ~Figueras                      \orcit{0000-0002-3393-0007} \inst{\ref{inst:0044},\ref{inst:0045},\ref{inst:0046}}
\and Y.        ~Fournier                      \orcit{0000-0002-6633-9088} \inst{\ref{inst:0144}}
\and C.        ~Fouron                                                    \inst{\ref{inst:0178}}
\and M.        ~Gai                           \orcit{0000-0001-9008-134X} \inst{\ref{inst:0061}}
\and M.        ~Galinier                      \orcit{0000-0001-7920-0133} \inst{\ref{inst:0039}}
\and A.        ~Garcia-Gutierrez                                          \inst{\ref{inst:0046},\ref{inst:0044},\ref{inst:0045}}
\and M.        ~Garc\'{i}a-Torres             \orcit{0000-0002-6867-7080} \inst{\ref{inst:0281}}
\and A.        ~Garofalo                      \orcit{0000-0002-5907-0375} \inst{\ref{inst:0101}}
\and E.        ~Gerlach                       \orcit{0000-0002-9533-2168} \inst{\ref{inst:0047}}
\and R.        ~Geyer                         \orcit{0000-0001-6967-8707} \inst{\ref{inst:0047}}
\and P.        ~Giacobbe                      \orcit{0000-0001-7034-7024} \inst{\ref{inst:0061}}
\and G.        ~Gilmore                       \orcit{0000-0003-4632-0213} \inst{\ref{inst:0041},\ref{inst:0287}}
\and S.        ~Girona                        \orcit{0000-0002-1975-1918} \inst{\ref{inst:0288}}
\and G.        ~Giuffrida                     \orcit{0000-0002-8979-4614} \inst{\ref{inst:0123}}
\and R.        ~Gomel                                                     \inst{\ref{inst:0175}}
\and A.        ~Gomez                         \orcit{0000-0002-3796-3690} \inst{\ref{inst:0107}}
\and J.        ~Gonz\'{a}lez-N\'{u}\~{n}ez    \orcit{0000-0001-5311-5555} \inst{\ref{inst:0292}}
\and I.        ~Gonz\'{a}lez-Santamar\'{i}a   \orcit{0000-0002-8537-9384} \inst{\ref{inst:0107}}
\and E.        ~Gosset                                                    \inst{\ref{inst:0021},\ref{inst:0295}}
\and M.        ~Granvik                       \orcit{0000-0002-5624-1888} \inst{\ref{inst:0126},\ref{inst:0297}}
\and V.        ~Gregori Barrera                                           \inst{\ref{inst:0046},\ref{inst:0044},\ref{inst:0045}}
\and R.        ~Guti\'{e}rrez-S\'{a}nchez     \orcit{0009-0003-1500-4733} \inst{\ref{inst:0091}}
\and M.        ~Haywood                       \orcit{0000-0003-0434-0400} \inst{\ref{inst:0017}}
\and A.        ~Helmer                                                    \inst{\ref{inst:0211}}
\and A.        ~Helmi                         \orcit{0000-0003-3937-7641} \inst{\ref{inst:0304}}
\and K.        ~Henares                                                   \inst{\ref{inst:0163}}
\and S.L.      ~Hidalgo                       \orcit{0000-0002-0002-9298} \inst{\ref{inst:0306},\ref{inst:0307}}
\and T.        ~Hilger                        \orcit{0000-0003-1646-0063} \inst{\ref{inst:0047}}
\and D.        ~Hobbs                         \orcit{0000-0002-2696-1366} \inst{\ref{inst:0049}}
\and C.        ~Hottier                       \orcit{0000-0002-3498-3944} \inst{\ref{inst:0017}}
\and H.E.      ~Huckle                                                    \inst{\ref{inst:0027}}
\and M.        ~Jab\l{}o\'{n}ska              \orcit{0000-0001-6962-4979} \inst{\ref{inst:0312},\ref{inst:0313}}
\and F.        ~Jansen                                                    \inst{\ref{inst:0314}}
\and \'{O}.    ~Jim\'{e}nez-Arranz            \orcit{0000-0001-7434-5165} \inst{\ref{inst:0044},\ref{inst:0045},\ref{inst:0046}}
\and J.        ~Juaristi Campillo                                         \inst{\ref{inst:0038}}
\and S.        ~Khanna                        \orcit{0000-0002-2604-4277} \inst{\ref{inst:0061},\ref{inst:0304}}
\and G.        ~Kordopatis                    \orcit{0000-0002-9035-3920} \inst{\ref{inst:0039}}
\and \'{A}     ~K\'{o}sp\'{a}l                \orcit{0000-0001-7157-6275} \inst{\ref{inst:0146},\ref{inst:0058},\ref{inst:0147}}
\and Z.        ~Kostrzewa-Rutkowska                                       \inst{\ref{inst:0023}}
\and M.        ~Kun                           \orcit{0000-0002-7538-5166} \inst{\ref{inst:0146}}
\and S.        ~Lambert                       \orcit{0000-0001-6759-5502} \inst{\ref{inst:0093}}
\and A.F.      ~Lanza                         \orcit{0000-0001-5928-7251} \inst{\ref{inst:0119}}
\and J.-F.     ~Le Campion                                                \inst{\ref{inst:0040}}
\and Y.        ~Lebreton                      \orcit{0000-0002-4834-2144} \inst{\ref{inst:0330},\ref{inst:0331}}
\and S.        ~Leccia                        \orcit{0000-0001-5685-6930} \inst{\ref{inst:0332}}
\and G.        ~Lecoutre                                                  \inst{\ref{inst:0136}}
\and S.        ~Liao                          \orcit{0000-0002-9346-0211} \inst{\ref{inst:0334},\ref{inst:0061},\ref{inst:0336}}
\and L.        ~Liberato                      \orcit{0000-0003-3433-6269} \inst{\ref{inst:0039},\ref{inst:0338}}
\and E.        ~Licata                        \orcit{0000-0002-5203-0135} \inst{\ref{inst:0061}}
\and H.E.P.    ~Lindstr{\o}m                  \orcit{0009-0004-8864-5459} \inst{\ref{inst:0061},\ref{inst:0341},\ref{inst:0342}}
\and T.A.      ~Lister                        \orcit{0000-0002-3818-7769} \inst{\ref{inst:0343}}
\and E.        ~Livanou                       \orcit{0000-0003-0628-2347} \inst{\ref{inst:0112}}
\and C.        ~Loup                                                      \inst{\ref{inst:0020}}
\and L.        ~Mahy                          \orcit{0000-0003-0688-7987} \inst{\ref{inst:0018}}
\and R.G.      ~Mann                          \orcit{0000-0002-0194-325X} \inst{\ref{inst:0102}}
\and M.        ~Manteiga                      \orcit{0000-0002-7711-5581} \inst{\ref{inst:0348}}
\and J.M.      ~Marchant                      \orcit{0000-0002-3678-3145} \inst{\ref{inst:0349}}
\and M.        ~Marconi                       \orcit{0000-0002-1330-2927} \inst{\ref{inst:0332}}
\and D.        ~Mar\'{i}n Pina                \orcit{0000-0001-6482-1842} \inst{\ref{inst:0044},\ref{inst:0045},\ref{inst:0046}}
\and S.        ~Marinoni                      \orcit{0000-0001-7990-6849} \inst{\ref{inst:0123},\ref{inst:0124}}
\and D.J.      ~Marshall                      \orcit{0000-0003-3956-3524} \inst{\ref{inst:0356}}
\and J.        ~Mart\'{i}n Lozano             \orcit{0009-0001-2435-6680} \inst{\ref{inst:0100}}
\and J.M.      ~Mart\'{i}n-Fleitas            \orcit{0000-0002-8594-569X} \inst{\ref{inst:0043}}
\and G.        ~Marton                        \orcit{0000-0002-1326-1686} \inst{\ref{inst:0146}}
\and N.        ~Mary                                                      \inst{\ref{inst:0211}}
\and A.        ~Masip                         \orcit{0000-0003-1419-0020} \inst{\ref{inst:0046},\ref{inst:0044},\ref{inst:0045}}
\and D.        ~Massari                       \orcit{0000-0001-8892-4301} \inst{\ref{inst:0101}}
\and A.        ~Mastrobuono-Battisti          \orcit{0000-0002-2386-9142} \inst{\ref{inst:0017}}
\and T.        ~Mazeh                         \orcit{0000-0002-3569-3391} \inst{\ref{inst:0175}}
\and P.J.      ~McMillan                      \orcit{0000-0002-8861-2620} \inst{\ref{inst:0049}}
\and J.        ~Meichsner                     \orcit{0000-0002-9900-7864} \inst{\ref{inst:0047}}
\and S.        ~Messina                       \orcit{0000-0002-2851-2468} \inst{\ref{inst:0119}}
\and D.        ~Michalik                      \orcit{0000-0002-7618-6556} \inst{\ref{inst:0033}}
\and N.R.      ~Millar                                                    \inst{\ref{inst:0041}}
\and A.        ~Mints                         \orcit{0000-0002-8440-1455} \inst{\ref{inst:0144}}
\and D.        ~Molina                        \orcit{0000-0003-4814-0275} \inst{\ref{inst:0045},\ref{inst:0044},\ref{inst:0046}}
\and R.        ~Molinaro                      \orcit{0000-0003-3055-6002} \inst{\ref{inst:0332}}
\and L.        ~Moln\'{a}r                    \orcit{0000-0002-8159-1599} \inst{\ref{inst:0146},\ref{inst:0378},\ref{inst:0147}}
\and G.        ~Monari                        \orcit{0000-0002-6863-0661} \inst{\ref{inst:0020}}
\and M.        ~Mongui\'{o}                   \orcit{0000-0002-4519-6700} \inst{\ref{inst:0044},\ref{inst:0045},\ref{inst:0046}}
\and P.        ~Montegriffo                   \orcit{0000-0001-5013-5948} \inst{\ref{inst:0101}}
\and A.        ~Montero                                                   \inst{\ref{inst:0100}}
\and R.        ~Mor                           \orcit{0000-0002-8179-6527} \inst{\ref{inst:0386},\ref{inst:0045},\ref{inst:0046}}
\and A.        ~Mora                                                      \inst{\ref{inst:0043}}
\and R.        ~Morbidelli                    \orcit{0000-0001-7627-4946} \inst{\ref{inst:0061}}
\and T.        ~Morel                         \orcit{0000-0002-8176-4816} \inst{\ref{inst:0021}}
\and D.        ~Morris                        \orcit{0000-0002-1952-6251} \inst{\ref{inst:0102}}
\and D.        ~Munoz                                                     \inst{\ref{inst:0211}}
\and T.        ~Muraveva                      \orcit{0000-0002-0969-1915} \inst{\ref{inst:0101}}
\and C.P.      ~Murphy                                                    \inst{\ref{inst:0009}}
\and I.        ~Musella                       \orcit{0000-0001-5909-6615} \inst{\ref{inst:0332}}
\and Z.        ~Nagy                          \orcit{0000-0002-3632-1194} \inst{\ref{inst:0146}}
\and S.        ~Nieto                                                     \inst{\ref{inst:0010}}
\and L.        ~Noval                                                     \inst{\ref{inst:0211}}
\and A.        ~Ogden                                                     \inst{\ref{inst:0041}}
\and C.        ~Ordenovic                                                 \inst{\ref{inst:0039}}
\and C.        ~Pagani                        \orcit{0000-0001-5477-4720} \inst{\ref{inst:0402}}
\and I.        ~Pagano                        \orcit{0000-0001-9573-4928} \inst{\ref{inst:0119}}
\and L.        ~Palaversa                     \orcit{0000-0003-3710-0331} \inst{\ref{inst:0266}}
\and P.A.      ~Palicio                       \orcit{0000-0002-7432-8709} \inst{\ref{inst:0039}}
\and L.        ~Pallas-Quintela               \orcit{0000-0001-9296-3100} \inst{\ref{inst:0107}}
\and A.        ~Panahi                        \orcit{0000-0001-5850-4373} \inst{\ref{inst:0175}}
\and C.        ~Panem                                                     \inst{\ref{inst:0024}}
\and S.        ~Payne-Wardenaar                                           \inst{\ref{inst:0038}}
\and L.        ~Pegoraro                                                  \inst{\ref{inst:0024}}
\and A.        ~Penttil\"{ a}                 \orcit{0000-0001-7403-1721} \inst{\ref{inst:0126}}
\and P.        ~Pesciullesi                                               \inst{\ref{inst:0010}}
\and A.M.      ~Piersimoni                    \orcit{0000-0002-8019-3708} \inst{\ref{inst:0245}}
\and M.        ~Pinamonti                     \orcit{0000-0002-4445-1845} \inst{\ref{inst:0061}}
\and F.-X.     ~Pineau                        \orcit{0000-0002-2335-4499} \inst{\ref{inst:0020}}
\and E.        ~Plachy                        \orcit{0000-0002-5481-3352} \inst{\ref{inst:0146},\ref{inst:0378},\ref{inst:0147}}
\and G.        ~Plum                                                      \inst{\ref{inst:0017}}
\and E.        ~Poggio                        \orcit{0000-0003-3793-8505} \inst{\ref{inst:0039},\ref{inst:0061}}
\and D.        ~Pourbaix$^\dagger$            \orcit{0000-0002-3020-1837} \inst{\ref{inst:0140},\ref{inst:0295}}
\and A.        ~Pr\v{s}a                      \orcit{0000-0002-1913-0281} \inst{\ref{inst:0424}}
\and L.        ~Pulone                        \orcit{0000-0002-5285-998X} \inst{\ref{inst:0123}}
\and E.        ~Racero                        \orcit{0000-0002-6101-9050} \inst{\ref{inst:0100},\ref{inst:0427}}
\and M.        ~Rainer                        \orcit{0000-0002-8786-2572} \inst{\ref{inst:0054},\ref{inst:0429}}
\and C.M.      ~Raiteri                       \orcit{0000-0003-1784-2784} \inst{\ref{inst:0061}}
\and P.        ~Ramos                         \orcit{0000-0002-5080-7027} \inst{\ref{inst:0431},\ref{inst:0044},\ref{inst:0046}}
\and M.        ~Ramos-Lerate                  \orcit{0009-0005-4677-8031} \inst{\ref{inst:0091}}
\and M.        ~Ratajczak                     \orcit{0000-0002-3218-2684} \inst{\ref{inst:0312}}
\and P.        ~Re Fiorentin                  \orcit{0000-0002-4995-0475} \inst{\ref{inst:0061}}
\and S.        ~Regibo                        \orcit{0000-0001-7227-9563} \inst{\ref{inst:0149}}
\and C.        ~Reyl\'{e}                     \orcit{0000-0003-2258-2403} \inst{\ref{inst:0136}}
\and V.        ~Ripepi                        \orcit{0000-0003-1801-426X} \inst{\ref{inst:0332}}
\and A.        ~Riva                          \orcit{0000-0002-6928-8589} \inst{\ref{inst:0061}}
\and H.-W.     ~Rix                           \orcit{0000-0003-4996-9069} \inst{\ref{inst:0058}}
\and G.        ~Rixon                         \orcit{0000-0003-4399-6568} \inst{\ref{inst:0041}}
\and N.        ~Robichon                      \orcit{0000-0003-4545-7517} \inst{\ref{inst:0017}}
\and C.        ~Robin                                                     \inst{\ref{inst:0211}}
\and M.        ~Romero-G\'{o}mez              \orcit{0000-0003-3936-1025} \inst{\ref{inst:0044},\ref{inst:0045},\ref{inst:0046}}
\and N.        ~Rowell                        \orcit{0000-0003-3809-1895} \inst{\ref{inst:0102}}
\and F.        ~Royer                         \orcit{0000-0002-9374-8645} \inst{\ref{inst:0017}}
\and D.        ~Ruz Mieres                    \orcit{0000-0002-9455-157X} \inst{\ref{inst:0041}}
\and K.A.      ~Rybicki                       \orcit{0000-0002-9326-9329} \inst{\ref{inst:0451}}
\and G.        ~Sadowski                      \orcit{0000-0002-3411-1003} \inst{\ref{inst:0140}}
\and A.        ~S\'{a}ez N\'{u}\~{n}ez        \orcit{0009-0001-6078-0868} \inst{\ref{inst:0046},\ref{inst:0044},\ref{inst:0045}}
\and A.        ~Sagrist\`{a} Sell\'{e}s       \orcit{0000-0001-6191-2028} \inst{\ref{inst:0038}}
\and J.        ~Sahlmann                      \orcit{0000-0001-9525-3673} \inst{\ref{inst:0010}}
\and V.        ~Sanchez Gimenez               \orcit{0000-0003-1797-3557} \inst{\ref{inst:0046},\ref{inst:0044},\ref{inst:0045}}
\and N.        ~Sanna                         \orcit{0000-0001-9275-9492} \inst{\ref{inst:0054}}
\and R.        ~Santove\~{n}a                 \orcit{0000-0002-9257-2131} \inst{\ref{inst:0107}}
\and M.        ~Sarasso                       \orcit{0000-0001-5121-0727} \inst{\ref{inst:0061}}
\and C.        ~Sarrate Riera                                             \inst{\ref{inst:0067},\ref{inst:0044},\ref{inst:0046}}
\and E.        ~Sciacca                       \orcit{0000-0002-5574-2787} \inst{\ref{inst:0119}}
\and J.C.      ~Segovia                                                   \inst{\ref{inst:0100}}
\and D.        ~S\'{e}gransan                 \orcit{0000-0003-2355-8034} \inst{\ref{inst:0002}}
\and S.        ~Shahaf                        \orcit{0000-0001-9298-8068} \inst{\ref{inst:0451}}
\and A.        ~Siebert                       \orcit{0000-0001-8059-2840} \inst{\ref{inst:0020},\ref{inst:0472}}
\and L.        ~Siltala                       \orcit{0000-0002-6938-794X} \inst{\ref{inst:0126}}
\and E.        ~Slezak                                                    \inst{\ref{inst:0039}}
\and R.L.      ~Smart                         \orcit{0000-0002-4424-4766} \inst{\ref{inst:0061},\ref{inst:0235}}
\and O.N.      ~Snaith                        \orcit{0000-0003-1414-1296} \inst{\ref{inst:0017},\ref{inst:0478}}
\and E.        ~Solano                        \orcit{0000-0003-1885-5130} \inst{\ref{inst:0479}}
\and F.        ~Solitro                                                   \inst{\ref{inst:0079}}
\and D.        ~Souami                        \orcit{0000-0003-4058-0815} \inst{\ref{inst:0330},\ref{inst:0482}}
\and J.        ~Souchay                                                   \inst{\ref{inst:0093}}
\and L.        ~Spina                         \orcit{0000-0002-9760-6249} \inst{\ref{inst:0032}}
\and E.        ~Spitoni                       \orcit{0000-0001-9715-5727} \inst{\ref{inst:0039},\ref{inst:0486}}
\and F.        ~Spoto                         \orcit{0000-0001-7319-5847} \inst{\ref{inst:0487}}
\and L.A.      ~Squillante                                                \inst{\ref{inst:0079}}
\and I.A.      ~Steele                        \orcit{0000-0001-8397-5759} \inst{\ref{inst:0349}}
\and H.        ~Steidelm\"{ u}ller                                        \inst{\ref{inst:0047}}
\and J.        ~Surdej                        \orcit{0000-0002-7005-1976} \inst{\ref{inst:0021}}
\and L.        ~Szabados                      \orcit{0000-0002-2046-4131} \inst{\ref{inst:0146}}
\and F.        ~Taris                                                     \inst{\ref{inst:0093}}
\and M.B.      ~Taylor                        \orcit{0000-0002-4209-1479} \inst{\ref{inst:0494}}
\and R.        ~Teixeira                      \orcit{0000-0002-6806-6626} \inst{\ref{inst:0248}}
\and K.        ~Tisani\'{c}                   \orcit{0000-0001-6382-4937} \inst{\ref{inst:0266}}
\and L.        ~Tolomei                       \orcit{0000-0002-3541-3230} \inst{\ref{inst:0079}}
\and F.        ~Torra                         \orcit{0000-0002-8429-299X} \inst{\ref{inst:0067},\ref{inst:0044},\ref{inst:0046}}
\and G.        ~Torralba Elipe                \orcit{0000-0001-8738-194X} \inst{\ref{inst:0107},\ref{inst:0502},\ref{inst:0503}}
\and M.        ~Tsantaki                      \orcit{0000-0002-0552-2313} \inst{\ref{inst:0054}}
\and A.        ~Ulla                          \orcit{0000-0001-6424-5005} \inst{\ref{inst:0505},\ref{inst:0506}}
\and N.        ~Unger                         \orcit{0000-0003-3993-7127} \inst{\ref{inst:0002}}
\and O.        ~Vanel                         \orcit{0000-0002-7898-0454} \inst{\ref{inst:0017}}
\and A.        ~Vecchiato                     \orcit{0000-0003-1399-5556} \inst{\ref{inst:0061}}
\and D.        ~Vicente                       \orcit{0000-0002-1584-1182} \inst{\ref{inst:0288}}
\and S.        ~Voutsinas                                                 \inst{\ref{inst:0102}}
\and M.        ~Weiler                                                    \inst{\ref{inst:0046},\ref{inst:0044},\ref{inst:0045}}
\and \L{}.     ~Wyrzykowski                   \orcit{0000-0002-9658-6151} \inst{\ref{inst:0312}}
\and H.        ~Zhao                          \orcit{0000-0003-2645-6869} \inst{\ref{inst:0039},\ref{inst:0517}}
\and J.        ~Zorec                         \orcit{0000-0003-1257-6915} \inst{\ref{inst:0518}}
\and T.        ~Zwitter                       \orcit{0000-0002-2325-8763} \inst{\ref{inst:0519}}
\and L.        ~Balaguer-N\'{u}\~{n}ez        \orcit{0000-0001-9789-7069} \inst{\ref{inst:0046},\ref{inst:0044},\ref{inst:0045}}
\and N.        ~Leclerc                       \orcit{0009-0001-5569-6098} \inst{\ref{inst:0017}}
\and S.        ~Morgenthaler                  \orcit{0009-0005-6349-3716} \inst{\ref{inst:0555}}
\and G.        ~Robert                                                    \inst{\ref{inst:0211}}
\and S.        ~Zucker                        \orcit{0000-0003-3173-3138} \inst{\ref{inst:0192}}
}
\institute{
     Department of Physics and Astronomy G. Galilei, University of Padova, Vicolo dell'Osservatorio 3, 35122, Padova, Italy\relax                                                                                                                               \label{inst:0001}
\and Department of Astronomy, University of Geneva, Chemin Pegasi 51, 1290 Versoix, Switzerland\relax                                                                                                                                                                \label{inst:0002}
\and University of Vienna, Department of Astrophysics, T\"{ u}rkenschanzstra{\ss}e 17, A1180 Vienna, Austria\relax                                                                                                                                                   \label{inst:0004}
\and Department of Astronomy, University of Geneva, Chemin d'Ecogia 16, 1290 Versoix, Switzerland\relax                                                                                                                                                              \label{inst:0005}
\and European Space Agency (ESA), European Space Astronomy Centre (ESAC), Camino bajo del Castillo, s/n, Urbanizaci\'{o}n Villafranca del Castillo, Villanueva de la Ca\~{n}ada, 28692 Madrid, Spain\relax                                                           \label{inst:0009}
\and RHEA for European Space Agency (ESA), Camino bajo del Castillo, s/n, Urbanizaci\'{o}n Villafranca del Castillo, Villanueva de la Ca\~{n}ada, 28692 Madrid, Spain\relax                                                                                          \label{inst:0010}
\and Sednai S\`{a}rl, Geneva, Switzerland\relax                                                                                                                                                                                                                      \label{inst:0014}
\and GEPI, Observatoire de Paris, Universit\'{e} PSL, CNRS, 5 Place Jules Janssen, 92190 Meudon, France\relax                                                                                                                                                        \label{inst:0017}
\and Royal Observatory of Belgium, Ringlaan 3, 1180 Brussels, Belgium\relax                                                                                                                                                                                          \label{inst:0018}
\and Universit\'{e} de Strasbourg, CNRS, Observatoire astronomique de Strasbourg, UMR 7550,  11 rue de l'Universit\'{e}, 67000 Strasbourg, France\relax                                                                                                              \label{inst:0020}
\and Institut d'Astrophysique et de G\'{e}ophysique, Universit\'{e} de Li\`{e}ge, 19c, All\'{e}e du 6 Ao\^{u}t, B-4000 Li\`{e}ge, Belgium\relax                                                                                                                      \label{inst:0021}
\and CRAAG - Centre de Recherche en Astronomie, Astrophysique et G\'{e}ophysique, Route de l'Observatoire Bp 63 Bouzareah 16340 Algiers, Algeria\relax                                                                                                               \label{inst:0022}
\and Leiden Observatory, Leiden University, Niels Bohrweg 2, 2333 CA Leiden, The Netherlands\relax                                                                                                                                                                   \label{inst:0023}
\and CNES Centre Spatial de Toulouse, 18 avenue Edouard Belin, 31401 Toulouse Cedex 9, France\relax                                                                                                                                                                  \label{inst:0024}
\and Mullard Space Science Laboratory, University College London, Holmbury St Mary, Dorking, Surrey RH5 6NT, United Kingdom\relax                                                                                                                                    \label{inst:0027}
\and INAF - Osservatorio astronomico di Padova, Vicolo Osservatorio 5, 35122 Padova, Italy\relax                                                                                                                                                                     \label{inst:0032}
\and European Space Agency (ESA), European Space Research and Technology Centre (ESTEC), Keplerlaan 1, 2201AZ, Noordwijk, The Netherlands\relax                                                                                                                      \label{inst:0033}
\and Univ. Grenoble Alpes, CNRS, IPAG, 38000 Grenoble, France\relax                                                                                                                                                                                                  \label{inst:0036}
\and Astronomisches Rechen-Institut, Zentrum f\"{ u}r Astronomie der Universit\"{ a}t Heidelberg, M\"{ o}nchhofstr. 12-14, 69120 Heidelberg, Germany\relax                                                                                                           \label{inst:0038}
\and Universit\'{e} C\^{o}te d'Azur, Observatoire de la C\^{o}te d'Azur, CNRS, Laboratoire Lagrange, Bd de l'Observatoire, CS 34229, 06304 Nice Cedex 4, France\relax                                                                                                \label{inst:0039}
\and Laboratoire d'astrophysique de Bordeaux, Univ. Bordeaux, CNRS, B18N, all{\'e}e Geoffroy Saint-Hilaire, 33615 Pessac, France\relax                                                                                                                               \label{inst:0040}
\and Institute of Astronomy, University of Cambridge, Madingley Road, Cambridge CB3 0HA, United Kingdom\relax                                                                                                                                                        \label{inst:0041}
\and Aurora Technology for European Space Agency (ESA), Camino bajo del Castillo, s/n, Urbanizaci\'{o}n Villafranca del Castillo, Villanueva de la Ca\~{n}ada, 28692 Madrid, Spain\relax                                                                             \label{inst:0043}
\and Institut de Ci\`{e}ncies del Cosmos (ICCUB), Universitat  de  Barcelona  (UB), Mart\'{i} i  Franqu\`{e}s  1, 08028 Barcelona, Spain\relax                                                                                                                       \label{inst:0044}
\and Departament de F\'{i}sica Qu\`{a}ntica i Astrof\'{i}sica (FQA), Universitat de Barcelona (UB), c. Mart\'{i} i Franqu\`{e}s 1, 08028 Barcelona, Spain\relax                                                                                                      \label{inst:0045}
\and Institut d'Estudis Espacials de Catalunya (IEEC), c. Gran Capit\`{a}, 2-4, 08034 Barcelona, Spain\relax                                                                                                                                                         \label{inst:0046}
\and Lohrmann Observatory, Technische Universit\"{ a}t Dresden, Mommsenstra{\ss}e 13, 01062 Dresden, Germany\relax                                                                                                                                                   \label{inst:0047}
\and Lund Observatory, Division of Astrophysics, Department of Physics, Lund University, Box 43, 22100 Lund, Sweden\relax                                                                                                                                            \label{inst:0049}
\and INAF - Osservatorio Astrofisico di Arcetri, Largo Enrico Fermi 5, 50125 Firenze, Italy\relax                                                                                                                                                                    \label{inst:0054}
\and Nicolaus Copernicus Astronomical Center, Polish Academy of Sciences, ul. Bartycka 18, 00-716 Warsaw, Poland\relax                                                                                                                                               \label{inst:0055}
\and Max Planck Institute for Astronomy, K\"{ o}nigstuhl 17, 69117 Heidelberg, Germany\relax                                                                                                                                                                         \label{inst:0058}
\and INAF - Osservatorio Astrofisico di Torino, via Osservatorio 20, 10025 Pino Torinese (TO), Italy\relax                                                                                                                                                           \label{inst:0061}
\and University of Turin, Department of Physics, Via Pietro Giuria 1, 10125 Torino, Italy\relax                                                                                                                                                                      \label{inst:0063}
\and DAPCOM Data Services, c. dels Vilabella, 5-7, 80500 Vic, Barcelona, Spain\relax                                                                                                                                                                                 \label{inst:0067}
\and ALTEC S.p.a, Corso Marche, 79,10146 Torino, Italy\relax                                                                                                                                                                                                         \label{inst:0079}
\and Gaia DPAC Project Office, ESAC, Camino bajo del Castillo, s/n, Urbanizaci\'{o}n Villafranca del Castillo, Villanueva de la Ca\~{n}ada, 28692 Madrid, Spain\relax                                                                                                \label{inst:0086}
\and Telespazio UK S.L. for European Space Agency (ESA), Camino bajo del Castillo, s/n, Urbanizaci\'{o}n Villafranca del Castillo, Villanueva de la Ca\~{n}ada, 28692 Madrid, Spain\relax                                                                            \label{inst:0091}
\and SYRTE, Observatoire de Paris, Universit\'{e} PSL, CNRS, Sorbonne Universit\'{e}, LNE, 61 avenue de l'Observatoire 75014 Paris, France\relax                                                                                                                     \label{inst:0093}
\and IMCCE, Observatoire de Paris, Universit\'{e} PSL, CNRS, Sorbonne Universit{\'e}, Univ. Lille, 77 av. Denfert-Rochereau, 75014 Paris, France\relax                                                                                                               \label{inst:0094}
\and Serco Gesti\'{o}n de Negocios for European Space Agency (ESA), Camino bajo del Castillo, s/n, Urbanizaci\'{o}n Villafranca del Castillo, Villanueva de la Ca\~{n}ada, 28692 Madrid, Spain\relax                                                                 \label{inst:0100}
\and INAF - Osservatorio di Astrofisica e Scienza dello Spazio di Bologna, via Piero Gobetti 93/3, 40129 Bologna, Italy\relax                                                                                                                                        \label{inst:0101}
\and Institute for Astronomy, University of Edinburgh, Royal Observatory, Blackford Hill, Edinburgh EH9 3HJ, United Kingdom\relax                                                                                                                                    \label{inst:0102}
\and CIGUS CITIC - Department of Computer Science and Information Technologies, University of A Coru\~{n}a, Campus de Elvi\~{n}a s/n, A Coru\~{n}a, 15071, Spain\relax                                                                                               \label{inst:0107}
\and ATG Europe for European Space Agency (ESA), Camino bajo del Castillo, s/n, Urbanizaci\'{o}n Villafranca del Castillo, Villanueva de la Ca\~{n}ada, 28692 Madrid, Spain\relax                                                                                    \label{inst:0108}
\and Kavli Institute for Cosmology Cambridge, Institute of Astronomy, Madingley Road, Cambridge, CB3 0HA\relax                                                                                                                                                       \label{inst:0111}
\and Department of Astrophysics, Astronomy and Mechanics, National and Kapodistrian University of Athens, Panepistimiopolis, Zografos, 15783 Athens, Greece\relax                                                                                                    \label{inst:0112}
\and Donald Bren School of Information and Computer Sciences, University of California, Irvine, CA 92697, USA\relax                                                                                                                                                  \label{inst:0117}
\and CENTRA, Faculdade de Ci\^{e}ncias, Universidade de Lisboa, Edif. C8, Campo Grande, 1749-016 Lisboa, Portugal\relax                                                                                                                                              \label{inst:0118}
\and INAF - Osservatorio Astrofisico di Catania, via S. Sofia 78, 95123 Catania, Italy\relax                                                                                                                                                                         \label{inst:0119}
\and Dipartimento di Fisica e Astronomia ""Ettore Majorana"", Universit\`{a} di Catania, Via S. Sofia 64, 95123 Catania, Italy\relax                                                                                                                                 \label{inst:0120}
\and INAF - Osservatorio Astronomico di Roma, Via Frascati 33, 00078 Monte Porzio Catone (Roma), Italy\relax                                                                                                                                                         \label{inst:0123}
\and Space Science Data Center - ASI, Via del Politecnico SNC, 00133 Roma, Italy\relax                                                                                                                                                                               \label{inst:0124}
\and Department of Physics, University of Helsinki, P.O. Box 64, 00014 Helsinki, Finland\relax                                                                                                                                                                       \label{inst:0126}
\and Finnish Geospatial Research Institute FGI, Vuorimiehentie 5, 02150 Espoo, Finland\relax                                                                                                                                                                         \label{inst:0127}
\and Institut UTINAM CNRS UMR6213, Universit\'{e} de Franche-Comt\'{e}, OSU THETA Franche-Comt\'{e} Bourgogne, Observatoire de Besan\c{c}on, BP1615, 25010 Besan\c{c}on Cedex, France\relax                                                                          \label{inst:0136}
\and HE Space Operations BV for European Space Agency (ESA), Keplerlaan 1, 2201AZ, Noordwijk, The Netherlands\relax                                                                                                                                                  \label{inst:0137}
\and Dpto. de Inteligencia Artificial, UNED, c/ Juan del Rosal 16, 28040 Madrid, Spain\relax                                                                                                                                                                         \label{inst:0138}
\and Institut d'Astronomie et d'Astrophysique, Universit\'{e} Libre de Bruxelles CP 226, Boulevard du Triomphe, 1050 Brussels, Belgium\relax                                                                                                                         \label{inst:0140}
\and Leibniz Institute for Astrophysics Potsdam (AIP), An der Sternwarte 16, 14482 Potsdam, Germany\relax                                                                                                                                                            \label{inst:0144}
\and Konkoly Observatory, Research Centre for Astronomy and Earth Sciences, E\"{ o}tv\"{ o}s Lor{\'a}nd Research Network (ELKH), MTA Centre of Excellence, Konkoly Thege Mikl\'{o}s \'{u}t 15-17, 1121 Budapest, Hungary\relax                                       \label{inst:0146}
\and ELTE E\"{ o}tv\"{ o}s Lor\'{a}nd University, Institute of Physics, 1117, P\'{a}zm\'{a}ny P\'{e}ter s\'{e}t\'{a}ny 1A, Budapest, Hungary\relax                                                                                                                   \label{inst:0147}
\and Instituut voor Sterrenkunde, KU Leuven, Celestijnenlaan 200D, 3001 Leuven, Belgium\relax                                                                                                                                                                        \label{inst:0149}
\and Department of Astrophysics/IMAPP, Radboud University, P.O.Box 9010, 6500 GL Nijmegen, The Netherlands\relax                                                                                                                                                     \label{inst:0150}
\and Institute of Physics, Ecole Polytechnique F\'ed\'erale de Lausanne (EPFL), Observatoire de Sauverny, 1290 Versoix, Switzerland\relax                                                                                                                            \label{inst:0159}
\and Quasar Science Resources for European Space Agency (ESA), Camino bajo del Castillo, s/n, Urbanizaci\'{o}n Villafranca del Castillo, Villanueva de la Ca\~{n}ada, 28692 Madrid, Spain\relax                                                                      \label{inst:0163}
\and LASIGE, Faculdade de Ci\^{e}ncias, Universidade de Lisboa, Edif. C6, Campo Grande, 1749-016 Lisboa, Portugal\relax                                                                                                                                              \label{inst:0170}
\and School of Physics and Astronomy , University of Leicester, University Road, Leicester LE1 7RH, United Kingdom\relax                                                                                                                                             \label{inst:0171}
\and School of Physics and Astronomy, Tel Aviv University, Tel Aviv 6997801, Israel\relax                                                                                                                                                                            \label{inst:0175}
\and Cavendish Laboratory, JJ Thomson Avenue, Cambridge CB3 0HE, United Kingdom\relax                                                                                                                                                                                \label{inst:0176}
\and Telespazio for CNES Centre Spatial de Toulouse, 18 avenue Edouard Belin, 31401 Toulouse Cedex 9, France\relax                                                                                                                                                   \label{inst:0178}
\and National Observatory of Athens, I. Metaxa and Vas. Pavlou, Palaia Penteli, 15236 Athens, Greece\relax                                                                                                                                                           \label{inst:0181}
\and Depto. Estad\'istica e Investigaci\'on Operativa. Universidad de C\'adiz, Avda. Rep\'ublica Saharaui s/n, 11510 Puerto Real, C\'adiz, Spain\relax                                                                                                               \label{inst:0185}
\and EURIX S.r.l., Corso Vittorio Emanuele II 61, 10128, Torino, Italy\relax                                                                                                                                                                                         \label{inst:0191}
\and Porter School of the Environment and Earth Sciences, Tel Aviv University, Tel Aviv 6997801, Israel\relax                                                                                                                                                        \label{inst:0192}
\and ATOS for CNES Centre Spatial de Toulouse, 18 avenue Edouard Belin, 31401 Toulouse Cedex 9, France\relax                                                                                                                                                         \label{inst:0193}
\and HE Space Operations BV for European Space Agency (ESA), Camino bajo del Castillo, s/n, Urbanizaci\'{o}n Villafranca del Castillo, Villanueva de la Ca\~{n}ada, 28692 Madrid, Spain\relax                                                                        \label{inst:0195}
\and LFCA/DAS,Universidad de Chile,CNRS,Casilla 36-D, Santiago, Chile\relax                                                                                                                                                                                          \label{inst:0197}
\and SISSA - Scuola Internazionale Superiore di Studi Avanzati, via Bonomea 265, 34136 Trieste, Italy\relax                                                                                                                                                          \label{inst:0202}
\and University of Turin, Department of Computer Sciences, Corso Svizzera 185, 10149 Torino, Italy\relax                                                                                                                                                             \label{inst:0210}
\and Thales Services for CNES Centre Spatial de Toulouse, 18 avenue Edouard Belin, 31401 Toulouse Cedex 9, France\relax                                                                                                                                              \label{inst:0211}
\and Dpto. de Matem\'{a}tica Aplicada y Ciencias de la Computaci\'{o}n, Univ. de Cantabria, ETS Ingenieros de Caminos, Canales y Puertos, Avda. de los Castros s/n, 39005 Santander, Spain\relax                                                                     \label{inst:0213}
\and Institut de F\'{i}sica d'Altes Energies (IFAE), The Barcelona Institute of Science and Technology, Campus UAB, 08193 Bellaterra (Barcelona), Spain\relax                                                                                                        \label{inst:0219}
\and Port d'Informaci\'{o} Cient\'{i}fica (PIC), Campus UAB, C. Albareda s/n, 08193 Bellaterra (Barcelona), Spain\relax                                                                                                                                              \label{inst:0220}
\and Instituto de Astrof\'{i}sica, Universidad Andres Bello, Fernandez Concha 700, Las Condes, Santiago RM, Chile\relax                                                                                                                                              \label{inst:0228}
\and Centre for Astrophysics Research, University of Hertfordshire, College Lane, AL10 9AB, Hatfield, United Kingdom\relax                                                                                                                                           \label{inst:0235}
\and University of Turin, Mathematical Department ""G.Peano"", Via Carlo Alberto 10, 10123 Torino, Italy\relax                                                                                                                                                       \label{inst:0239}
\and University of Antwerp, Onderzoeksgroep Toegepaste Wiskunde, Middelheimlaan 1, 2020 Antwerp, Belgium\relax                                                                                                                                                       \label{inst:0242}
\and INAF - Osservatorio Astronomico d'Abruzzo, Via Mentore Maggini, 64100 Teramo, Italy\relax                                                                                                                                                                       \label{inst:0245}
\and Instituto de Astronomia, Geof\`{i}sica e Ci\^{e}ncias Atmosf\'{e}ricas, Universidade de S\~{a}o Paulo, Rua do Mat\~{a}o, 1226, Cidade Universitaria, 05508-900 S\~{a}o Paulo, SP, Brazil\relax                                                                  \label{inst:0248}
\and M\'{e}socentre de calcul de Franche-Comt\'{e}, Universit\'{e} de Franche-Comt\'{e}, 16 route de Gray, 25030 Besan\c{c}on Cedex, France\relax                                                                                                                    \label{inst:0254}
\and Ru{\dj}er Bo\v{s}kovi\'{c} Institute, Bijeni\v{c}ka cesta 54, 10000 Zagreb, Croatia\relax                                                                                                                                                                       \label{inst:0266}
\and Astrophysics Research Centre, School of Mathematics and Physics, Queen's University Belfast, Belfast BT7 1NN, UK\relax                                                                                                                                          \label{inst:0268}
\and Data Science and Big Data Lab, Pablo de Olavide University, 41013, Seville, Spain\relax                                                                                                                                                                         \label{inst:0281}
\and Institute of Astrophysics, FORTH, Crete, Greece\relax                                                                                                                                                                                                           \label{inst:0287}
\and Barcelona Supercomputing Center (BSC), Pla\c{c}a Eusebi G\"{ u}ell 1-3, 08034-Barcelona, Spain\relax                                                                                                                                                            \label{inst:0288}
\and ETSE Telecomunicaci\'{o}n, Universidade de Vigo, Campus Lagoas-Marcosende, 36310 Vigo, Galicia, Spain\relax                                                                                                                                                     \label{inst:0292}
\and F.R.S.-FNRS, Rue d'Egmont 5, 1000 Brussels, Belgium\relax                                                                                                                                                                                                       \label{inst:0295}
\and Asteroid Engineering Laboratory, Lule\aa{} University of Technology, Box 848, S-981 28 Kiruna, Sweden\relax                                                                                                                                                     \label{inst:0297}
\and Kapteyn Astronomical Institute, University of Groningen, Landleven 12, 9747 AD Groningen, The Netherlands\relax                                                                                                                                                 \label{inst:0304}
\and IAC - Instituto de Astrofisica de Canarias, Via L\'{a}ctea s/n, 38200 La Laguna S.C., Tenerife, Spain\relax                                                                                                                                                     \label{inst:0306}
\and Department of Astrophysics, University of La Laguna, Via L\'{a}ctea s/n, 38200 La Laguna S.C., Tenerife, Spain\relax                                                                                                                                            \label{inst:0307}
\and Astronomical Observatory, University of Warsaw,  Al. Ujazdowskie 4, 00-478 Warszawa, Poland\relax                                                                                                                                                               \label{inst:0312}
\and Research School of Astronomy \& Astrophysics, Australian National University, Cotter Road, Weston, ACT 2611, Australia\relax                                                                                                                                     \label{inst:0313}
\and European Space Agency (ESA, retired), European Space Research and Technology Centre (ESTEC), Keplerlaan 1, 2201AZ, Noordwijk, The Netherlands\relax                                                                                                             \label{inst:0314}
\and LESIA, Observatoire de Paris, Universit\'{e} PSL, CNRS, Sorbonne Universit\'{e}, Universit\'{e} de Paris, 5 Place Jules Janssen, 92190 Meudon, France\relax                                                                                                     \label{inst:0330}
\and Universit\'{e} Rennes, CNRS, IPR (Institut de Physique de Rennes) - UMR 6251, 35000 Rennes, France\relax                                                                                                                                                        \label{inst:0331}
\and INAF - Osservatorio Astronomico di Capodimonte, Via Moiariello 16, 80131, Napoli, Italy\relax                                                                                                                                                                   \label{inst:0332}
\and Shanghai Astronomical Observatory, Chinese Academy of Sciences, 80 Nandan Road, Shanghai 200030, People's Republic of China\relax                                                                                                                               \label{inst:0334}
\and University of Chinese Academy of Sciences, No.19(A) Yuquan Road, Shijingshan District, Beijing 100049, People's Republic of China\relax                                                                                                                         \label{inst:0336}
\and S\~{a}o Paulo State University, Grupo de Din\^{a}mica Orbital e Planetologia, CEP 12516-410, Guaratinguet\'{a}, SP, Brazil\relax                                                                                                                                \label{inst:0338}
\and Niels Bohr Institute, University of Copenhagen, Juliane Maries Vej 30, 2100 Copenhagen {\O}, Denmark\relax                                                                                                                                                      \label{inst:0341}
\and DXC Technology, Retortvej 8, 2500 Valby, Denmark\relax                                                                                                                                                                                                          \label{inst:0342}
\and Las Cumbres Observatory, 6740 Cortona Drive Suite 102, Goleta, CA 93117, USA\relax                                                                                                                                                                              \label{inst:0343}
\and CIGUS CITIC, Department of Nautical Sciences and Marine Engineering, University of A Coru\~{n}a, Paseo de Ronda 51, 15071, A Coru\~{n}a, Spain\relax                                                                                                            \label{inst:0348}
\and Astrophysics Research Institute, Liverpool John Moores University, 146 Brownlow Hill, Liverpool L3 5RF, United Kingdom\relax                                                                                                                                    \label{inst:0349}
\and IRAP, Universit\'{e} de Toulouse, CNRS, UPS, CNES, 9 Av. colonel Roche, BP 44346, 31028 Toulouse Cedex 4, France\relax                                                                                                                                          \label{inst:0356}
\and MTA CSFK Lend\"{ u}let Near-Field Cosmology Research Group, Konkoly Observatory, MTA Research Centre for Astronomy and Earth Sciences, Konkoly Thege Mikl\'{o}s \'{u}t 15-17, 1121 Budapest, Hungary\relax                                                      \label{inst:0378}
\and Pervasive Technologies s.l., c. Saragossa 118, 08006 Barcelona, Spain\relax                                                                                                                                                                                     \label{inst:0386}
\and School of Physics and Astronomy, University of Leicester, University Road, Leicester LE1 7RH, United Kingdom\relax                                                                                                                                              \label{inst:0402}
\and Villanova University, Department of Astrophysics and Planetary Science, 800 E Lancaster Avenue, Villanova PA 19085, USA\relax                                                                                                                                   \label{inst:0424}
\and Departmento de F\'{i}sica de la Tierra y Astrof\'{i}sica, Universidad Complutense de Madrid, 28040 Madrid, Spain\relax                                                                                                                                          \label{inst:0427}
\and INAF - Osservatorio Astronomico di Brera, via E. Bianchi, 46, 23807 Merate (LC), Italy\relax                                                                                                                                                                    \label{inst:0429}
\and National Astronomical Observatory of Japan, 2-21-1 Osawa, Mitaka, Tokyo 181-8588, Japan\relax                                                                                                                                                                   \label{inst:0431}
\and Department of Particle Physics and Astrophysics, Weizmann Institute of Science, Rehovot 7610001, Israel\relax                                                                                                                                                   \label{inst:0451}
\and Centre de Donn\'{e}es Astronomique de Strasbourg, Strasbourg, France\relax                                                                                                                                                                                      \label{inst:0472}
\and University of Exeter, School of Physics and Astronomy, Stocker Road, Exeter, EX2 7SJ, United Kingdom\relax                                                                                                                                                      \label{inst:0478}
\and Departamento de Astrof\'{i}sica, Centro de Astrobiolog\'{i}a (CSIC-INTA), ESA-ESAC. Camino Bajo del Castillo s/n. 28692 Villanueva de la Ca\~{n}ada, Madrid, Spain\relax                                                                                        \label{inst:0479}
\and naXys, Department of Mathematics, University of Namur, Rue de Bruxelles 61, 5000 Namur, Belgium\relax                                                                                                                                                           \label{inst:0482}
\and INAF. Osservatorio Astronomico di Trieste, via G.B. Tiepolo 11, 34131, Trieste, Italy\relax                                                                                                                                                                     \label{inst:0486}
\and Harvard-Smithsonian Center for Astrophysics, 60 Garden St., MS 15, Cambridge, MA 02138, USA\relax                                                                                                                                                               \label{inst:0487}
\and H H Wills Physics Laboratory, University of Bristol, Tyndall Avenue, Bristol BS8 1TL, United Kingdom\relax                                                                                                                                                      \label{inst:0494}
\and Escuela de Arquitectura y Polit\'{e}cnica - Universidad Europea de Valencia, Spain\relax                                                                                                                                                                        \label{inst:0502}
\and Escuela Superior de Ingenier\'{i}a y Tecnolog\'{i}a - Universidad Internacional de la Rioja, Spain\relax                                                                                                                                                        \label{inst:0503}
\and Applied Physics Department, Universidade de Vigo, 36310 Vigo, Spain\relax                                                                                                                                                                                       \label{inst:0505}
\and Instituto de F{'i}sica e Ciencias Aeroespaciais (IFCAE), Universidade de Vigo‚ \'{A} Campus de As Lagoas, 32004 Ourense, Spain\relax                                                                                                                            \label{inst:0506}
\and Purple Mountain Observatory, Chinese Academy of Sciences, Nanjing 210023, China\relax                                                                                                                                                                           \label{inst:0517}
\and Sorbonne Universit\'{e}, CNRS, UMR7095, Institut d'Astrophysique de Paris, 98bis bd. Arago, 75014 Paris, France\relax                                                                                                                                           \label{inst:0518}
\and Faculty of Mathematics and Physics, University of Ljubljana, Jadranska ulica 19, 1000 Ljubljana, Slovenia\relax                                                                                                                                                 \label{inst:0519}
\and Institute of Mathematics, Ecole Polytechnique F\'ed\'erale de Lausanne (EPFL), Switzerland\relax
\label{inst:0555}
}

\date{Received 26 June 2023 / Accepted 11 August 2023}

\abstract{
    The third \Gaia Data Release (DR3) provided photometric time series of more than 2 million long-period variable (LPV) candidates. Anticipating the publication of full radial-velocity data planned with Data Release 4, this Focused Product Release (FPR) provides radial-velocity time series for a selection of LPV candidates with high-quality observations.
}{
    We describe the production and content of the \Gaia catalog of LPV radial-velocity time series, and the methods used to compute the variability parameters published as part of the \Gaia FPR.
}{
    Starting from the DR3 catalog of LPV candidates, we applied several filters to construct a sample of sources with high-quality radial-velocity measurements. We modeled their radial-velocity and photometric time series to derive their periods and amplitudes, and further refined the sample by requiring compatibility between the radial-velocity period and at least one of the $G$, $\gbp$, or $\grp$ photometric periods.
}{
    The catalog includes radial-velocity time series and variability parameters for 9\,614 sources in the magnitude range $6\lesssim G/\mags\lesssim 14$, including a flagged top-quality subsample of 6\,093 stars whose radial-velocity periods are fully compatible with the values derived from the $G$, $\gbp$, and $\grp$ photometric time series.
    The radial-velocity time series contain a mean of 24 measurements per source taken unevenly over a duration of about three years.
    We identify the great majority of the sources (88\%) as genuine LPV candidates, with about half of them showing a pulsation period and the other half displaying a long secondary period. The remaining 12\% of the catalog consists of candidate ellipsoidal binaries. Quality checks against radial velocities available in the literature show excellent agreement. We provide some illustrative examples and cautionary remarks.
}{
    The publication of radial-velocity time series for almost ten thousand LPV candidates constitutes, by far, the largest such database available to date in the literature. The availability of simultaneous photometric measurements gives a unique added value to the \Gaia catalog.
}

\keywords{stars: variables - stars: AGB and post-AGB - stars: carbon - catalogs - methods: data analysis - techniques: radial velocities}

\maketitle

\section{Introduction}
\label{sec:introduction}

Evolved stars of low and intermediate mass show various kinds of light variability summarized in the class of long period variables (LPVs). Within this class, there are radially pulsating objects showing small to large light amplitudes in various pulsation modes and with various degrees of periodicity, but also stars whose light variability is due to the presence of a binary or due to eclipses of orbiting dust clouds. For disentangling the various causes for variability in these stars, sometimes even occurring in combinations, contemporaneous monitoring of radial-velocity (RV) variations has proven to be a useful approach.

Early measurements of RV variations in LPVs date back to the 1920s \citep{joy_1926}. It was noted already then that emission and absorption lines in Miras show different kinds of velocity variations. This was supported by several further studies, all using lines in the blue part of the spectrum, but the observed variability pattern did not allow for a conclusive description of the pulsation in these stars \citep{joy_1954, reid_dickinson_1976}. A major step forward was achieved by the first monitoring of RV changes in the near-infrared. The landmark paper by \citet{hhr_1982} revealed the photospheric kinematics for the Mira \object{$\chi$\,Cyg}, allowing for components related to stellar pulsation and to mass outflow to be identified, respectively. Line doubling of high-excitation CO lines was found near light maximum and, together with the appearance of hydrogen emission lines at those phases, interpreted as a trace of shock fronts. Combining velocity data from the violet to the radio regime allowed for a stratigraphy of a Mira's atmosphere to be constructed out to its circumstellar layers \citep{wallerstein_1985}.

Measurements of velocity amplitudes in Mira variables have played a key role in the discussion on the pulsation mode of these stars \citep[see][for a summary]{wood_sebo_1996}. In addition, these observations constrained dynamical models of LPV atmospheres and led to today's understanding that the levitation of the outer layers of the stellar atmosphere driven by pulsation is essential for driving an efficient mass loss during this evolutionary phase \citep[e.g.,][]{hoefner_olofsson_2018}. 

Since the periods of LPVs can reach values of a few hundred days, obtaining velocity curves at high resolution with a good phase coverage remained challenging. The total number of Miras with such datasets available in the literature is still limited to a few tens \citep{hsh_1984, hinkle_barnbaum_1996, lhh_1999, alvarez_etal_2001, lebzelter_etal_2005a, lebzelter_etal_2005b}. However, this sample covers a wide range in period, metallicity, and chemistry, revealing a consistent pattern in the velocity variations with s-shaped velocity curves in the near-infrared and peak-to-peak velocity amplitudes, depending on the lines used to trace the variation, between 20 and 30 $\kms$ \citep{lh_2002,nowotny_etal_2010}.

For physical and observational reasons, most of these studies were done in the 1.6 or 2.2\,$\mu$m range relying on the first and second overtone lines of CO. These lines trace parts of the stellar atmosphere close to the pulsation driving zone \citep{nowotny_etal_2010}. Within the spectrum, they are located close to the maximum of the spectral energy distribution of Miras and in an area with comparably low line blending and telluric absorption. Atomic lines in the same spectral range show a behavior very similar to the molecular lines \citep{hinkle_barnes_1979}. Velocity time series from the 4000\,{\AA} region show a much less expressed pattern with an amplitude of only 8 $\kms$. In the 4000 to 6800\,{\AA} range covered in the study of \citet{alvarez_etal_2001}, amplitudes around 20 $\kms$ were measured, and thus the lines in this range compare well with the near-infrared range.

The semiregular variables (SRVs) show significantly smaller light amplitudes and most of them are pulsating in an overtone mode \citep{wood_sebo_1996}. Consequently, RV amplitudes are expected to be smaller for these stars, which has been confirmed observationally for SRVs with light amplitudes ranging from 0.1 to more than 2.5 mag \citep{lebzelter_1999,lebzelter_etal_2005a}. For the small amplitude and short period end, velocities of 1 to 5 $\kms$ have been reported. Some stars have characteristics somehow between SRVs and Miras, such as \object{W Hya} (with a period of 390 d and an amplitude of more than 2 mag in $V$), and reach velocity amplitudes around 10 $\kms$. Semiregular light variability is typically reflected in the velocity variations.

From the point of view of RV variations, the ellipsoidal variables form a group of special interest among the LPVs \citep{soszynski_etal_2004}. From their location in the period-luminosity diagram of LPVs, these stars are also known as sequence E stars. They are close binaries with one object being a red giant and the other one typically being a main sequence star. While there is no visible eclipse, regardless of it being due to an angle of orbital inclination that is too steep or due to the red giant being orders of magnitude brighter than the companion, there is a gravitational distortion of the red giant, which fills the Roche lobe. This produces an elongated shape of the object, and as the star rotates, brightness variations are observed due to this asymmetry. 

As a consequence, the light and RV curves of these stars show two light cycles, but only one velocity cycle within one orbital period \citep{nicholls_wood_2010}. \citet{nie_wood_2014} presented an extensive database of RV curves for 81 ellipsoidal variables. About 20\% of these systems show eccentric orbits, a fraction twice as high as derived from light-curve analysis alone \citep{nie_etal_2017}, which stresses the importance of RV data for the understanding of these variables. During their further evolution, the unseen companion will gain mass from the red giant leading to a common envelope system at some point. Ellipsoidal variables are assumed to be precursors of close binary planetary nebulae \citep{nicholls_wood_2010}.

Another group of binaries among the LPVs are the symbiotic stars consisting of a red giant and a degenerated star such as a white dwarf or a neutron star. In the case of D-type symbiotics, the evolved star is a Mira \citep{hinkle_etal_2013}. Radial-velocity changes thus combine pulsation and orbital motion. However, orbital periods of these systems are typically longer than decades \citep{seaquist_taylor_1990} and they are therefore difficult to detect even in long velocity time series.

Finally, RV curves play a critical role in the explanation of the mysterious sequence-D stars. These LPVs show radial pulsation in some overtone modes combined with a  secondary period that is typically ten times longer. Fundamental mode pulsation has been excluded as the cause of this secondary period \citep{wood_2000}. Binarity and strange modes were suggested as alternative solutions. Interestingly, these long periods seem to form a period-luminosity sequence by themselves.

The origin of this kind of variability remains a matter of debate. \citet{nicholls_wood_2010} showed that sequence-D stars are not ellipsoidals.  From an attempt to model the velocity curves of a small sample of sequence-D stars, \citet{hinkle_etal_2002} concluded that binarity is unlikely the cause of the variation because almost all of the objects analyzed show extremely similar values for the orbital parameters K, e, and $\omega$.  \citet{soszynski_udalski_2014} and \citet{soszynski_etal_2021} suggest from a careful analysis of light curves and infrared data that sequence-D variability can be explained by an orbiting dust cloud in combination with a low-mass companion in a close circular orbit. On the other hand, \citet{saio_etal_2015} show that the sequence-D period-luminosity relation agrees with expectations from oscillatory convective modes. 

The observation of reliable RV curves of LPVs plays an important role for interpreting various aspects of these stars and their evolution. Observational challenges have limited the collection of large datasets up to now. Considering the variety of objects found among LPVs, the small existing dataset remains insufficient.

Since its second data release, \Gaia has provided high-quality data for the study of the variability of LPVs, with the publication of photometric time series in the $G$, $\gbp$, and $\grp$ bands of $\sim$150\,000 candidate LPVs in the second data release and over 2 million candidate LPVs in the \Gaia Data Release 3 (DR3), respectively \citep{dr2_lpv,dr3_lpv}. Moreover, \Gaia has the unique capability of simultaneously obtaining photometric and spectroscopic measurements owing to its \textit{Radial Velocity Spectrometer} (RVS), thereby substantially boosting the possibility to investigate stellar variability. This feature was first exploited with the publication of RV time series of Cepheids and RR Lyrae as part of \Gaia DR3 \citep{dr3_cep,dr3_rrl}. Here we extend this dataset to an additional 9\,614 sources that are part of the \Gaia DR3 catalog of LPV candidates. In Sect.~\ref{sec:catalog_construction} we describe the procedures involved in the construction of this FPR catalog, while we present its content in Sect.~\ref{sec:catalog_content} and discuss its quality in Sect.~\ref{sec:catalog_quality}. In Sect.~\ref{sec:catalog_overview} we give an overview of the catalog, while Sect.~\ref{sec:summary_and_conclusions} is dedicated to a summary and to conclusions.

Several Appendices complete the main body of the text. Appendix~\ref{asec:lsp} gives additional details on the classification of LPV candidates presented in Sect.~\ref{sec:catalog_content}. Appendix~\ref{app:median_RV_in_unevenly_sampled_time_series} illustrates cases where the median RV differs significantly from the systemic RV. Appendix~\ref{asec:numerical_differences_with_respect_to_gaia_dr3} analyzes the impact of the Java bug mentioned in Sect.~\ref{sec:catalog_construction:time_series_processing} on the LPV results published in DR3. Finally, Appendix~\ref{app:catalog_retrieval} gives some example queries to retrieve the data of the present catalog from the \Gaia archive.

\begin{table*}
\caption{Summary of the steps involved in the construction of the catalog. The exact criterion by which two periods are considered ``similar'' (indicated by $P_1\simeq P_2$) is described in Sect.~\ref{sec:catalog_construction:post_filtering:filter23}}
\label{tab:filter}
\centering
\begin{tabular}{lllr}
\hline\hline
Steps & Sub-steps & Selection & $\#$ sources \\
\hline
\multirow{3}{*}{Starting sample}
    & \texttt{DR3-LPV} & 2$^{nd}$ \Gaia LPV catalog & 1\,720\,588 \\
    & \texttt{DR3-RV} & radial velocity published in DR3 & 33\,812\,183 \\
    & \hspace{3mm} \texttt{DR3-LPV-RV} & \texttt{DR3-LPV} and \texttt{DR3-RV} & 501\,308 \\
\hline
\multirow{4}{*}{Pre-filter}
    & \texttt{bright} & \texttt{DR3-LPV-RV} and $\grvs<12\,\mags$ & 249\,600 \\
    & \texttt{vis-periods} & \texttt{DR3-LPV-RV} and \texttt{rv\_visibility\_periods\_used} $\geq12$ & 180\,850 \\
    & \texttt{small-err-rv} & \texttt{DR3-LPV-RV} and  $\erv<0.175\,\times$ \texttt{rv\_amplitude\_robust} & 224\,696 \\
    & \hspace{3mm} \texttt{filter-0} & \texttt{bright} and \texttt{vis-periods} and \texttt{small-err-rv} & 110\,654 \\
\hline
\texttt{processing} & \multicolumn{3}{c}{Outliers removal, computation of best-fit model and period of the RV time series.} \\
\hline
\multirow{16}{*}{Post-filter}
    & \texttt{num-outliers} & \texttt{filter-0} and number of outliers in the RV time series $\noutrv\leq1$ & 105\,715 \\
    & \texttt{no-trend} & \texttt{filter-0} and RV model's polynomial degree $D_p=0$ & 90\,942 \\
    & \texttt{high-snrv} & \texttt{filter-0} and $\snrv>1.5$ & 58\,725 \\
    & \hspace{3mm} \texttt{filter-1} & \texttt{no-trend} and \texttt{num-outliers} and \texttt{high-snrv} & 44\,216 \\
    \cmidrule(lr){2-4}
    & \texttt{Prv-low-limit} & \texttt{filter-1} and $\prv>35\,{\rm days}$ & 24\,118 \\
    & \texttt{Prv-up-limit} & \texttt{filter-1} and $\prv<\durationrv$ & 43\,621 \\
    & \hspace{3mm} \texttt{filter-2} & \texttt{Prv-low-limit} and \texttt{Prv-up-limit} & 23\,523 \\
    \cmidrule(lr){2-4}
    & \texttt{Prv-sim-Pg} & \texttt{filter-2} and $\prv\simeq\pg$ & 6\,392 \\
    & \texttt{Prv-sim-Pbp} & \texttt{filter-2} and $\prv\simeq\pbp$ & 5\,768 \\
    & \texttt{Prv-sim-Prp} & \texttt{filter-2} and $\prv\simeq\prp$ & 6\,534 \\
    & \texttt{Prv-sim-any-Pph} & \texttt{filter-2} and ($\prv\simeq\pg$ or $\prv\simeq\pbp$ or $\prv\simeq\prp$) & 7\,551 \\
    & \texttt{Prv-sim-2Pg} & \texttt{filter-2} and $\prv\simeq2\pg$ & 1\,701 \\
    & \texttt{Prv-sim-2Pbp} & \texttt{filter-2} and $\prv\simeq2\pbp$ & 1\,646 \\
    & \texttt{Prv-sim-2Prp} & \texttt{filter-2} and $\prv\simeq2\prp$ & 1\,703 \\
    & \texttt{Prv-sim-any-2Pph} & \texttt{filter-2} and ($\prv\simeq2\pg$ or $\prv\simeq2\pbp$ or $\prv\simeq2\prp$) & 2\,242 \\
    & \hspace{3mm} \texttt{filter-3}$^{(a)}$ & (\texttt{Prv-sim-any-Pph} or \texttt{Prv-sim-any-2Pph}) & 9\,614 \\
\hline
    & \texttt{Prv-sim-Pph+2Pph} & \texttt{filter-3} and (\texttt{Prv-sim-any-Pph} and \texttt{Prv-sim-any-2Pph}) & 179 \\
    & \texttt{Prv-sim-Pph-only} & \texttt{filter-3} and (\texttt{Prv-sim-any-Pph} and (not \texttt{Prv-sim-any-2Pph})) & 7\,372 \\
Additional 
    & \texttt{Prv-sim-2Pph-only} & \texttt{filter-3} and ((not \texttt{Prv-sim-any-Pph}) and \texttt{Prv-sim-any-2Pph}) & 2\,063 \\
statistics 
    & \texttt{Prv-sim-all-Pph} & \texttt{filter-3} and ($\prv\simeq\pg$ and $\prv\simeq\pbp$ and $\prv\simeq\prp$) & 4\,899 \\
    & \texttt{Prv-sim-all-2Pph} & \texttt{filter-3} and ($\prv\simeq2\pg$ and $\prv\simeq2\pbp$ and $\prv\simeq2\prp$) & 1\,194 \\
    & \hspace{3mm} \texttt{top-quality}$^{(b)}$ & (\texttt{Prv-sim-all-Pph} or \texttt{Prv-sim-all-2Pph}) & 6\,093 \\\hline
\end{tabular}
\tablefoot{$^{(a)}$ All sources belonging to the \texttt{filter-3} subset are published as part of the FPR. $^{(b)}$ The sources belonging to the subset \texttt{top-quality} are identified by the flag \rvflag.}
\end{table*}

\section{Catalog construction}
\label{sec:catalog_construction}

Our starting dataset is the 2$^{\rm nd}$ \Gaia catalog of LPV candidates \citep{dr3_lpv} published as part of the \Gaia DR3 \citep{dr3}. More precisely, we consider the sources that appear in the table \texttt{gaiadr3.vari\_long\_period\_variable} of the \Gaia Archive.
All these sources have their photometric time series already published in DR3.
More than 70\% of them do not have median RV published in \Gaia DR3, most likely because they are too faint \citep[see][]{dr3_katz_etal_2023}. Therefore, we discard these sources, and focus on the remaining 501\,308 LPV candidates having RV data in \Gaia DR3. Hereinafter we adopt the notation $\pubrv$ to indicate the median RV published as part of \Gaia DR3 (it corresponds to the field \texttt{radial\_velocity} of the \texttt{gaiadr3.gaia\_source} table in the \Gaia archive\footnote{
    \url{https://gea.esac.esa.int/archive/}
}).

For the construction of the catalog, we proceed in three main steps. To begin with, we employ the quantities derived from the \Gaia RVS, and published as part of \Gaia DR3, to refine the input source list to be fed to the processing pipeline. We refer to this first step as ``pre-filtering,'' and describe it in Sect.~\ref{sec:catalog_construction:pre_filtering}.

We then analyze the time series of the selected sources with an updated version of the pipeline used for variability processing in \Gaia DR3 \citep{dr3_var,dr3_lpv}, as we describe in Sect.~\ref{sec:catalog_construction:time_series_processing}. Both the RV time series and the three photometric time series (in the \Gaia $G$, $\gbp$, and $\grp$ bands) undergo this ``processing'' step, that involves the detection and removal of outlier epochs, the computation of time series statistics, and the determination of the best-fit model.

Lastly, we employ the resulting quantities to further refine the sample of sources for publication. This final step is referred to as ``post-filtering,'' and is described in Sect.~\ref{sec:catalog_construction:post_filtering}. The filtering conditions and number of selected sources of each step and the corresponding sub-steps are summarized in Table~\ref{tab:filter}.

\subsection{Pre-filtering}
\label{sec:catalog_construction:pre_filtering}

\begin{figure}
\centering
\includegraphics[width=\columnwidth]{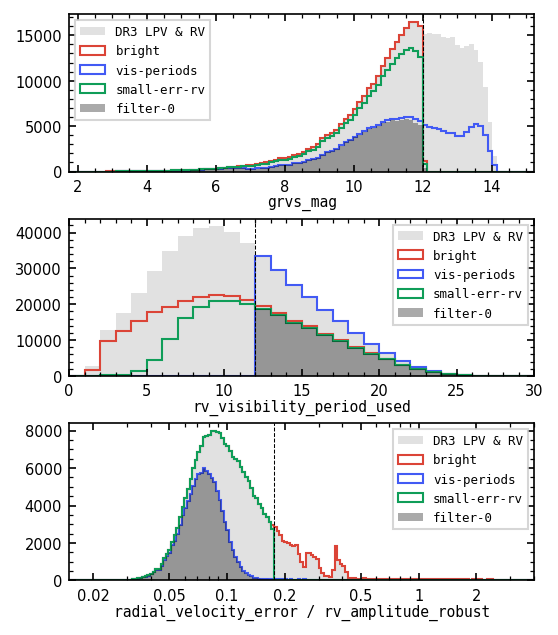}
\caption{Distribution of the pre-filtering parameters $\grvs$ (top panel), \texttt{rv\_visibility\_periods\_used} (middle panel), and $\erv/\texttt{rv\_amplitude\_robust}$ (bottom panel), with vertical dashed lines indicating the filter limits. Different colors indicate the starting set (\texttt{DR3-LPV-RV}, gray filled histogram) and the individual pre-filters, labeled \texttt{bright} (red curve), \texttt{vis-periods} (blue curve), and \texttt{small-err-rv} (green curve) as in Table~\ref{tab:filter}. The black filled histogram corresponds to the application of all three pre-filters (\texttt{filter-0}). We note that 50\,349 sources in the starting sample of 501\,308 sources lack a published value of $\grvs$ as it would be fainter than $14.1\,\mags$, and that the quantity \texttt{rv\_amplitude\_robust} is not provided for sources with $\grvs>12\,\mags$ \citep{dr3_sartoretti_etal_2023}.}
\label{fig:hist_prefilter}
\end{figure}

At this stage we aim to limit the sample to the objects with the highest-quality RV measurements by taking advantage of the information available from \Gaia DR3 (namely in the \texttt{gaiadr3.gaia\_source} table of the \Gaia Archive). This is achieved by retaining only sources with large enough RVS flux, a sufficient number of RV measurements, and relatively small RV uncertainty $\erv$. The relevant quantities and corresponding cuts involved in this process are illustrated in Fig.~\ref{fig:hist_prefilter}.

We begin by applying a filter that excludes the faintest objects in our dataset, using the median value $\grvs$ of the epoch $G_{\rm RVS}^t$ magnitudes \citep[\texttt{grvs\_mag} in the \Gaia Archive, see][]{dr3_sartoretti_etal_2023}, which are obtained by integration of the RVS epoch spectra. By requiring that $\grvs<12\,\mags$ we limit our sample to ``bright'' stars (top panel of Fig.~\ref{fig:hist_prefilter}) following the distinction made for the DR3 RVS processing \citep{dr3_katz_etal_2023}. Almost 250\,000 sources meet this criterion. It is worth pointing out that several RV-related quantities published in DR3, such as the median RV and its uncertainty, are computed with different methods depending on whether the sources are brighter or fainter than $\grvs=12\,\mags$. Having required that $\grvs<12\,\mags$, these quantities are defined unequivocally for all sources in our sample. Namely, the RV is obtained as the median of the single-transit RVs, while the RV error is the uncertainty on the median of the epoch RVs, with a constant offset accounting for a calibration floor contribution \citep{sartoretti_etal_2022_dr3_doc_ch6}.

Then, we apply a condition to the number of data points in each RV time series. We note that the actual number of RV observations is not necessarily appropriate for this filtering step, as they often come in groups that span a relatively short period of time (often shorter than several days) because of the \Gaia scanning law \citep[see][]{dr1_var}. This issue is often overcome through the concept of visibility period, that is a group of transits separated from other such groups by a gap of at least 4 days. The number of visibility periods used in the derivation of radial velocities is a parameter available from DR3, and we employ it to set the condition \texttt{rv\_visibility\_periods\_used} $\geq12$. As will be explained in Sect.~\ref{sec:catalog_construction:time_series_processing}, a minimum number of 9 data points is necessary to obtain a time series model, but that may still not be enough for the model to be well-constrained. At the same time, raising too much the threshold would lead to the exclusion of too many sources, as can be appreciated from the middle panel of Fig.~\ref{fig:hist_prefilter}. We found that a good compromise could be attained by setting the threshold at 12. The condition on visibility periods is fulfilled by about 180\,000 sources in our starting dataset.

The uncertainty $\erv$ on the median RV is provided in the \texttt{radial\_velocity\_error} data field of the \texttt{gaiadr3.gaia\_source} table. Instead of setting an absolute upper limit to $\erv$, we rather compare it with the amplitude of the RV curve (\texttt{rv\_amplitude\_robust}) estimated during DR3 processing after outlier removal\footnote{
    We remark that the outlier removal procedure employed during DR3 processing is different from the one adopted in the pipeline used in the present work (Sect.~\ref{sec:catalog_construction:time_series_processing}), therefore the results are not necessarily the same (see Sect.~\ref{sec:catalog_quality:median_radial_velocity}).
}.
We inspected visually the distribution of said parameters for the sources in our sample, before and after the application of the conditions on $\grvs$ and number of visibility periods (bottom panel of Fig.~\ref{fig:hist_prefilter}), and decided to construct the filtering condition in the form
\begin{equation}
    \erv < 0.175\,\times\,\texttt{rv\_amplitude\_robust}\,,
\end{equation}
which retains almost 225\,000 objects from the starting sample. The combination of the three conditions described above results in a pre-filtered sample of 110\,654 RV time series, that we input into the variability pipeline.

\subsection{Time series processing}
\label{sec:catalog_construction:time_series_processing}

Overall, the processing of the RV and photometric time series is performed in a very similar manner as it was done for the photometric time series of LPVs in DR3. Therefore, we briefly summarize the procedure, focusing on the specific parameters for RV analysis and the few differences due to updates to the pipeline, and refer the reader to \citet[][and references therein]{dr3_lpv} for more details. The processing operations involve the detection and removal of outliers, followed by the calculation of time series statistics, and the derivation of the best-fit model.

The setup for detecting outliers in the photometric time series are unchanged with respect to DR3. For the RV time series, we exclude epochs with RV that:
\begin{itemize}
    \item   have an uncertainty larger than $5\,\kms$;
    \item   deviate from the median of the time series by more than $100\,\kms$;
    \item   deviate from the median of the time series by more than 10 times the median absolute deviation of the time series.
\end{itemize}
The choice of these parameters was guided by physical considerations concerning the typical RV amplitude for pulsation in LPVs, that is not expected to exceed several tens of $\kms$. However, we quickly realized that the sample contains a non-negligible fraction of high-quality RV curves likely originating from binarity, that we did not want to reject. Therefore, we have launched a few runs of the pipeline using rather permissive values, and tuned them by visual inspection of the distribution of the resulting time series statistics.

To describe the RV and photometric time series we adopt the same kind of mono-periodic model with frequency $f_X$ (where the subscript $X \in \{{\rv}, G, \gbp, \grp\}$ indicates the type of time series), that consists of the sum of a polynomial trend of degree $D_{p,X}\leq1$ (i.e., no trend or a linear trend) and a Fourier series with up to $N_{h,X}=3$ components (i.e., up to the second harmonic). Using a notation similar to that of \citet{dr1_var}, the model is defined as
\begin{equation}\label{eq:time_series_model}
    y = \sum_{k=1}^{N_{h,X}} A_{k,X}\cos\left[2\pi k f_X (t-t_{0,X}) + \psi_{k,X}\right] + \sum_{i=0}^{D_{p,X}} c_{i,X}(t-t_{0,X})^i \,
\end{equation}
where $A_{k,X}$ and $\psi_{k,X}$ are the amplitude and phase of the $k$-th Fourier component, respectively, and $t_{0,X}$ is a reference epoch. To avoid overfitting, the number of Fourier components is limited by the condition $N_{h,X}<\pi\,/\,\Delta\phi_{{\max},X}$ on the maximum phase gap $\Delta\phi_{{\max},X}$ of the folded time series \citep[cf. ][]{dr1_var}. While this approach is effective in most cases, it may fail for the few time series that end up having large and repeated gaps, and hence lack coverage of specific phase intervals, which makes them especially exposed to overfitting. A similar effect may result if the best-fit period is longer than the duration of the time series (see Sect.~\ref{sec:catalog_construction:post_filtering:filter23}).

For each source, the RV time series and the three photometric time series are processed independently of each other. For each time series, we begin by computing the periodogram. It is computed over the frequency range [$7 \cdot 10^{-4}$, 0.1]~d$^{-1}$, with an even spacing in frequency of $0.33 \times 10^{-4}$~d$^{-1}$. We take the period of the time series to be equal to the value corresponding to the highest peak of the periodogram. After a first determination of the best model in the form given by Eq.~\ref{eq:time_series_model}, we employ a nonlinear Levenberg-Marquardt optimization algorithm to improve the result.

It should be clear that the main peak of the periodogram identifies the strongest periodic signal in a time series, which is not necessarily the same as the period of the underlying physical process. In particular, in the case of ellipsoidal red giants, the light curve shows two minima per cycle of possibly different depths, that mimic a variation with a period half as long as the true orbital period. This effect is not present in the RV time series, so that the occurrence is not uncommon of sources whose RV period is twice the photometric period as determined from the periodogram. This will be further discussed in Sect.~\ref{sec:catalog_content}.

We performed safety checks by comparing the newly derived photometric periods with the ones derived from $G$-band light curves and published in \citet{dr3_lpv}. To our surprise, despite the pipeline setup and sequence of operations being identical, we found that in several cases the results are not exactly the same. We traced this discrepancy to a bug of Java version 8 affecting the nonlinear modeling of the time series, that disappeared during the upgrade to Java 17 performed after the conclusion of \Gaia DR3 operations. We remark that the deviations are small, and affect a minimal fraction of the sources. We provide a deeper analysis of this issue in Appendix~\ref{asec:numerical_differences_with_respect_to_gaia_dr3}.

\subsection{Post-filtering}
\label{sec:catalog_construction:post_filtering}

We tackle the post-filtering in two successive sub-steps. The first one involves the properties of the cleaned RV curves (i.e., after outlier removal) revealed by the time series statistics as well as the parameters of the best-fit model, with the exception of the frequency. The latter quantity is considered in the second sub-step, aimed at excluding the objects whose best RV period ($\prv$) is uncertain. We construct a filtering criterion by comparing with each other the periods derived from the RV and photometric time series.

\subsubsection{Selection on RV time series properties}
\label{sec:catalog_construction:post_filtering:filter1}

To begin with, we assess the impact of outlier removal on the RV time series of the pre-filtered sample, and examine the number $\noutrv=\nobsrvraw-\nobsrv$ of rejected epochs as a parameter for constructing an additional filter, where $\nobsrv$ is the number of epochs in the cleaned RV time series and $\nobsrvraw$ is the number of valid measurements in the original RV time series (i.e., excluding \texttt{NaN} values). The majority of the time series (about 88\%) are unaffected, while $\noutrv=1$ for about 7.5\% of the sources, and the remaining 4.5\% of RV curves had at least two rejected epochs. Visual inspection of time series and folded RV curves with varying number of outliers reveals satisfying results for $\noutrv\leq1$, as well as a rapid degradation with increasing $\noutrv$. We therefore restrict our sample by requiring that no more than one RV epoch is excluded during outliers removal, a condition that selects 105\,715 sources from the pre-filtered sample.

One of the properties that we found to be often associated with low-quality fits is the adoption of a first-degree polynomial in the RV curve model. About 18\% of the RV time series in the pre-filtered sample are modeled this way. This is done automatically by the variability pipeline when the inclusion of a linear trend results in a better fit with the underlying data. However, the combination of a relatively small number of epochs and characteristic time scales comparable with the duration of the time series make this approach poorly suited for LPVs. In contrast with variable objects with shorter periods, for which the inclusion of a linear trend can significantly improve the characterization of the time series, in the case of LPVs it tends to pick up the signal associated with long periods, while in some cases erroneously detrends RV curves with poor phase coverage. We therefore retain only the time series modeled without a linear trend, which correspond to 90\,952 sources in the pre-filtered dataset.

\begin{figure}
\centering
\includegraphics[width=\columnwidth]{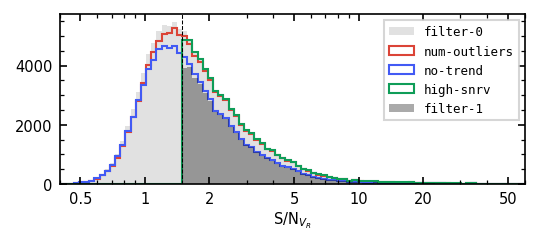}
\caption{Distribution of the RV signal-to-noise ratio, $\snrv$, for the pre-filtered sample (gray histogram) and with the conditions involved in the \texttt{filter-1} post-filtering step, labeled \texttt{num-outliers} (red curve), \texttt{no-trend} (blue curve), and \texttt{high-snrv} (green curve) and described in Table~\ref{tab:filter}. The dark gray histogram shows the combination of the three conditions. The vertical dashed line indicates the $\snrv=1.5$ threshold.}
\label{fig:hist_filter1}
\end{figure}

Finally, we apply a threshold to the signal-to-noise ratio of the cleaned RV time series at $\snrv=1.5$ (see Fig.~\ref{fig:hist_filter1}). In the pre-filtered sample there are 58\,725 sources above that limit. By combining the three conditions described above, we reduce the pre-filtered sample down to 44\,216 sources.

\subsubsection{Selection on the periods}
\label{sec:catalog_construction:post_filtering:filter23}

\begin{figure}
\centering
\includegraphics[width=\columnwidth]{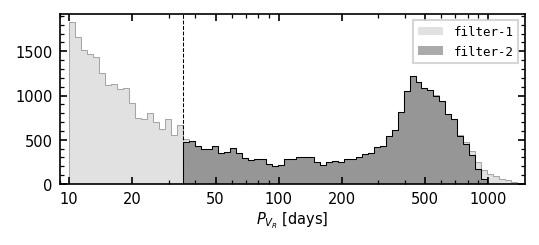}
\caption{Distribution of the RV periods of the sample after applying the post-filtering conditions \texttt{filter-1} (gray histogram, see Table~\ref{tab:filter}) and after applying the conditions on the periods themselves (\texttt{filter-2}, black histogram).
The vertical dashed line indicates the lower period limit at 35 days.}
\label{fig:hist_filter2}
\end{figure}

We follow the approach described in \citet{dr3_lpv} to bound the range of RV periods, setting a fixed lower limit to $35\,\days$, and excluding the cases in which $\prv$ is longer than the duration $\durationrv$ of the RV time series. We recall that the adoption of such a lower limit by \citet{dr3_lpv} for the photometric time series was aimed at minimizing the contamination from spurious signals. These conditions further reduce our dataset to 23\,523 sources. The distribution of RV periods before and after the application of these conditions is shown in Fig.~\ref{fig:hist_filter2}.

Finally, we examine how close the RV period is to the periods obtained from modeling the photometric time series. To do so, we follow the method described by \citet{dr3_eb} (their Sect.~4.1). We quantify the similarity between the RV period $\prv$ and a photometric period $\pph$ (either $\pg$, $\pbp$, or $\prp$) by the quantity
\begin{equation}\label{eq:phasedev}
    r_{\rv,{\rm ph}} = \frac{|\prv - \pph|}{\prv}\frac{\durationrv}{\prv} \,,
\end{equation}
which represents the maximum phase deviation a signal with period $\pph$ can accumulate with respect to $\prv$ during the observation duration $\durationrv$. In order to better understand the meaning of Eq.~\ref{eq:phasedev}, we note that $\delta\phi_{\rv,{\rm ph}}=|\prv - \pph|\,/\,\prv$ is the difference between the two periods normalized to $\prv$, whereas $\durationrv\,/\,\prv=\ncycrv$ is the number of cycles with period $\prv$ covered by the RV time series. Let us assume that the RV and photometric curves are in phase at the very beginning of the RV time series. Unless the two periods are identical, after one RV cycle $\prv$ the two curves show a phase offset that is exactly equal to $\delta\phi_{\rv,{\rm ph}}$. After two RV cycles the phase deviation is twice as large, and so on. At the very end of the RV time series, the phase offset is $\delta\phi_{\rv,{\rm ph}}\times \ncycrv = r_{\rv,{\rm ph}}$. It is easy to see that this is also the maximum possible phase deviation for given $\prv$, $\pph$, and $\durationrv$.

From Eq.~\ref{eq:phasedev} it is clear that $r_{\rv,{\rm ph}}$ is defined asymmetrically, and that $r_{\rv,{\rm ph}}\neq r_{{\rm ph},\rv}$. However, the closer the values of the period being compared, and the smaller the asymmetry is. The distribution of the values of $r_{\rv,{\rm ph}}$ and $r_{{\rm ph},\rv}$ for all three photometric periods show that the two quantities rapidly converge when they are smaller than unity. We thus construct our ``period similarity'' condition in the form
\begin{equation}\label{eq:simperiod1}
    r_{\rv,{\rm ph}}^{\max} = r_{{\rm ph},\rv}^{\max} = \max(r_{\rv,{\rm ph}}, r_{{\rm ph},\rv}) < 1\,.
\end{equation}

\begin{figure}
\centering
\includegraphics[width=\columnwidth]{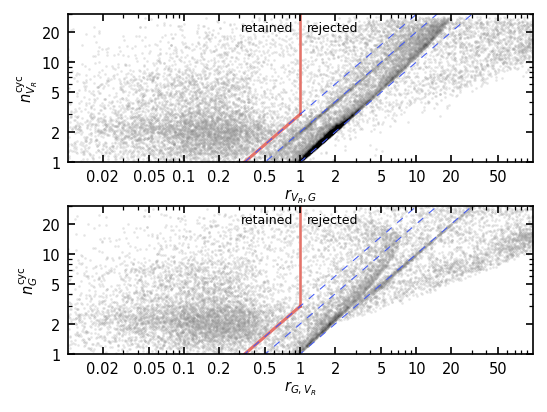}
\caption{Number of observed cycles versus phase deviation at the last cycle, comparing the RV and $G$-band curve models for the sample post-filtered down to \texttt{filter-2} (see Table~\ref{tab:filter}). More precisely, the top panel shows the phase deviation $r_{\rv,G}$ with respect to the last RV cycle and the number $\ncycrv$ of RV cycles, while the same quantities are referred to the $G$-band time series in the bottom panel. In each panel, the thick red line marks the upper limit to the phase difference employed in post-filtering (Eq.~\ref{eq:simperiod1}), while the dashed lines indicate $n^{\rm cyc}/r=3$, 2, and 1. Data points to the right of the thick red line are rejected. A similar picture emerges when the $\gbp$ or $\grp$ time series are considered in place of $G$.}
\label{fig:pdGRV_ncRV_filter3ex}
\end{figure}

Fig.~\ref{fig:pdGRV_ncRV_filter3ex} shows the distribution of $r_{\rv,{\rm ph}}$ versus the number of cycles $\ncycrv$ covered by the RV time series. A large number of sources accumulate along two slanted stripes in the diagram, that are only partially rejected by the condition defined by Eq.~\ref{eq:simperiod1}. These stripes correspond to $r_{\rv,{\rm ph}}\simeq \ncycrv$ and $r_{\rv,{\rm ph}}\simeq 2\ncycrv$, respectively. It is easy to show that the former case corresponds to $\pph\simeq2\prv$ or $\pph\ll\prv$, and the latter to $\pph\simeq3\prv$ (or $\pph\simeq-\prv$, which is not possible as periods have positive values). They clearly indicate situations of incompatibility between pairs of periods, and should be excluded. To do so, we require that
\begin{equation}\label{eq:simperiod2}
    \delta P_{\rv,{\rm ph}} = \delta P_{{\rm ph},\rv} = \frac{|\prv-\pph|}{\min(\prv,\pph)} < \frac{1}{3}\,,
\end{equation}
which completes our condition on period similarity (Eq.~\ref{eq:simperiod1}). The combination of Eq.~\ref{eq:simperiod1} and~\ref{eq:simperiod2} for the $G$-band and RV periods is equivalent to taking only the data points that are on the left of the red lines in both panels of Fig.~\ref{fig:pdGRV_ncRV_filter3ex}. Therefore, our definition of period similarity is given by
\begin{equation}\label{eq:simperiod}
    \prv\simeq\pph\;\;\;\Longleftrightarrow\;\;\;\left[\left(r_{\rv,{\rm ph}}^{\max} < 1\right)\;{\rm and}\;\left(\delta P_{\rv,{\rm ph}}<\frac{1}{3}\right)\right] \,.
\end{equation}

It should be noted that, if a source displays variability due to binarity, the main peak in the periodogram of any one of its photometric time series can be half of the true orbital period, and hence of the RV period. Therefore, the requirement that $\prv\simeq\pph$ could lead to exclude these sources. In order to avoid this, we are also interested in using a requirement in the form $\prv\simeq2\,\pph$, which means that any occurrence of $\pph$ in Eq.~\ref{eq:simperiod} is replaced by $2\,\pph$. Our final requirement is therefore
\begin{equation}\label{eq:simperiod_bands}
    (\prv\simeq\pph\;{\rm or}\;\prv\simeq2\,\pph)\;\;{\rm for\;any\;}\;\pph\in\{\pg,\pbp,\prp\}\,.
\end{equation}

\begin{figure}
\centering
\includegraphics[width=\columnwidth]{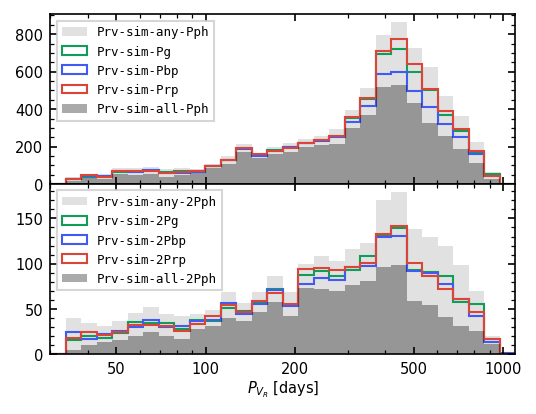}    
\caption{Period distribution, after applying \texttt{filter-2}, for the subsets with RV period similar to one or more photometric periods (top panel), or twice of it (bottom panel). The light gray and dark gray histograms represent the sets in which the RV period is similar to at least one photometric period or to all of them, respectively. The colored curves represent the sets in which the RV period is similar to $\pg$ (green), $\pbp$ (blue), or $\prp$ (red).}
\label{fig:hist_Prv_sim}
\end{figure}

Figure~\ref{fig:hist_Prv_sim} displays the period distributions for the sources displaying compatibility between the RV period and one or more photometric periods according to Eq.~\ref{eq:simperiod}. Comparing with the distribution in Fig.~\ref{fig:hist_filter2} we note the effectiveness of this type of selection at rejecting periods shorter than $\sim200\,\days$, where a higher rate of occurrence of spurious frequencies is expected \citep[see][]{holl_etal_2023}. This is also true when the RV period is compared with twice one of the photometric periods.

We note that the number of sources with $\prv\simeq\prp$ is slightly higher than that with $\prv\simeq\pg$, which in turn are more numerous than the objects with $\prv\simeq\pbp$. The same trend is present when the comparison is made against twice the photometric period, but is less pronounced. This could be indicative of a color-dependence of the photometric variability features, typical of pulsating stars. The fact that this feature becomes less conspicuous when $\prv\simeq2\pph$ would support the interpretation that the variability of these sources is extrinsic and associated with binarity.

We construct the final filter by applying the condition defined by Eq.~\ref{eq:simperiod_bands} inclusively to the three photometric periods, that is, we require that the RV period is similar to at least one of them. The final dataset consists of 9\,614 sources.

\subsection{Top-quality sample}
\label{sec:catalog_construction:top_quality_sample}

\begin{figure}
\centering
\includegraphics[width=\columnwidth]{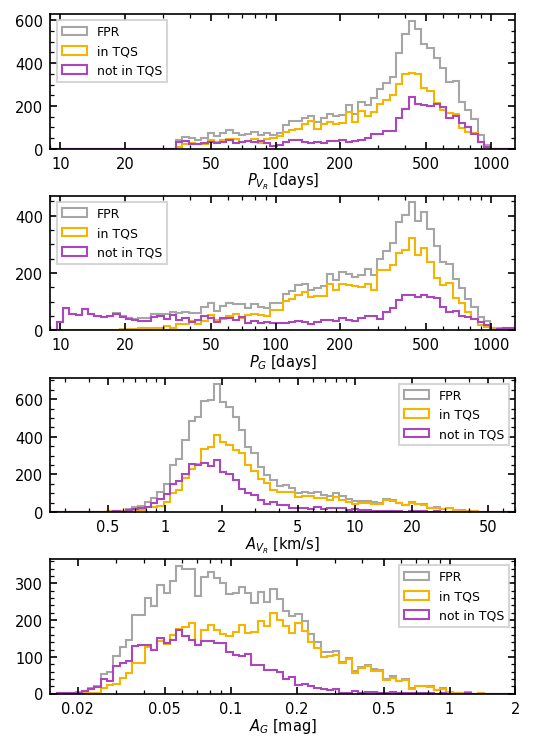}
\caption{Distribution of the RV and $G$-band variability parameters for the all the FPR sources (gray curves), and distinguishing between whether they are in the top-quality sample (TQS, orange curves) or not (purple curves). Panels from top to bottom show the distributions of RV periods ($\prv$), $G$-band periods ($\pg$), RV semi-amplitudes ($\amprv$) and $G$-band semi-amplitudes ($\ampg$).}
\label{fig:hist_P_amp_tqs_fpr}
\end{figure}

By the criteria defined above, we identified a subset of the FPR sample consisting of sources displaying a high degree of compatibility between the RV and photometric variability. Namely, there are 6\,093 sources whose RV period is consistent with each one of the three periods derived from the photometric time series. This means that these sources fullfil the condition
\begin{equation}
    (\prv\simeq\pph\;{\rm or}\;\prv\simeq2\,\pph)\;\;{\rm for\;all\;}\;\pph\in\{\pg,\pbp,\prp\}\,.
\end{equation}
These sources are identified by the field \rvflag\texttt{=True} in the \Gaia Archive (see Sect.~\ref{sec:catalog_construction:data_fields}), and form a subset that we dubbed the ``top-quality sample'' (TQS).

Such a high consistency between the RV and photometric periods is a strong indication that a signal originating from the same physical process is being detected in all four time series, with two important consequences. On the one hand, these sources can be used to investigate a given type of variability in its different aspects (physical motion, changes in brightness and color) with a good degree of confidence that they all trace the same phenomenon. Given the multi-periodic nature of LPVs, this is far from trivial. On the other hand, there is a comparatively small probability that the periodic signal picked up by the variability processing pipeline is spurious.

Other than this self-consistency, the sources in the TQS have on the average the same properties as the remaining FPR sources, with the only exception that they include a larger fraction of sources identified as binary variables (see Sect.~\ref{sec:catalog_content}). Indeed, binary candidates are assigned to the TQS with a higher frequency ($\sim80\%$) than other sources in the FPR ($\sim60\%$). These trends are likely to be attributed to the fact that, owing to its geometric nature, binary-induced variability shows smooth variations compared with the pulsation of LPVs, known to display irregularities. 

We compared the TQS and the other FPR sources in terms of the distributions of several quantities from \Gaia DR3. The sources in the former set display slightly better astrometry (smaller errors in sky coordinates and proper motions), but the two sets are equivalent in terms of relative parallax uncertainty. These properties are to be attributed to a slightly higher number of visibility periods used in the astrometric solution. The uncertainty associated with both photometric and RV measurements as reported in the \Gaia DR3 source table is slightly higher for the TQS sources, which simply reflects the fact that they tend to exhibit larger variability amplitudes. The TQS sources follow essentially the same brightness distributions of all other FPR sources in all three \Gaia bands, except they are slightly brighter in $\gbp$ and fainter in $\grp$, and as a result they appear to have a slightly bluer color which reflects the larger fraction of binaries in the TQS compared to other FPR sources, see Sect.~\ref{sec:catalog_overview}. The two sets do not show any particular difference in their RV distributions.

Some more significant differences between the two sets are found in terms of the variability parameters (Fig.~\ref{fig:hist_P_amp_tqs_fpr}). The requirement of period consistency effectively excludes from the TQS sources with short $G$-band periods (due to the lower limit at 35 days on $\prv$). Moreover, the amplitude distribution of TQS sources tends to be skewed toward slightly larger values compared with the full FPR sample, as they are associated with a higher signal-to-noise ratio (and hence a higher chance of picking the same periodicity in RV and in photometry). Other amplitude-related parameters (such as standard deviation, interquartile range, or Stetson variability index) show similar trends. For similar reasons the TQS sources display smaller Abbe values \citep{mowlavi_2014,mowlavi_etal_2017} than other sources in all \Gaia bands, indicating smoother light curves. However, such a difference is not present for the Abbe value computed for the RV time series.

Finally, we inspected the mean value of the uncertainties associated with single epochs (either of the RV or photometric time series) and the mean of the absolute residuals of the time series models, and found differences between the distribution associated with the TQS and with other sources that are consistent with the different amplitude distributions. Therefore, we remark that we consider this subsample to be of superior quality within the FPR because of its content of coherent physical information rather than in terms of actual quality of measurements.

\subsection{Data fields}
\label{sec:catalog_construction:data_fields}

The present catalog follows the same scheme as the 2${\rm nd}$ \Gaia catalog of LPV candidates \citep{dr3_lpv}, and has therefore the same data fields, with the addition of the corresponding fields for the RV variability. More precisely, the fields \texttt{solution\_id}, \texttt{source\_id}, \texttt{median\_delta\_wl\_rp}, and \texttt{isCstar} are left unchanged, while the fields \texttt{frequency}, \texttt{frequency\_error}, and \texttt{amplitude} have their values replaced with the newly derived parameters of the best-fit model for the $G$-band time series (see Sect.~\ref{sec:catalog_construction:time_series_processing} for the reason of the updated values). Finally, the following four data fields are added.

\vspace{2mm}
\textbf{\scshape \large frequency\_rv \hypertarget{?}}: Frequency of the RV curve (double, Frequency[$day^{-1}$])

This field provides the frequency determined from the RV time series.

\vspace{2mm}
\textbf{\scshape \large frequency\_error\_rv \hypertarget{?}}: Uncertainty on the RV frequency (float, Frequency[$day^{-1}$])

This field provides the uncertainty on the frequency of the RV time series.

\vspace{2mm}
\textbf{\scshape \large amplitude\_rv \hypertarget{?}}: Amplitude of the RV curve (float, Velocity[$\kms$])

This field gives the half peak-to-peak amplitude (semi-amplitude in $\kms$, based on the best-fit model (see Sect.~\ref{sec:catalog_construction:time_series_processing}).

\vspace{2mm}
\textbf{\scshape \large flag\_rv \hypertarget{?}}: Flag identifying the top-quality subsample (boolean)

This field identifies the sources whose RV period is fully compatible with all three photometric periods (see Sect.~\ref{sec:catalog_construction:top_quality_sample}).

The full RV time series for all sources in this FPR are available for download from the table \texttt{gaiafpr.vari\_epoch\_radial\_velocity} in the \Gaia archive, while the statistics for the cleaned RV time series are provided in the table \texttt{gaiafpr.vari\_rad\_vel\_statistics}, following the same scheme adopted in \Gaia DR3 for the RV time series of Cepheids and RR~Lyrae \citep[cf.][]{dr3_cep,dr3_rrl}. In Appendix~\ref{app:catalog_retrieval} we provide some instructions on how to retrieve the FPR data.

Hereinafter we adopt the notation $\ampg$ and $\amprv$ to indicate the quantities \texttt{amplitude} and \texttt{amplitude\_rv} published in this FPR, corresponding to the semi-amplitude of the fundamental component of the best-fit Fourier model of the $G$-band and RV time series, respectively. We note that roughly half of the $G$-band time series and more than 80\% of the RV time series have been modeled with a single-component Fourier series, so the published value is exactly the semi-amplitude of the model. The remaining time series have harmonic components whose amplitude is typically much smaller than that of the fundamental component, so that the semi-amplitude of the latter is still representative to the semi-amplitude of the full Fourier model. Therefore, for simplicity, we often refer to $\ampg$ (not to be confused with the $G$-band extinction) and $\amprv$ as ``semi-amplitude of the time series models,'' whereas their formal meaning should be clear.

\section{Catalog content}
\label{sec:catalog_content}

\begin{figure}
\centering
\includegraphics[width=\columnwidth]{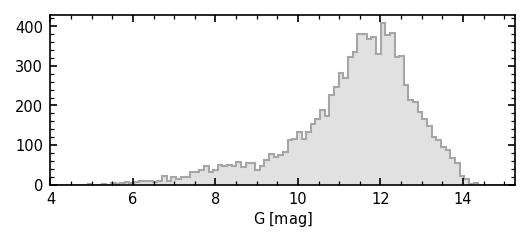}
\caption{$G$-band brightness distribution of the FPR sample.}
\label{fig:hist_gmag}
\end{figure}

\begin{figure}
\centering
\includegraphics[width=\columnwidth]{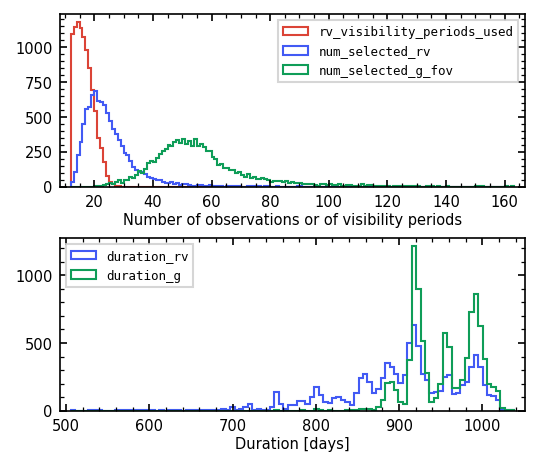}
\caption{Distribution of the number of observations (top) and duration (bottom) of the RV and $G$-band time series of the FPR sources. The red line indicates the number visibility periods used to derive the median RV published in DR3 (a single visibility period may contain multiple epochs, see Sect.~\ref{sec:catalog_construction:pre_filtering}). The blue and green lines indicate the number of measurements in the cleaned RV and $G$-band time series (top) or their duration (bottom).}
\label{fig:hist_nobs_duration}
\end{figure}

This FPR provides epoch RVs for 9\,614 candidate LPVs, of which the $G$, $\gbp$ and $\grp$ time series are available in DR3 as part of the second \Gaia catalog of LPV candidates. Figure~\ref{fig:hist_gmag} shows the $G$-band distribution of these sources, which cover the range $6\lesssim G/\mags\lesssim 14$. The RV time series have between 12 and 90 measurements, with an average of 24 epochs, unevenly sampling a time interval of about 3 years. More precisely, the RV time series have a mean duration of 905 days, spanning a range between about 500 and 1000 days, but with a distribution skewed toward longer durations (typically $\gtrsim800$ days). The number of epochs in the RV and $G$-band time series, as well as the number of visibility periods adopted for deriving median RVs in DR3 (see Sect.~\ref{sec:catalog_construction:pre_filtering}), are illustrated in Fig.~\ref{fig:hist_nobs_duration}. We note that these numbers correspond to the cleaned time series, that is after outlier removal, and are the same values given in the \Gaia archive summary tables, whereas the published \Gaia light curves include the outlier epochs as well (flagged to indicate whether they have been rejected by the variability pipeline).

\begin{figure*}
\centering
\includegraphics[width=0.925\textwidth]{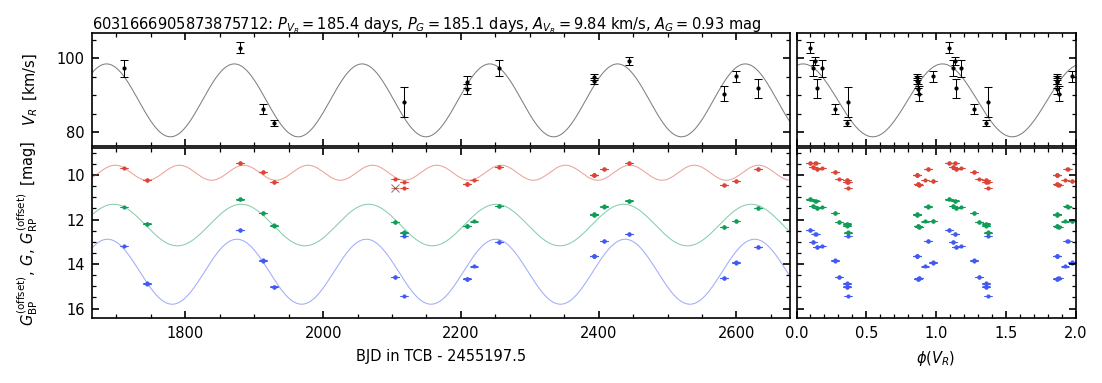}
\caption{Example time series for a source with mixed consistency between the photometric and RV time series. This source has $\pg\simeq\pbp\simeq\prv$, while $\prp \simeq 0.5 \prv$. The panels in the top row show the RV data and model, while the photometric data and corresponding models are shown in the panels in the bottom row (in red, green, and blue for the $\grp$, $G$, and $\gbp$ bands, respectively). For visualization purposes, an arbitrary offset is applied to the $\grp$ and $\gbp$ time series. The \Gaia DR3 source ID of this object is indicated in the title, together with the period and semi-amplitude of the best-fit $G$-band and RV time series models. The panels on the right show the four time series folded by the RV period.}
\label{fig:RVTS_ex_mixPph}
\end{figure*}

Besides the filtering steps, in Table~\ref{tab:filter} we provide a summary of a few interesting subsets of the final sample, obtained by further comparing the periods derived from the RV and photometric time series. Following the criteria defined in Sect.~\ref{sec:catalog_construction:post_filtering:filter1}, the period comparison for a given source has three possible outcomes: (1) $\prv\simeq\pph$, (2) $\prv\simeq2\,\pph$, or (3) $\prv\not\simeq\pph$ and $\prv\not\simeq2\,\pph$. These conditions are not necessarily the same for each of the three photometric periods. For instance, a source might be such that $\prv\simeq\pg$, $\prv\simeq2\,\prp$, and at the same time $\prv\not\simeq\pbp$ and $\prv\not\simeq2\,\pbp$. However, due to the filters we applied, either conditions (1) or (2) must be verified for at least one of $\pph\in\{\pg,\pbp,\prp\}$ for all sources published in the FPR.
These conditions allow us to distinguish between three types of sources:
\begin{itemize}
    \item   only 1:1 compatibility: $\prv\simeq\pph$ for at least one photometric period, but none of the other periods meets the condition $\prv\simeq2\,\pph$;
    \item   only 2:1 compatibility: $\prv\simeq2\,\pph$ for at least one photometric period, but none of the other periods meets the condition $\prv\simeq\pph$;
    \item   ``mixed'' compatibility: $\prv\simeq\pph$ for at least one photometric period, and $\prv\simeq2\,\pph$ for at least one of the other photometric periods (as in the example above).
\end{itemize}

The majority of the FPR sources fall in the first category, consisting of 7\,372 sources (about 77\%). There is no direct indication that the variability of these sources results from binarity, as none of the photometric period is close to a 2:1 ratio with respect to the RV period. Of course, this does not prove that they are not binary variables. However, it is reasonable to assume that most of these sources are probably pulsating stars, at least for the purpose of assessing the relative fractions of these types of variables in the FPR. Similarly, the 2\,063 sources (about 21\%) belonging to the second category in the list above are probably binary variables. More precisely, as they are selected among bright red giants, these sources are most likely ellipsoidal variables (ELL), and will be referred as such hereinafter. Further evidence supporting this statement will be provided in Sect.~\ref{sec:catalog_overview}.

Finally, there exist 179 sources such that their RV period is consistent with one or two of the photometric periods, and twice the value of the remaining ones. We examined visually the time series of a random sample of these sources, and found that the cleaned light curves often show large phase gaps when folded with the RV period. Figure~\ref{fig:RVTS_ex_mixPph} shows a clear example with a lack of data near minimum light. All time series show a similar trend, and the best-fit model to the RV, $G$, and $\gbp$ time series is visually convincing, yet the $\grp$ has a best-fit model with half the period found in the other time series. A similar situation can arise when the time series covers a small number of RV cycles, so it becomes difficult to constrain the period precisely. It is clear that this kind of mixed consistency between photometric and RV periods has artificial causes and, in principle, none of the two periods can be confidently taken to be the correct one. It is not possible to make any inference on the nature of these sources based only on their periods. However, the fact that they represent less than 2\% of the FPR is encouraging.

These three categories give us a general idea of the fractions of ellipsoidal and pulsating variables in the FPR based on weak conditions on period consistency. Stronger conditions can be imposed by restricting the analysis to the TQS, which includes 4\,899 probable pulsators ($\prv\simeq\pph$ for each $\pph$) and 1\,194 probable ELL ($\prv\simeq2\pph$ for each $\pph$). The two kinds of sources make up for about 80\% and 20\% of the TQS, respectively. These percentages are fully compatible with the values found in the previous paragraph.

\begin{figure}
\centering
\includegraphics[width=\columnwidth]{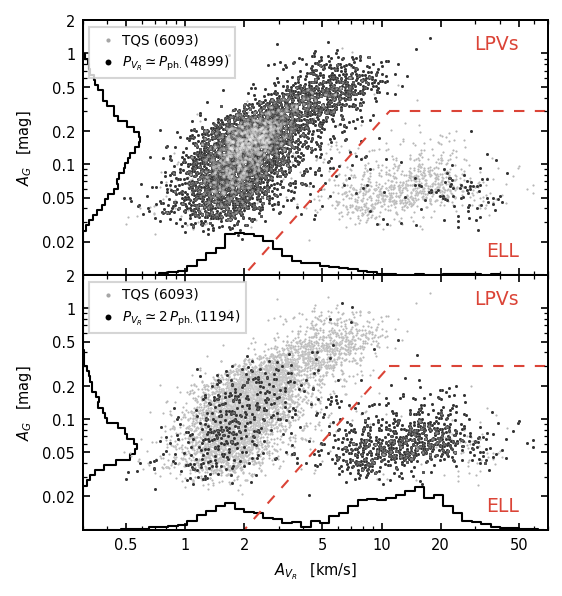}
\caption{Comparison between the semi-amplitudes $\ampg$, $\amprv$ of the best-fit models of the $G$-band and RV time series for the TQS (light gray symbols in the background). The darker symbols indicate sources whose RV period is consistent with the photometric periods in a 1:1 ratio (top panel) or in a 2:1 ratio (bottom panel). The dashed red line corresponds to Eq.~\ref{eq:amp_condition_ell}, and the size of each sample is indicated in the legend.}
\label{fig:drv_dg_2panels}
\end{figure}

\subsection{Candidate ellipsoidal variables}
\label{sec:catalog_content:candidate_ellipsoidal_variables}

Classifying the types of variables discussed above based only on the ratio between the RV period and photometric periods is not necessarily a good approach. In particular, it might be inappropriate if one or more of the cleaned time series end up having a small number of measurements, so that the corresponding period is poorly constrained. Therefore, we use the semi-amplitude $\amprv$ of the RV time series model, and the corresponding value $\ampg$ for the $G$-band model, to perform a deeper analysis. In doing so we consider only the TQS in the rest of this section, so to obtain as clean a picture as possible.

The $G$-band and RV semi-amplitudes derived for the sources in this sample are displayed in Fig.~\ref{fig:drv_dg_2panels}. Two groups are clearly separated in this diagram (in either panel). The first group shows $G$-band variations over a wide range ($0.02\lesssim\ampg\,/\,\mags<2$), but is limited to relatively small RV amplitudes ($\amprv\lesssim10\,\kms$, with only a few exceptions). The second group is characterized by large RV variations ($\amprv\gtrsim5\,\kms$) and relatively small light amplitudes ($\ampg\lesssim0.2\,\mags$). We can readily interpret the former group as consisting of pulsating stars, whose brightness changes can become very large \citep[owing to strong absorption by molecules that form efficiently in the expanding phase of the cycle, ][]{reid_goldston_2002} while they can hardly attain pulsation velocities larger than $\sim20\,\kms$ \citep{nowotny_etal_2010}. In contrast, orbital velocities can easily exceed that value in binaries, but their $G$-band variations do not exceed a few tenths of magnitude. This interpretation is supported by the fact that the vast majority of sources with $\prv\simeq\pph$ are found in the former group (black points in the top panel of Fig.~\ref{fig:drv_dg_2panels}), whereas most sources in the latter group have $\prv\simeq2\,\pph$ (black points in the bottom panel), although some contamination is present in both.

Based on the distributions displayed in Fig.~\ref{fig:drv_dg_2panels}, we identify ELL candidates by the condition
\begin{equation}\label{eq:amp_condition_ell}
    \ampg < 0.3\,\mags\;\;\;{\rm and}\;\;\;\frac{\ampg}{\mags} < 2.5\cdot10^{-3}\cdot\left(\frac{\amprv}{\kms}\right)^2 \,
\end{equation}
which corresponds to the region in Fig.~\ref{fig:drv_dg_2panels} below and to the right of the dashed red line.
We prefer Eq.~\ref{eq:amp_condition_ell} to a condition based on the RV-to-photometric period ratios as it is based on physical arguments, and allows us to identify ellipsoidal variables more confidently. For instance, we note that there are several sources in Fig.~\ref{fig:drv_dg_2panels} (top panel) having $20\lesssim\amprv\,/\,\kms\lesssim50$ that are unlikely to be pulsators, but would be classified as such based only on the ratio between their RV period and photometric periods. Further evidence in support of this approach is given in Sect.~\ref{sec:catalog_overview}.

At the same time, there are sources with $\prv\simeq2\,\pph$ that end up outside of the region associated with ELLs in Fig.~\ref{fig:drv_dg_2panels} (bottom panel). While there is no a-priori reason why they should not be binaries, their distribution in this diagram is consistent with that of the sources with $\prv\simeq\pph$ (top panel of Fig.~\ref{fig:drv_dg_2panels}), suggesting that they display the same kind of variability. Visual inspection of their RV and light curves indicates that the 2:1 period ratio is probably artificial. This most likely results from the fact that many stars in this part of the diagram are semi-regular variables with multi-periodic variability, not necessarily well-described by a single-period model.

\subsection{Candidate LPVs: Pulsation and long secondary periods}
\label{sec:catalog_content:lpv}

\begin{figure}
\centering
\includegraphics[width=\columnwidth]{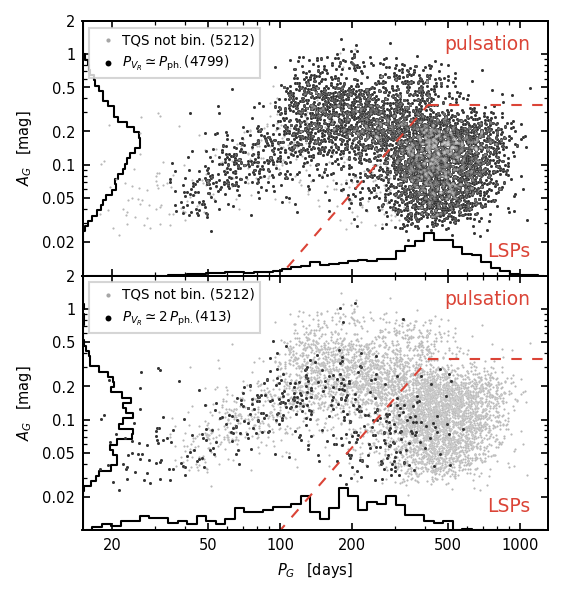}
\caption{Similar to Fig.~\ref{fig:drv_dg_2panels}, but comparing the period $\pg$ and semi-amplitude $\ampg$ derived from $G$-band time series for TQS sources that are probable LPVs (not identified as ellipsoidal variable candidates). The dashed red line corresponds to Eq.~\ref{eq:condition_lsp}, and the size of each sample is indicated in the legend.}
\label{fig:ampg_Pg_2panels}
\end{figure}

For pulsating LPV stars the photometric amplitude of variability increases with the pulsation period, a trend that can be identified in the left and top sides of the diagram in Fig.~\ref{fig:ampg_Pg_2panels} ($20\lesssim\pg\,/\,\days\lesssim500$). A second group of stars can be seen in the bottom-right corner of the diagram, characterized by long periods and comparatively small $G$-band amplitudes. While it is likely that these sources are also pulsating LPVs, the dominant period picked up by the variability processing pipeline is probably a long secondary period \citep[LSP; see, e.g., fig.~16 of][]{dr3_lpv}.

We tentatively identify LSPs in the period amplitude diagram by the condition
\begin{equation}\label{eq:condition_lsp}
    \ampg < 0.35\,\mags\;\;\;{\rm and}\;\;\;\frac{\ampg}{\mags} < 10^{-7}\cdot\left(\frac{\pg}{\days}\right)^{2.5} \,.
\end{equation}
We remark that this criterion and the resulting classification is necessarily approximate, and is adopted only for the purpose of characterizing the content of the FPR. In principle, a knowledge of the absolute brightness is required in order to accurately differentiate between pulsation periods and LSPs so that one can construct a period-luminosity diagram. This cannot be done for the entire FPR sample because of the relatively large uncertainties affecting the parallaxes of a number of sources, even though more than one third of them have relative parallax errors better than 10\% (they are examined in more detail in Sect.~\ref{sec:color_absolute_magnitude_diagram}). 

We note that certain LPVs, such as some relatively massive AGB stars or red supergiants, have long pulsation periods and relatively small photometric amplitude, and their variability could thus mimic the LSP variation. These stars overlap with LSPs in the period-amplitude diagram \citep[see e.g., fig.~18 of][]{lebzelter_etal_2019}, and might be misclassified by our criteria. However, given the rarity of these stars, this has a negligible impact in our quantification of the relative fraction of variability types in the FPR. They are further examined in Sect.~\ref{sec:color_absolute_magnitude_diagram}.

We also note that pulsation and LSP usually coexist in LPVs that exhibit the LSP phenomenon. The knowledge of multiperiodicity and the amplitude associated with each period can then improve the classification derived from the period-amplitude diagram. However, as only one period is extracted from each time series in this FPR, we consider our selection appropriate enough for our purposes. Additional evidence to support this is provided in Appendix~\ref{asec:lsp}.

\begin{table*}
\caption{Number of sources assigned to different types (LPVs showing pulsation or LSP variability, or ellipsoidal variables) in the TQS and in the full FPR, distinguishing between the stars showing compatibility between the RV and photometric periods in a 1:1 or 2:1 ratio.}
\label{tab:table_types}
\centering
\begin{tabular}{llccccccc}
\hline\hline
 & & \multicolumn{3}{c}{TQS} && \multicolumn{3}{c}{FPR} \\
\cline{3-5} \cline{7-9}
 & & all $\prv\simeq\pph$ & all $\prv\simeq2\pph$ & total && any $\prv\simeq\pph$ & any $\prv\simeq2\pph$ & total$^{(a)}$ \\
\hline
\multirow{2}{*}{LPV} & pulsation & 1\,992 &~~\,304 & 2\,296 && 3\,266 &~~\,906 & 4\,084 \\
                     & LSP       & 2\,807 &~~\,109 & 2\,916 && 4\,125 &~~\,370 & 4\,421 \\
\multicolumn{2}{l}{ELL}          &~~\,100 &~~\,781 &~~\,881 &&~~\,160 &~~\,966 & 1\,109 \\
\hline
\multicolumn{2}{l}{Total}        & 4\,899 & 1\,194 & 6\,093 && 7\,551 & 2\,242 & 9\,614 \\
\hline
\end{tabular}
\tablefoot{
  \tablefoottext{a}{Sources that do not belong to the TQS may show photometric periods that are simultaneously compatible with $\prv$ in both the 2:1 and 1:1 ratios, so the corresponding ``total'' column is not necessarily equal to the sum of the two previous columns.
  }
}
\end{table*}

\subsection{Summary}
\label{sec:catalog_content:summary}

Table~\ref{tab:table_types} provides a summary of the number of sources identified as LPVs (either showing pulsation or LSP variability) or as ellipsoidal variables. The TQS consists by about 14\% of ELLs, by about 38\% of pulsating LPVs, and by about 48\% of LPVs for which we detect LSP-like variability. If the conditions defined by Eqs.~\ref{eq:amp_condition_ell} and~\ref{eq:condition_lsp} are extended to the entire FPR sample, these percentages become about 12\%, 42\%, and 46\%, respectively.

The numbers in Table~\ref{tab:table_types} also show that, overall, the TQS includes roughly 60\% of the LPVs (regardless of whether they show pulsation or LSPs), and 80\% of the ELL candidates, indicating that the latter enter more easily in the TQS. This is consistent with the fact that the geometric origin of the variability of the latter results in much smoother and regular variations than those presented by the former, increasing the chances of consistency between photometric and RV periods.

\begin{figure*}
\centering
\includegraphics[width=0.925\textwidth]{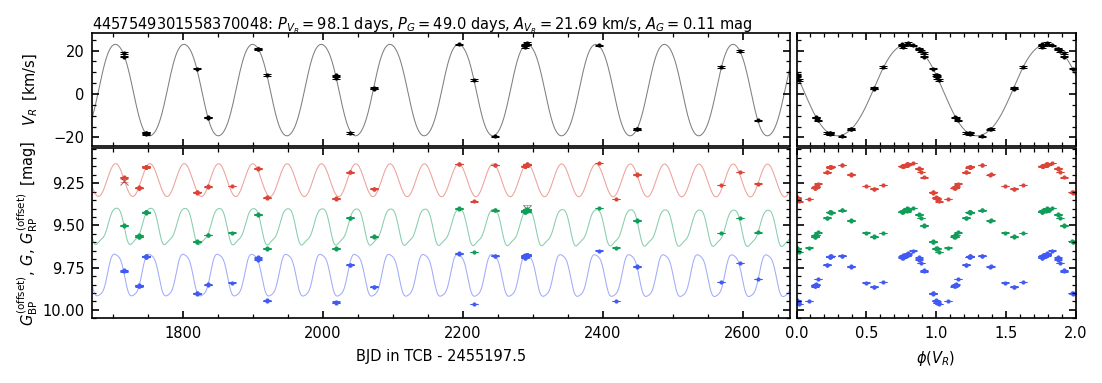}
\includegraphics[width=0.925\textwidth]{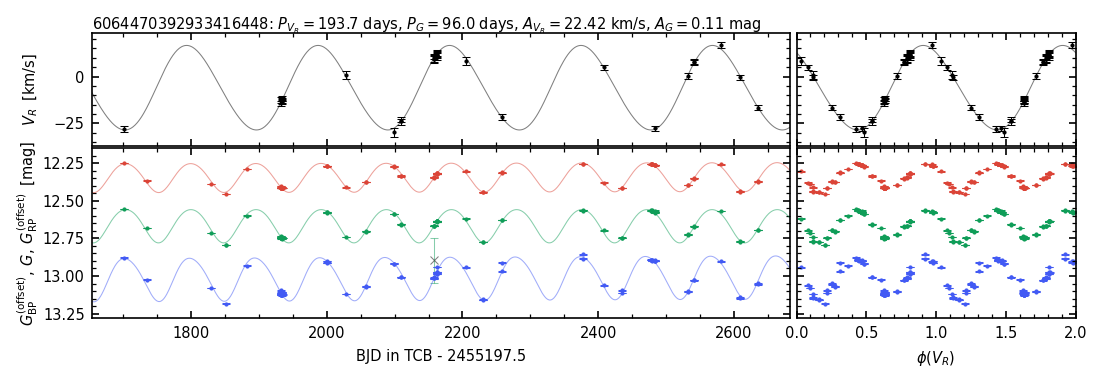}
\includegraphics[width=0.925\textwidth]{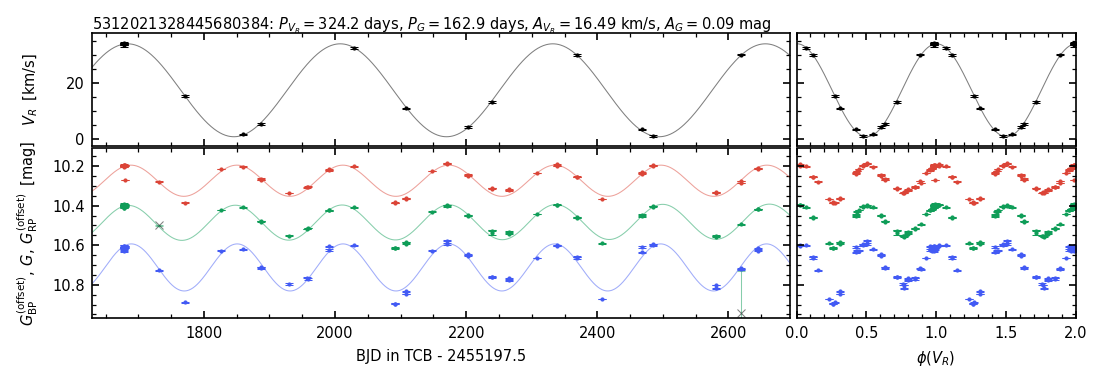}
\includegraphics[width=0.925\textwidth]{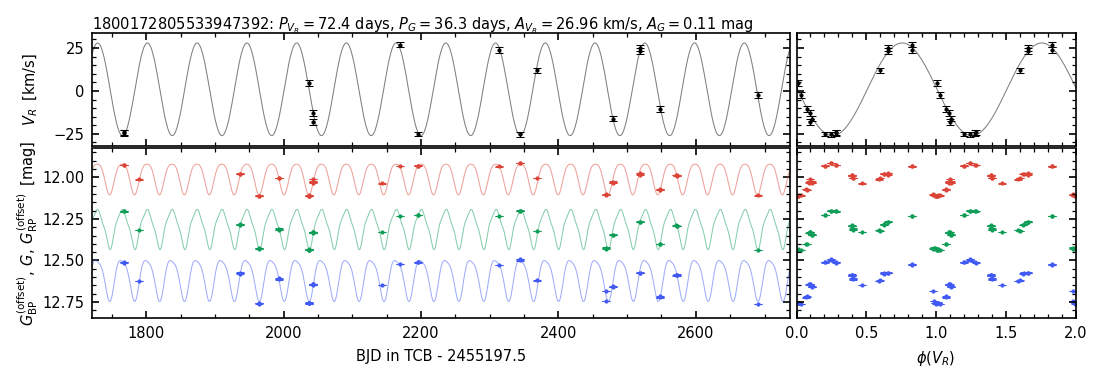}
\caption{Similar to Fig.~\ref{fig:RVTS_ex_mixPph}, but showing some example time series of ELL candidates. All data in the panels on the right-hand column are folded by the FPR RV period.}
\label{fig:RVTS_exBin}
\end{figure*}

\begin{figure*}
\centering
\includegraphics[width=0.925\textwidth]{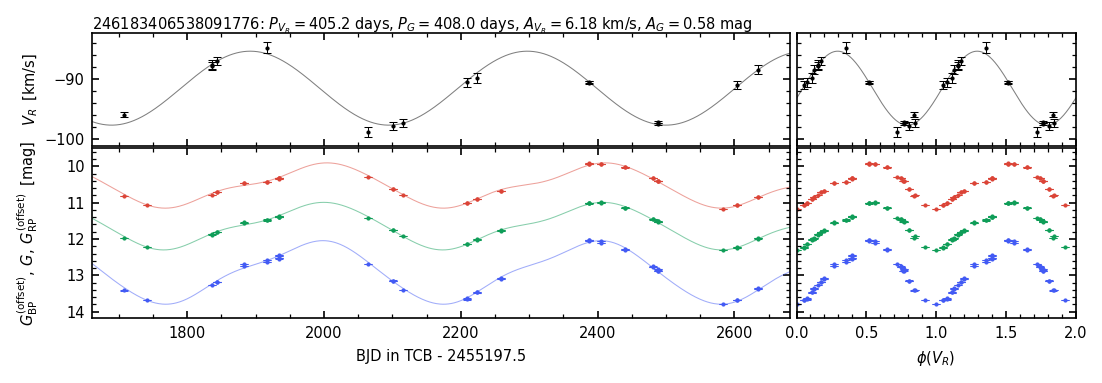}
\includegraphics[width=0.925\textwidth]{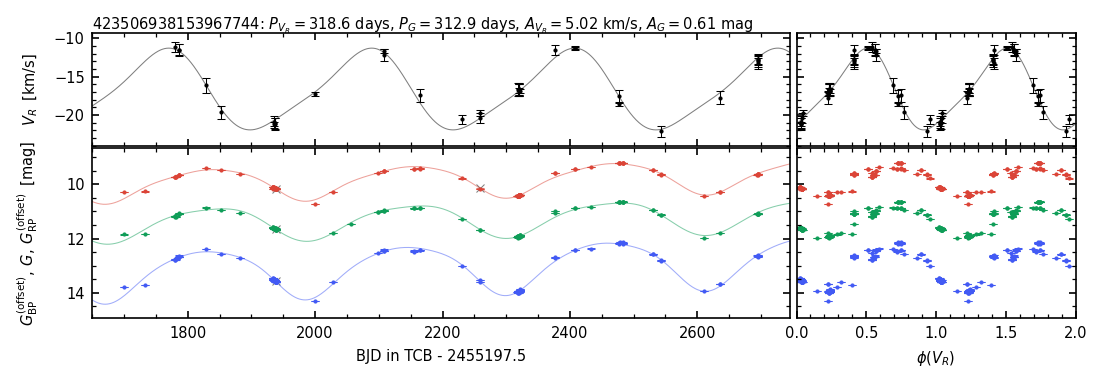}
\includegraphics[width=0.925\textwidth]{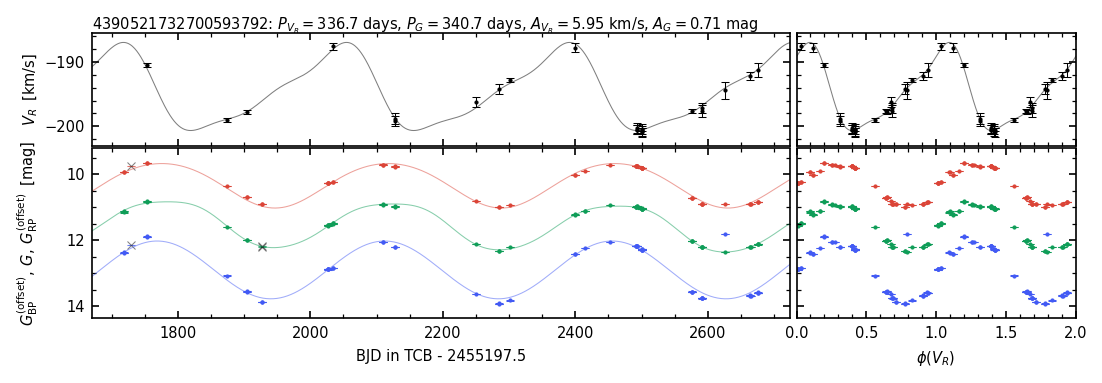}
\includegraphics[width=0.925\textwidth]{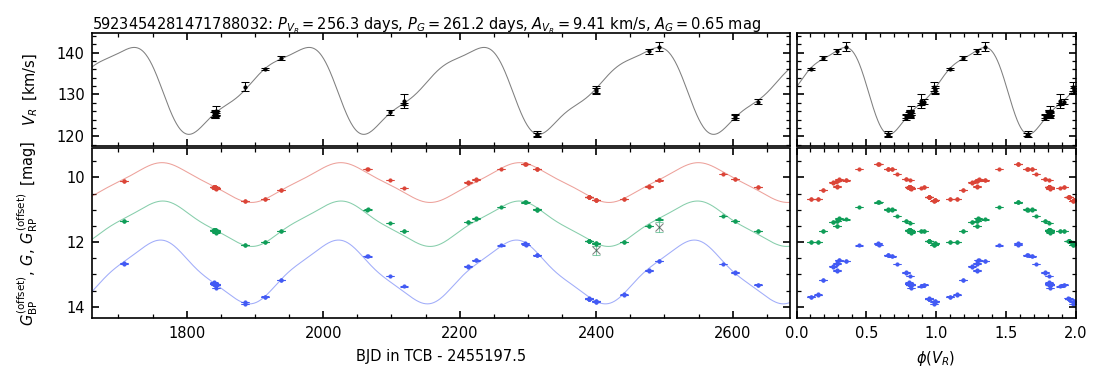}
\caption{Similar to Fig.~\ref{fig:RVTS_ex_mixPph}, but for pulsating LPV candidates. All data in the panels on the right-hand column are folded by the FPR RV period.}
\label{fig:RVTS_exPuls}
\end{figure*}

\begin{figure*}
\centering
\includegraphics[width=0.925\textwidth]{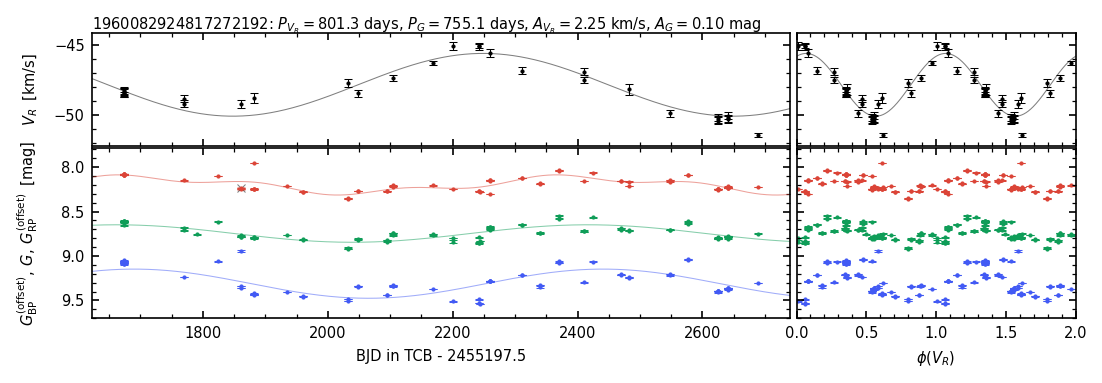}
\includegraphics[width=0.925\textwidth]{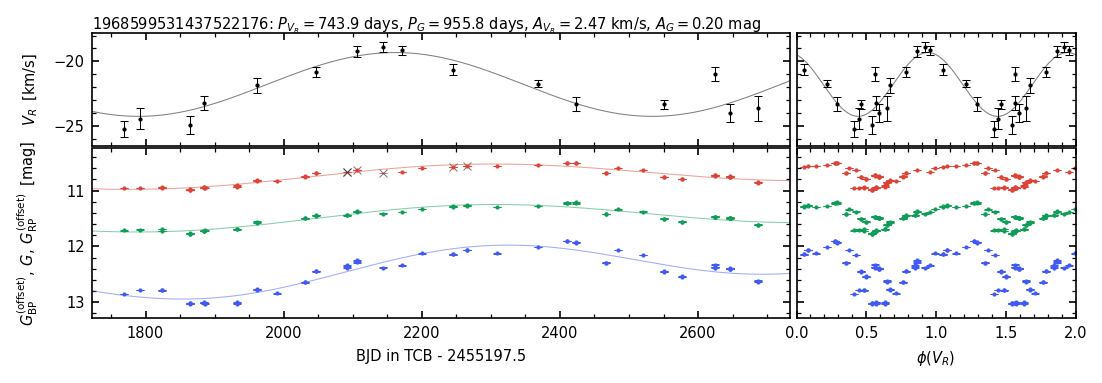}
\includegraphics[width=0.925\textwidth]{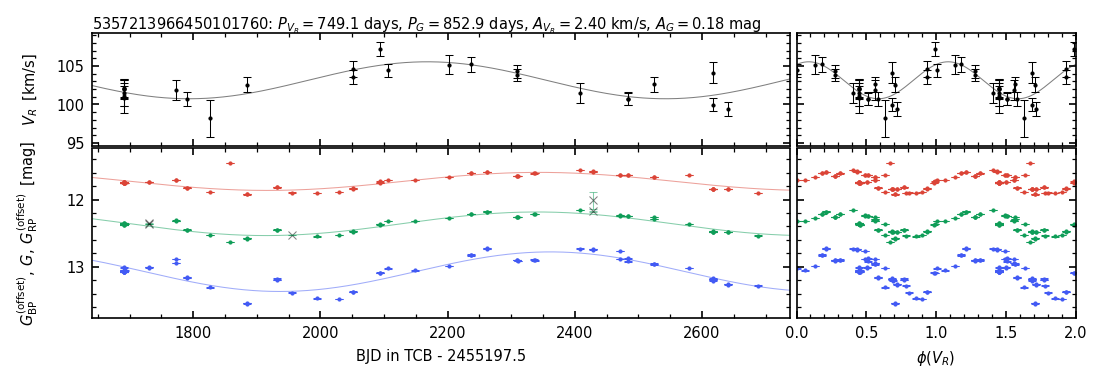}
\includegraphics[width=0.925\textwidth]{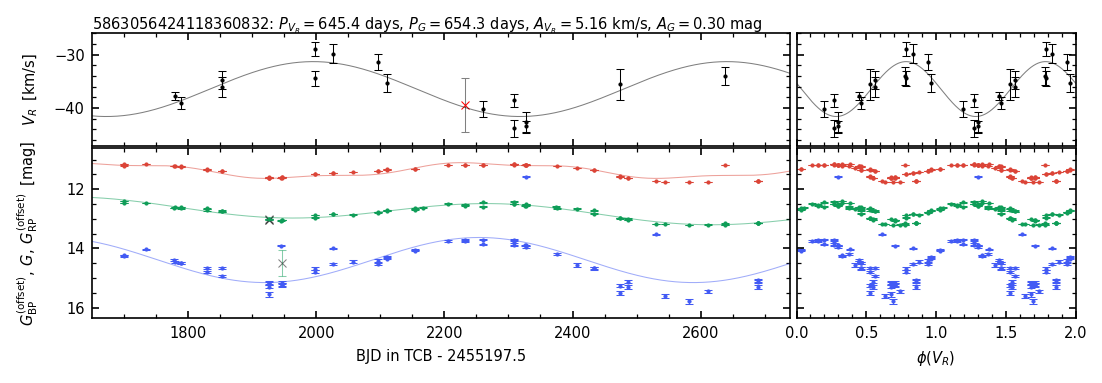}
\caption{Similar to Fig.~\ref{fig:RVTS_ex_mixPph}, but for LPV candidates for which we likely detect a LSP. All data in the panels on the right-hand column are folded by the FPR RV period.}
\label{fig:RVTS_exLSP}
\end{figure*}

It is worth noting that the classification we adopted is generally consistent with ratios between RV and photometric periods. Among the sources in the TQS, 89\% of the ones identified as ELLs show a 2:1 ratio between $\prv$ and $\pph$, whereas about 86\% and 96\% of the LPV candidates showing pulsation or LSP variability, respectively, are consistent with a 1:1 ratio. The corresponding percentages concerning the full FPR are roughly consistent with these values, although some differences arise due to the occurrence of sources showing mixed consistency. Some example time series of ELL, pulsating LPVs, and LSPs from the TQS are displayed in Figs.~\ref{fig:RVTS_exBin}, \ref{fig:RVTS_exPuls}, and~\ref{fig:RVTS_exLSP}, respectively.

\section{Catalog quality}
\label{sec:catalog_quality}

In the present section we check the quality of the FPR catalog. We examine the average RVs in Sect.~\ref{sec:catalog_quality:radial_velocity_estimates}, then we consider the RV variability in Sect.~\ref{sec:catalog_quality:radial_velocity_variability}, and finally we compare with literature data in Sect.~\ref{sec:catalog_quality:comparison_with_rv_data_from_literature}.

\subsection{Radial velocity estimates}
\label{sec:catalog_quality:radial_velocity_estimates}

\subsubsection{Median radial velocity: Consistency with \Gaia DR3}
\label{sec:catalog_quality:median_radial_velocity}

The values $\pubrv$ of RV published in \Gaia DR3 are derived with two different methods depending on the source brightness. In particular, if a source has $\grvs$ brighter than 12 mag (as is the case for all FPR sources), $\pubrv$ is computed as median values over the RV time series \citep{dr3_katz_etal_2023}. It is worth comparing these values with the median values $\medianrv$ resulting from the variability processing pipeline employed for the FPR sample (the field \texttt{median\_rv} in the table \texttt{gaiafpr.vari\_rad\_vel\_statistics}). Some small deviations are expected between the two values because of slight differences in terms of data and methods. Indeed, the definition of median value adopted by the \Gaia variability processing unit and by the \Gaia spectroscopic data processing unit are slightly different. To compute the median value of a dataset consisting of an odd number of values, both methods sort the data and take the middle value. In contrast, if the number of values is even, after sorting the variability processing pipeline takes the smaller of the two middle values, whereas the spectroscopic processing pipeline takes the mean of the two middle values. This means that the former method systematically results in smaller median values than the latter. Moreover, compared to the RV time series used to compute $\pubrv$, the time series published in this FPR can have one fewer epoch as a result of outlier removal during the variability processing (Sect.~\ref{sec:catalog_construction:time_series_processing}).

\begin{figure}
\centering
\includegraphics[width=\columnwidth]{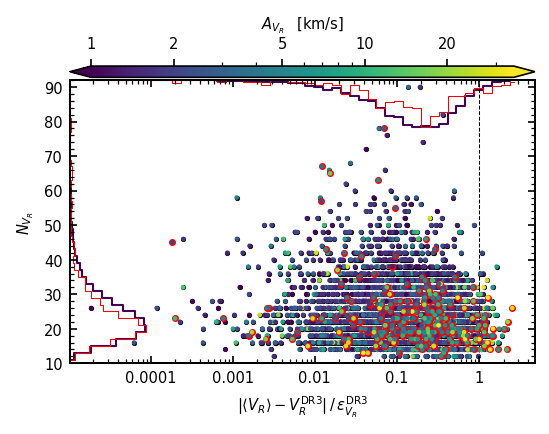}
\caption{Number $\nobsrv$ of epochs retained for RV variability processing against the absolute difference between the median RV derived by variability processing ($\medianrv$) and published in \Gaia DR3 ($\pubrv$), scaled to the RV uncertainty $\erv$ published in DR3. Data points are color-coded by the semi-amplitude of the RV time series model. The time series that had one RV epoch excluded by outlier removal during variability processing are circled in red. The value $\medianrv$ for these sources is computed from one fewer epoch compared to $\pubrv$.}
\label{fig:dif_med_dr3_rel_erdr3_nobs_amp_rv}
\end{figure}

In summary, we find differences between $\medianrv$ and $\pubrv$ for 305 sources whose RV time series had one outlier removed, and for 4\,672 sources whose time series had no outlier removed, but have an even number of epochs. These sources are displayed in Fig.~\ref{fig:dif_med_dr3_rel_erdr3_nobs_amp_rv}, where the absolute value of the difference between $\medianrv$ and $\pubrv$, normalized to the uncertainty $\erv$ on $\pubrv$, is shown against the number of epochs in the FPR time series. For the majority of the sources we find $|\medianrv-\pubrv|\simeq0.2\,\erv$, whereas only 104 sources have a difference between $\medianrv$ and $\pubrv$ that exceeds $\erv$. These sources are usually characterized by a relatively large RV amplitude or a small number of RV measurements. It is easy to see how such features can enhance the impact of outlier removal and methods differences on the calculation of the median RV, especially given the irregular time sampling of \Gaia observations.

\subsubsection{Mean and systemic radial velocities}
\label{sec:catalog_quality:mean_and_systemic_radial_velocities}

\begin{figure}
\centering
\includegraphics[width=\columnwidth]{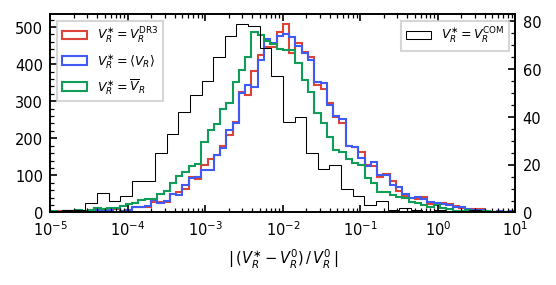}
\caption{Comparison of several average RV indicators with the zero point of the RV time series models. Different indicators are displayed in different colors (red: median value $\pubrv$ published in \Gaia DR3; blue: median value $\medianrv$ computed by variability processing; green: mean value $\meanrv$ computed by variability processing). The histograms show the distribution of absolute difference between each of the average values and $\zprv$, normalized to the latter. The thin black histogram, limited to a subset of the FPR sample, compares $\zprv$ with the center-of-mass velocity $V_R^{\rm COM}$ derived by the non-single stars processing pipeline for \Gaia DR3 (see Sect.~\ref{sec:catalog_quality:mean_and_systemic_radial_velocities} for more details), and refers to the scale on the right-hand side axis.}
\label{fig:hist_rv_reldif_zp_rv}
\end{figure}

\begin{figure}
\centering
\includegraphics[width=\columnwidth]{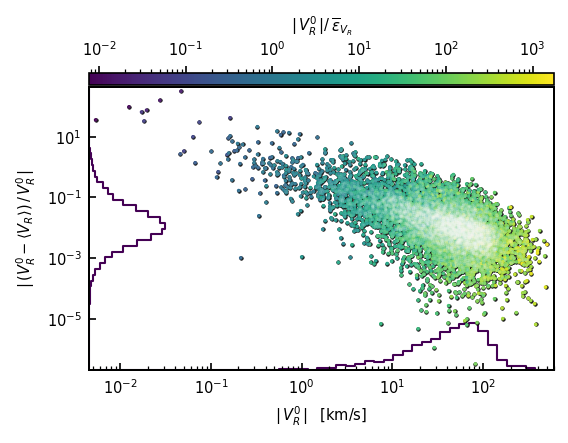}
\caption{Absolute difference between the median value $\medianrv$ and the zero point $\zprv$ RV estimates, scaled to the latter and shown against the absolute value $|\zprv|$ of the latter. Data points are color-coded by the ratio $|\zprv|/\meanerv$, showing that large discrepancies (top portion of the diagram) are associated with absolute values of the systemic RV comparable with or smaller than the RV uncertainty. A white shading indicates a more densely populated area of the diagram.}
\label{fig:reldif_median_zp_rv_zp_rv_div_merr_rv}
\end{figure}

\begin{figure}
\centering
\includegraphics[width=\columnwidth]{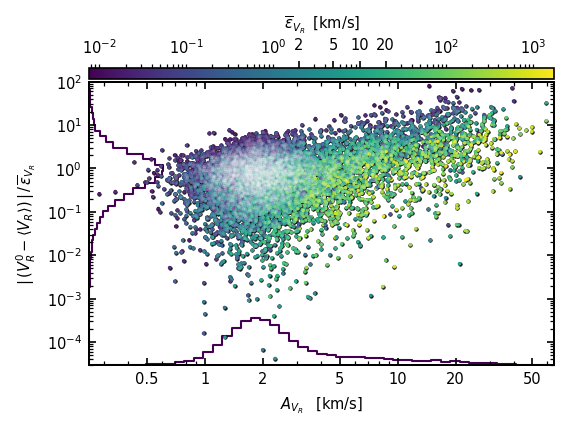}
\caption{Absolute difference between the median value $\medianrv$ and the zero point $\zprv$ RV estimates, scaled to the mean value of the uncertainties on individual RV epochs, and shown against the semi-amplitude $\amprv$ of the RV time series model. A white shading indicates a more densely populated area of the diagram.}
\label{fig:dif_median_zp_rv_div_merr_rv_vs_amp}
\end{figure}

\begin{figure}
\centering
\includegraphics[width=\columnwidth]{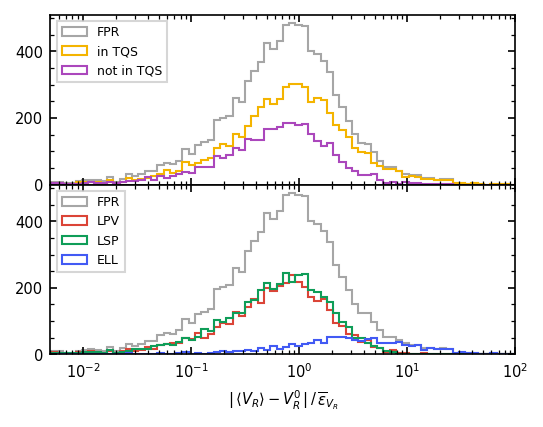}
\caption{Absolute difference between $\medianrv$ and $\zprv$, scaled by the mean of epoch RV uncertainties, for various subsets of the FPR sample. In the top panel, the orange curve corresponds to sources flagged for high consistency between RV and photometric periods, whereas all other sources are represented by the purple curve. In the bottom panel, the red, green and blue curve correspond to sources tentatively identified as pulsating LPVs, LPVs showing LSP variability, or ellipsoidal variables, respectively (see Sect.~\ref{sec:catalog_content}). The gray curves in both panels represent the whole FPR sample.}
\label{fig:hist_diff_median_zp_div_meanoferr_rv_types}
\end{figure}

\begin{figure}
\centering
\includegraphics[width=\columnwidth]{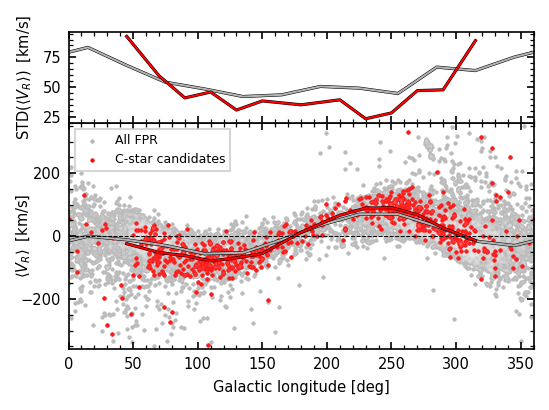}
\caption{Median RV ($\medianrv$) of FPR sources as a function of their galactic latitude. Red sources are C-star candidates. The solid lines indicate median values over bins of galactic latitude. The curves show the median value and standard deviation of $\medianrv$ in bins of galactic latitude (for C-stars these statistics are limited to galactic longitudes between 30$^{\circ}$ and 330$^{\circ}$, excluding regions where they are scarce).}
\label{fig:gall_std_median_rv_cstars}
\end{figure}

The uneven sampling of \Gaia time series also means that the median RV is not necessarily a good indicator of the systemic RV, that is the center-of-mass velocity, an accurate estimate of which generally requires the RV variability to be modeled. In principle, the zero-point $\zprv$ of the RV time series model (i.e., the constant term $c_{0,\rv}$ in Eq.~\ref{eq:time_series_model}, unpublished but computable with the published time series) is more representative of the systemic RV than an average over the time series. In order to assess this, we examine the fractional deviations between $\medianrv$ and $\zprv$ by taking the absolute value of their difference and scaling it to $|\zprv|$. The distribution of this quantity is displayed in Fig.~\ref{fig:hist_rv_reldif_zp_rv}, together with the fractional difference with respect to $\zprv$ of $\pubrv$ and of the mean value $\meanrv$ derived from variability processing. This diagram shows that in most cases the median values are compatible with $\zprv$ to within few percents, whereas the mean value performs slightly better, typically within less than a 1\%. Both indicators are consistent with $\zprv$ within about 10\% for the bulk of the sample.

As is discussed in Sect.~\ref{sec:catalog_quality:radial_velocity_variability}, a subset of the FPR sources are also present in the \Gaia DR3 non-single stars table, which provides center-of-mass velocities based on orbital solutions. We include in Fig.~\ref{fig:hist_rv_reldif_zp_rv} a comparison between these values and $\zprv$, that are found to be in excellent agreement, being usually compatible within 0.03\%. We leave further comparisons with the non-single stars results for the next section.

While such fractional difference is helpful to have a quick glance at the general degree of agreement between various indicators, it fails to characterize the stars having a very small systemic RVs (comparable with or smaller than the RV uncertainty) which lead to the tail of the distribution at large values, as can also be appreciated from Fig.~\ref{fig:reldif_median_zp_rv_zp_rv_div_merr_rv}. Therefore, we further examine the sample by scaling the difference $|\medianrv-\zprv|$ by the mean value $\meanerv$ of the uncertainties on individual RV epochs on a given time series. The resulting quantity is displayed in Fig.~\ref{fig:dif_median_zp_rv_div_merr_rv_vs_amp} as a function of the semi-amplitude of the RV time series model. Its distribution peaks around one, indicating that for the majority of the time series the value $\medianrv$ is compatible with $\zprv$ within the mean RV uncertainty. However, there is a clear tendency for this compatibility to degrade with increasing RV amplitude. This is primarily due to the uneven and irregular time sampling of \Gaia observations. While this is rarely an issue for sources that display little or no variability, it is something to be kept in mind when dealing with LPVs and ELLs, because:
\begin{itemize}
    \item   they have periods comparable with the observational baseline, so the \Gaia time series may cover only a small number of variability cycles;
    \item   they have large variability amplitudes, and so observations taken at a random phase of the variability cycle may be very far from the central value;
    \item   the \Gaia scanning law is such that a large number of observations may be concentrated within an interval of time much shorter than the typical time scales of LPVs and ELLs; these ``clusters'' of epochs have a very high statistical weight, but carry little more information than a single measurement taken in the middle of that time interval.
\end{itemize}
The combination of these effects increases the chance of selectively over-sampling or under-sampling a specific phase of the variability cycle, thereby skewing the median away from the mid-point of the RV curve. For similar reasons, the amplitude of variability may be underestimated by statistical indicators such as the standard deviation or the interquartile range. Some examples of RV curves showing these effects are provided in Appendix~\ref{app:median_RV_in_unevenly_sampled_time_series}.

We note that the top-quality sample shows a slightly larger difference, on the mean, between $\medianrv$ and $\zprv$ than the full FPR sample. This is due to the presence of a larger fraction of ellipsoidal variables in the former sample, which have large RV ranges (see Sect.~\ref{sec:catalog_content} and Fig.~\ref{fig:hist_diff_median_zp_div_meanoferr_rv_types}).

Finally, we recall that the epoch RVs are measured from RVS data by comparing observed spectra with synthetic stellar spectra used as templates \citep[see Sect.~6.4.8 of the \Gaia DR3 documentation][]{sartoretti_etal_2022_dr3_doc_ch6}. In this FPR the templates have $\Teff$ ranging from 3100 K to 7500 K (typically between 3300 K and 3800 K) and $\log(g)$ in the range -0.5 to 5.0, and are restricted to O-rich stars \citep{dr3_katz_etal_2023}. As LPVs can have stellar parameters outside these ranges and include C-stars, the matched template is not necessarily the most appropriate in terms of atmospheric parameters $\Teff$ and $\log(g)$, which might impact the derived zero-point and median RVs. Yet, the analysis of the overall features of the FPR sample does not highlight any clear systematic discrepancy with respect to what would be reasonably expected. For instance, the distribution of RV as a function of galactic longitude (Fig.~\ref{fig:gall_std_median_rv_cstars}) does not show a wider spread for C stars than for O-rich LPVs. Likewise, the sky map presented in Sect.~\ref{sec:catalog_overview} gives essentially the same picture as Fig.~5 of \citet{dr3_katz_etal_2023}, despite being limited to a selection of stars (i.e., cool red giants and including C-stars) that are substantially more exposed to the risk of template mismatch than the bulk of sources used by these authors in their figure. More importantly, the variability properties of the RV curves, that are the object of this FPR, are not expected to be significantly affected by the template mismatches, as all epoch RV measurements should be equally impacted by the mismatch. The LPVs undergoing large changes in $\Teff$ throughout the pulsation cycle may represent an exception as a different spectral template should be adopted for different epochs, but it is unclear how this might impact the RV variability data. Visual inspection of the RV curves and comparison with their photometric light curves give further support to these considerations.

\subsection{Radial velocity variability}
\label{sec:catalog_quality:radial_velocity_variability}

For the purpose of assessing the quality of the periods and amplitudes derived from RV time series of sources that are identified as ellipsoidal binary stars, we compare the FPR data with the results of non-single star (NSS) processing from \Gaia DR3. In particular, we consider the data from the \texttt{nss\_two\_body\_orbit} table from the \Gaia archive (Gosset et al. in prep.). We find that 855 of the FPR sources are also found in that table. Based on the classification outlined in Sect.~\ref{sec:catalog_content}, we identify 353 of them as ELL candidates, 296 as pulsating LPVs, and 205 as LPVs with a LSP.

\begin{figure}
\centering
\includegraphics[width=\columnwidth]{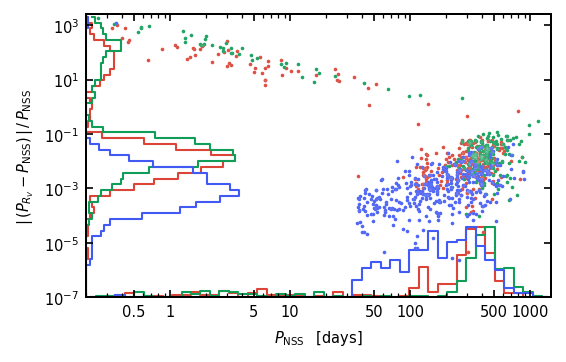}
\caption{Relative difference between the period $\prv$ estimated by variability processing and the value derived from the non-single star pipeline, relative to the latter, for the FPR sources that have a counterpart in the \Gaia DR3 table of orbital parameters of non-single stars. Data points are color-coded by variability type according to the classification presented in Sect.~\ref{sec:catalog_content} (blue: ellipsoidal variables; red: pulsating LPVs; green: LPVs with LSP).}
\label{fig:reldif_P_rv_nss}
\end{figure}

\begin{figure*}
\centering
\includegraphics[width=0.925\textwidth]{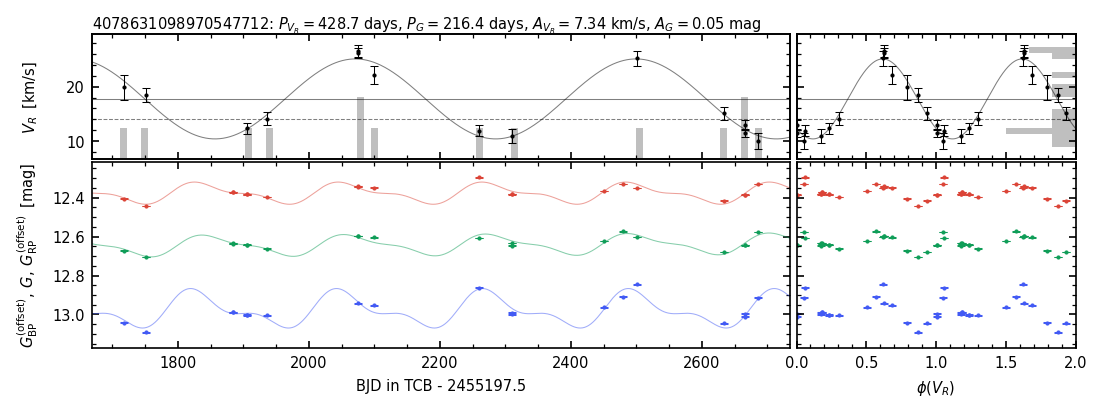}
\caption{Similar to Fig.~\ref{fig:RVTS_ex_mixPph}, showing the \Gaia RV and photometric curves of the ELL candidate showing a large mismatch between the FPR RV period (428.7 days) and the value published in the \Gaia DR3 NSS table (0.35 days). For the purpose of visualization, the $\gbp$ and $\grp$ time series are offset by an arbitrary amount. All data in the panels on the right-hand column are folded by the FPR RV period.}
\label{fig:RVTS_nssbin_Pmismatch}
\end{figure*}

We compare our RV periods with the NSS values in Fig.~\ref{fig:reldif_P_rv_nss}, which displays the absolute difference between the two periods scaled to the latter. In most cases we find a good degree of period compatibility (typically within a few percents), even for the sources that we do not classify as ELL candidates, although the agreement is better for the latter. In general, the comparison is slightly better for objects identified as pulsating LPVs compared with the LSP candidates. For sources that we identify as ellipsoidal variables, the periods are always within 10\% of each other, and typically compatible within 0.1\%. There is only one exception (whose time series are displayed in Fig.~\ref{fig:RVTS_nssbin_Pmismatch}) for which we derive a 428.7 days period in stark difference with the NSS period of 0.35 days. As the RV time series only covers 12 visibility periods, it is hard to conclude which period is more realistic. Both values correspond to a strong signal in the periodogram, but we do not detect the latter as our processing is limited to periods longer than 10 days. In any case, periods as short as 0.35~days are not expected for ellipsoidal red giants. A similar situation occurs for a number of other sources for which NSS results in relatively short periods, and hence there exists a large difference with respect to the FPR period (sequence of points in the upper part of Fig.~\ref{fig:reldif_P_rv_nss} extending to low NSS periods). However, none of these sources is identified as a ELL candidate according to our classification criteria, which casts some doubt on the validity of the orbital model adopted for modeling their RV time series for the NSS processing.

\begin{figure}
\centering
\includegraphics[width=\columnwidth]{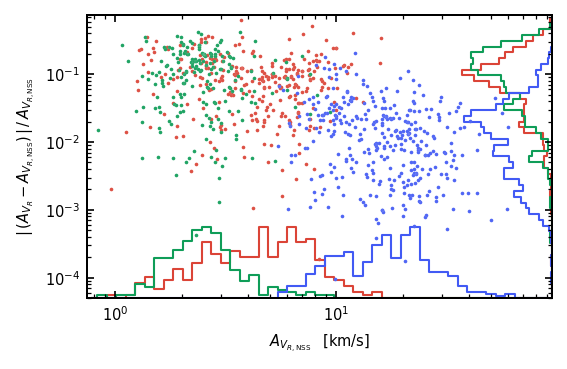}
\caption{Similar to Fig~\ref{fig:reldif_P_rv_nss}, but comparing the RV semi-amplitudes rather than the periods.}
\label{fig:reldif_amp_rv_nss}
\end{figure}

We perform a similar analysis in terms of the RV semi-amplitude, comparing the value derived by the variability pipeline with the NSS results (Fig.~\ref{fig:reldif_amp_rv_nss}). We find a qualitatively similar picture as for the periods, the compatibility between the NSS and our semi-amplitudes being typically within 2-3\% for ELL candidates, 10\% for pulsating LPV candidates, and 20\% for LSP candidates. It is interesting to observe that the RV curve asymmetries frequently displayed by pulsating LPVs are often accounted for by a large eccentricity in the binary models adopted by NSS.

Finally, we note that the NSS data for the matched sources result from the assumption of a single-lined spectroscopic binary model, with the only exception of the source \texttt{2567779977831471232} for which an orbital astrometric binary model was adopted, obtaining a 928.7$\pm$85.3 days period. According to the classification presented in Sect.~\ref{sec:catalog_content}, we also identify that source as a binary (ELL) candidate, but we find a RV period significantly shorter ($\prv=695.3\,\days$), although we note that the RV observation covers just about 800 days, hence the period cannot be accurately constrained.

\subsection{Comparison with RV data from literature}
\label{sec:catalog_quality:comparison_with_rv_data_from_literature}

We searched the literature for RV data to compare with this FPR, and found them to fall into two main categories. The first one concerns large-scale spectroscopic surveys providing extensive catalogs of RV data. The chances of finding matching objects against these source lists are relatively high, but at the same time they only allow for comparing average RVs, as they are based on single-epoch observations or a few epochs at most, and rarely provide RV time series. In Sect.~\ref{sec:catalog_quality:comparison_with_rv_data_from_literature:comparison_with_other_rv_surveys} we present a comparison with a few such catalogs.

Besides these surveys there exist a number of smaller-scale observational programs targeting specific types of stars or fields of the sky. These studies involve a more focused analysis and validation of the RV data of the targets, and often result in the publication of the time series. We attempted to cross-match the FPR catalog with the source lists from various such literature works, but found only a few matches, that are examined in Sect.~\ref{sec:catalog_quality:comparison_with_rv_data_from_literature:multi_epoch_rv_data}.

\subsubsection{Comparison with RV surveys}
\label{sec:catalog_quality:comparison_with_rv_data_from_literature:comparison_with_other_rv_surveys}

\begin{table}
\caption{Number of FPR sources cross-matched with external spectroscopic surveys.}
\label{tab:table_xms_types}
\centering
\begin{tabular}{lcccc}
\hline\hline
        & FPR    & APOGEE & GALAH$^{(a)}$ & RAVE   \\
\hline
LPV     & 4\,084 &     34 &       58 (65) &    503 \\
LSP     & 4\,421 &     41 &       89 (95) &    643 \\
ELL     & 1\,109 &     14 &       20 (21) &     80 \\
\hline
Total   & 9\,614 &     89 &     167 (181) & 1\,226 \\
in TQS  & 6\,093 &     60 &     108 (119) &    784 \\
\hline
\end{tabular}
\tablefoot{$^{(a)}$ Numbers indicate the sources with a published value of RV in GALAH, while the total number of matches is given in parentheses.}
\end{table}

\begin{figure}
\centering
\includegraphics[width=\columnwidth]{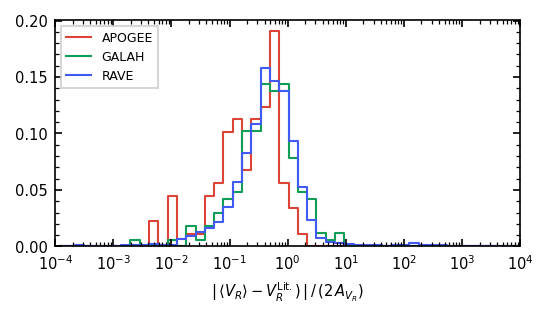}
\caption{Comparison between the median RV from this FPR with the average values provided by external catalogs (red: APOGEE; green: GALAH;, blue: RAVE). We consider the absolute difference between the two values, normalized to the peak-to-peak model amplitude. We note that the histograms are normalized to their area.}
\label{fig:hist_rv_dif_surveys}
\end{figure}

We compare the FPR results with the data published by three spectroscopic surveys: the Apache Point Observatory Galactic Evolution Experiment survey \citep[APOGEE, data release 16,][]{apogee_dr16}, the GALactic Archaeology with HERMES survey \citep[GALAH, data release 3,][]{galah_dr3}, and the RAdial Velocity Experiment survey \citep[RAVE, data release 6,][]{rave_dr6}. For the former two, we rely on the cross-match with \Gaia DR3 performed as part of their data release process (they provide best-match \Gaia DR3 source IDs), whereas for the latter we adopt the results of the \Gaia cross-match with external catalogs
(\texttt{gaiadr3.ravedr6\_best\_neighbour} table in the \Gaia archive).

A summary of the numbers of matched sources is provided in Table~\ref{tab:table_xms_types}. We find 89 matches with APOGEE, 167 with GALAH, and 1\,226 with RAVE (of which 60, 119, and 784, respectively, are in the TQS). In Fig.~\ref{fig:hist_rv_dif_surveys} we provide an overview of the comparison between the median RV $\medianrv$ from the FPR and the averages given by each of these surveys. Namely, that figure shows the distribution of the absolute difference between $\medianrv$ and the literature value, divided by the peak-to-peak amplitude of the \Gaia RV time series model. We employ the median RV rather than the zero-point RV for the purpose of comparison as the values published in each of the three surveys we compare with are also derived as averages (unless coming from single-epoch observations).

We find comparable results for all three surveys, with a slightly higher degree of compatibility with APOGEE. The differences in RV are usually comparable with or smaller than the amplitude of RV variability. As expected, larger deviations occur for sources having large RV amplitudes, and therefore typically for ellipsoidal variables, as well as for sources with few observations in the examined external surveys. Indeed, the GALAH RVs for the matched sources are all based on single-epoch measurements. Of the sources matched with APOGEE, 29 are RVs based on single-epoch data, 52 have between 2 and 4 epoch spectra, and the remaining 8 sources have at most 8 spectral observations. Similarly, the vast majority of the sources matched with RAVE (1\,141) has a single epoch, while 70 have been observed during two epochs, and only 15 have more than two epochs (at most 7).

While the poor time coverage in the external surveys we compare with is the most likely cause of discrepancy with our RV values, another possible cause could be related with spectral mismatch, whether in the \Gaia pipeline (as mentioned in Sect.~\ref{sec:catalog_quality:mean_and_systemic_radial_velocities}) or in the literature survey.
This seems to be the case, for example, for a few sources for which the RAVE catalog indicates effective temperatures in excess of 8\,000 K and that we identify as LPVs showing the largest RV differences compared to the FPR.
We found no evident correlation between the RV differences and the differences between the spectral parameters adopted by the \Gaia and external survey pipelines. A mismatch between the values of $\Teff$ or $\log(g)$ does not necessarily lead to large RV differences, at least as long as these values are not unreasonably far from the range of stellar parameters typical of LPVs.

\subsubsection{Comparison with literature multi-epoch RV data}
\label{sec:catalog_quality:comparison_with_rv_data_from_literature:multi_epoch_rv_data}

There exist a limited number of literature studies providing multi-epoch RV data of LPVs and ellipsoidal variables, hence we found only three FPR sources that we could compare with published RV time series. They are summarized in Table~\ref{tab:table_cf_rvcurves}, where the literature period is compared with the values $\pg$ and $\prv$ given in the FPR.

\begin{table*}
\caption{Properties of the sources whose RV time series are compared with literature multi-epoch RV studies.}
\label{tab:table_cf_rvcurves}
\centering
\begin{tabular}{cccccccc}
\hline\hline
Name   & \Gaia DR3 source ID & Var. type$^{(a)}$ & Sp. type$^{(a)}$ & $\pg$(FPR) & $\prv$(FPR) & $P_{\rm Lit.}$ & Reference \\
 & & & & $\days$ & $\days$ & $\days$ & \\
\hline
\object{AR Cep} & 2300884800185315712 &               SRb &            M4III &      575.2 &       550.4 &                & \citet{alvarez_etal_2001} \\
\object{RS CrB} & 1372316959598411520 &               SRa &               M7 &      330.9 &       305.9 &  328.3$\pm$2.6 & \citet{hinkle_etal_2002} \\
\object{R Nor}  & 5985676640941632384 &              Mira &         M3e-M6II &      505.5 &       496.7 &            507 & \citet{lebzelter_etal_2005a} \\
\hline
\end{tabular}
\tablefoot{
  \tablefoottext{a}{The variability and spectral types are taken from the corresponding reference paper.
  }
}
\end{table*}

\begin{figure*}
\centering
\includegraphics[width=0.925\textwidth]{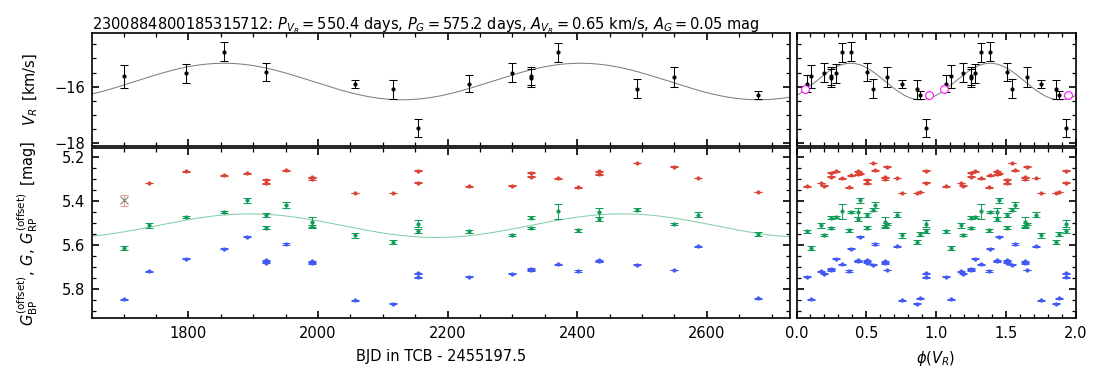}
\includegraphics[width=0.925\textwidth]{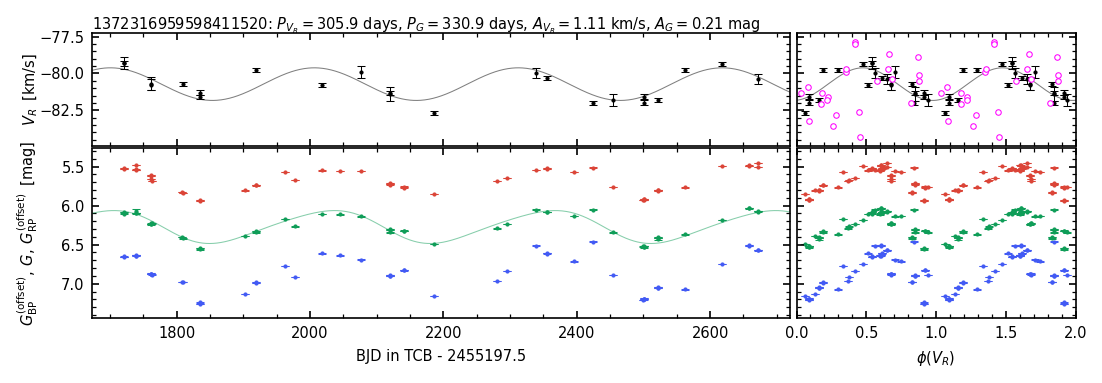}
\includegraphics[width=0.925\textwidth]{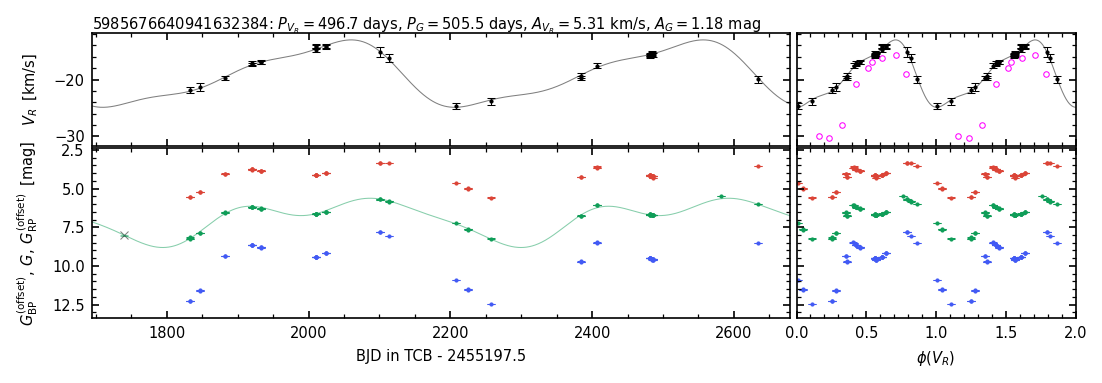}
\caption{Similar to Fig.~\ref{fig:RVTS_ex_mixPph}, but showing the \Gaia time series for the sources compared with literature as discussed in Sect.~\ref{sec:catalog_quality:comparison_with_rv_data_from_literature:multi_epoch_rv_data} (see also Table~\ref{tab:table_cf_rvcurves}). The sources displayed from top to bottom are the SRb star \object{AR Cep} \citep[compared with][]{alvarez_etal_2001}, the binary SRa star \object{RS CrB}  \citep[compared with][]{hinkle_etal_2002}, and the O-rich Mira \object{R Nor} \citep[compared with][]{lebzelter_etal_2005a}. Literature RV time series are displayed as magenta circles (with arbitrary phase offset) in the panels showing the folded RV curve. In each case, the \Gaia RV period (indicated in the header of each panel) is used for folding.}
\label{fig:RVTS_cf_lit}
\end{figure*}

\begin{figure}
\centering
\includegraphics[width=\columnwidth]{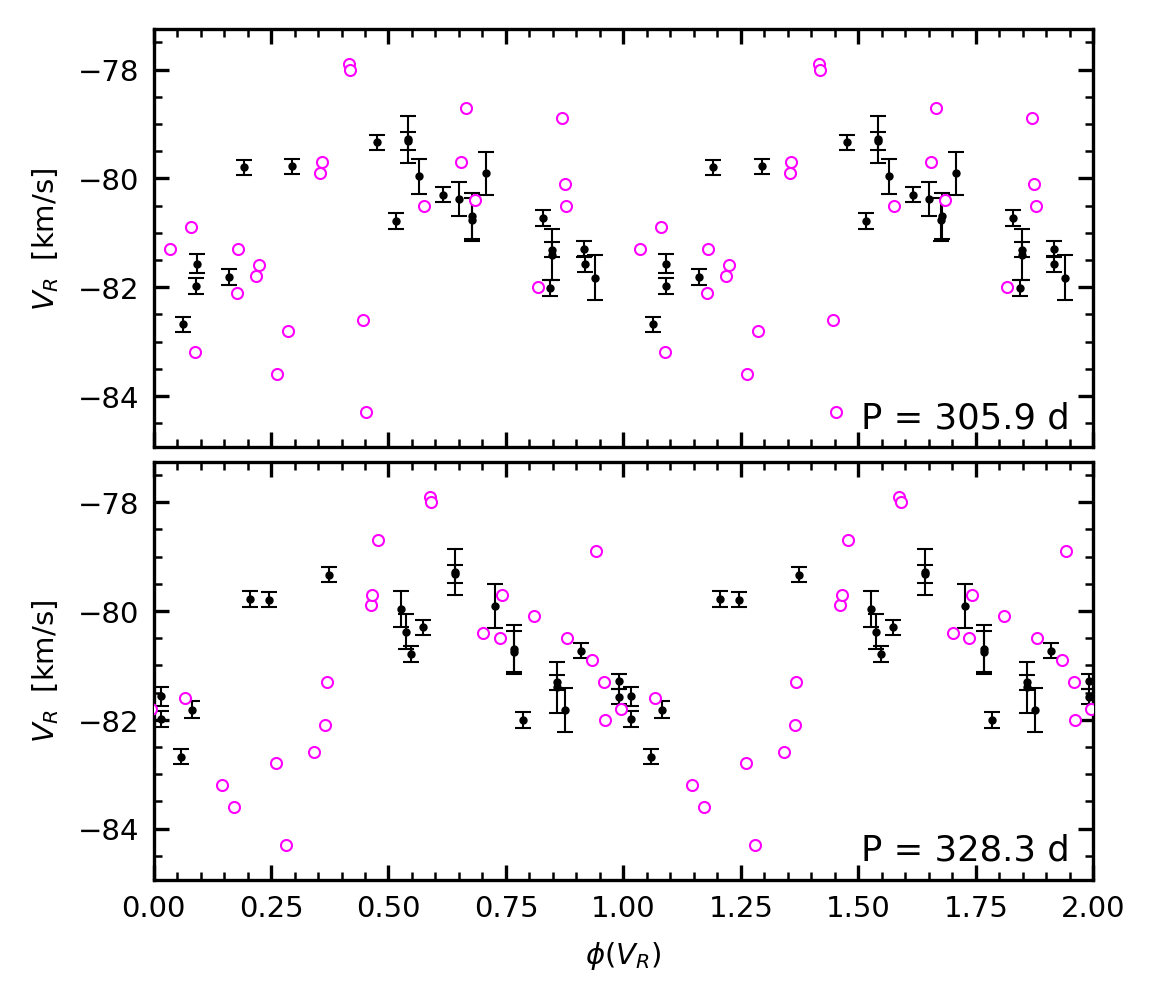}
\caption{Phased RV curve of \object{RS CrB} folded with the FPR RV period (top panel) and with the period derived by \citet{hinkle_etal_2002} (bottom panel). Symbols have the same meaning as in Fig.~\ref{fig:RVTS_cf_lit}.}
\label{fig:RVTS_RSCrB_cf_hinkle02_2}
\end{figure}

The first source for which we found a match is the SRb star \object{AR Cep} observed by \citet{alvarez_etal_2001}. They obtained spectral observations at optical wavelengths and derived RVs by cross-correlation with template spectra of types K0-III or M4-V. For the matched source they provide two RV epochs separated by about 2 months (roughly 10\% of the period found in the \Gaia time series). They found RVs between $-16.0\,\kms$ and $-16.5\,\kms$ (with little differences between the two spectral templates) that are in good agreement with our results, but they do not provide a period to compare with. Their RVs are shown on top of the \Gaia RV curve folded with the best-fit RV model in the top section of Fig.~\ref{fig:RVTS_cf_lit} (an arbitrary phase offset is applied for the purpose of visual comparison). This source is not part of the TQS, and according to the criteria defined in Sect.~\ref{sec:catalog_content} we identify it as a pulsating LPV.

Another source for which we found a match is the SRa star \object{RS CrB}, whose RV curve has been examined by \citet{hinkle_etal_2002} based on near-IR spectral observations. They obtained measurements for 23 epochs spanning 5 years, and derived a period of $328.3\pm1.6\,\days$ that agrees with the value $\pg=330.9\,\days$ we derived from the $G$-band time series, but is less consistent with our RV period, $\prv=305.9\,\days$. The time series of this source are compared in the middle section of Fig.~\ref{fig:RVTS_cf_lit}, and in Fig.~\ref{fig:RVTS_RSCrB_cf_hinkle02_2} we limit the comparison to the RV time series, folding them with both the FPR and literature RV periods. We note that both these periods are consistent with the longest of the three periods of RS CrB reported in literature, interpreted as resulting from binarity rather than pulsation. Indeed, the value given by \citet{hinkle_etal_2002} is based on an orbital solution. Interestingly, we classify this source as a LPV but it lies very close to the boundary line defined by Eq.~\ref{eq:condition_lsp}, so that its period is identified by pulsation. As clearly seen in Fig.~\ref{fig:RVTS_cf_lit}, the photometric and RV time series of this source are consistent with each other, as a result this source is part of the TQS.

Finally, we found that the O-rich Mira \object{R Nor} is present both in the FPR and in the list of sources investigated by \citet{lebzelter_etal_2005a}. They derived RV measurements from nine near-IR spectra covering about an entire pulsation period of the star, for which they report a value of 507 days. This is in good agreement with both the periods we derived from the $G$-band and RV time series ($\pg=505.5\,\days$, $\prv=496.7\,\days$). The time series of this star are compared in the bottom section of Fig.~\ref{fig:RVTS_cf_lit}. We identify this source as a pulsating LPV, but it is not part of the TQS.

It is worth noting that the RVs we are comparing for these three sources have been derived from different spectral ranges. Indeed, the observations by \citet{alvarez_etal_2001} are taken at short wavelengths, between 390.6 nm and 681.1 nm, whereas the \Gaia RVS covers the range between 846 nm and 870 nm. The spectral observations by \citet{hinkle_etal_2002} and by \citet{lebzelter_etal_2005a} cover an even more different range, being centered around $1.6\,\mu$m. This means that, if the observed variability results from pulsation, the RV measurements concern layers at different depths in the stellar atmosphere, and therefore the RV curves derived from different spectral ranges differ in terms of amplitude or show an offset between each other, which seems to be the case for R Nor (top-right panel in the bottom section of Fig.~\ref{fig:RVTS_cf_lit}).
While it should be kept in mind that the offset might be at least partially caused by a template mismatch in the processing of \Gaia RVs (see Sect.~\ref{sec:catalog_quality:mean_and_systemic_radial_velocities}), this figure clearly illustrates how such effects do not impact the quality of the RV variability parameters provided as part of this FPR.
In the case of \object{AR Cep} (top section of Fig.~\ref{fig:RVTS_cf_lit}) the small number of literature epochs to compare with makes it difficult to draw conclusions in this respect.
In contrast, we note that the RV curves of \object{RS CrB} (middle section of Fig.~\ref{fig:RVTS_cf_lit}) are well compatible in terms of average RV as well as amplitude (Fig.~\ref{fig:RVTS_RSCrB_cf_hinkle02_2}).
This tends to confirm the orbital nature of the variability we detect for this source.

\section{Catalog overview}
\label{sec:catalog_overview}

\begin{figure*}
\centering
\includegraphics[width=\textwidth]{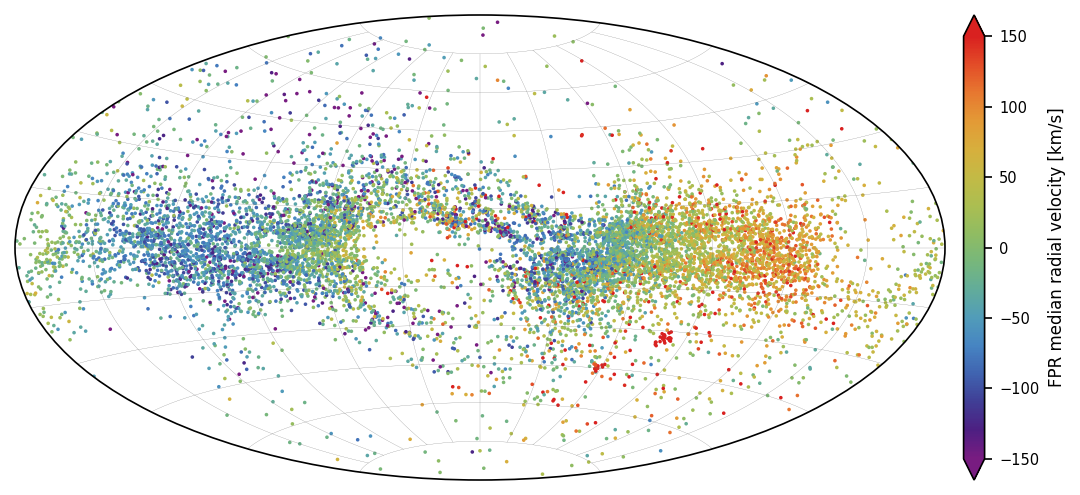}
\caption{Sky distribution in galactic coordinates of the sources in the FPR sample, color-coded by their median RV (the RV range is limited to $150\,\kms$ in absolute value for visibility). The velocity pattern resulting from the rotation of the Milky Way Galaxy is clearly visible, as well as several bright sources in the Magellanic Clouds having $\rv\gtrsim150\,\kms$.}
\label{fig:skymap_rv}
\end{figure*}

\begin{figure}
\centering
\includegraphics[width=\columnwidth]{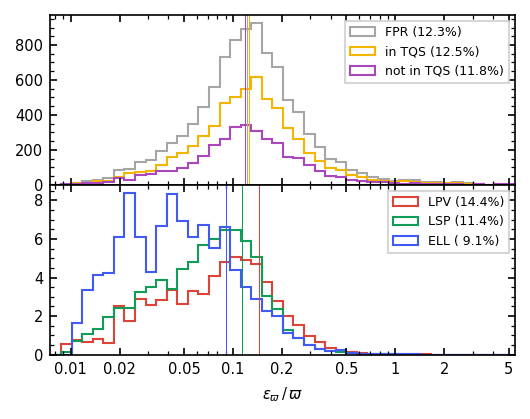}
\caption{Distribution of relative parallax errors in the FPR, excluding 92 sources with negative parallaxes. In the top panel, the histograms represent the full FPR (gray curve), the TQS (orange curve), and the sources that are not part of the TQS (purple). In the bottom panel, they indicate the sources identified as pulsating LPVs (red), LPVs showing a LSP (green), or ELL (blue). Note that the histograms in the bottom panel are normalized to their area. For each subset, the vertical line indicates the median relative error, which is reported as a percentage in the legends.}
\label{fig:hist_parallax_error_types}
\end{figure}

\begin{figure}
\centering
\includegraphics[width=\columnwidth]{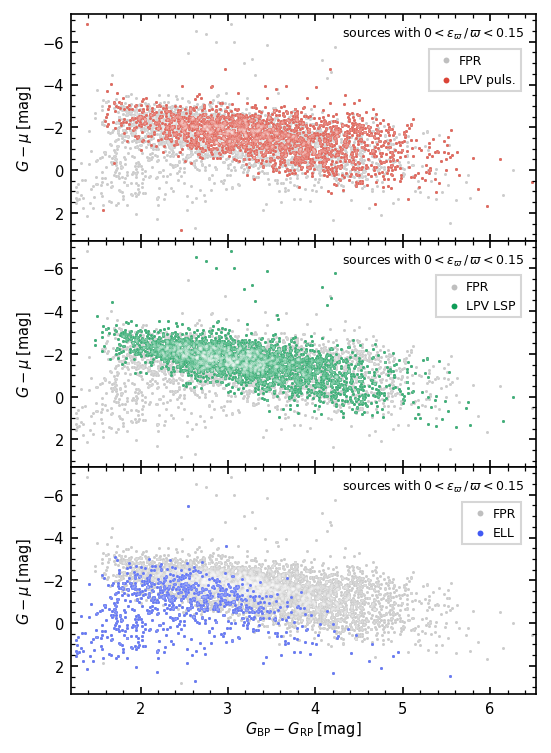}
\caption{\Gaia color~--~absolute magnitude diagram of the FPR sources with positive parallaxes and parallax errors smaller than 15\% (gray symbols), with $\mu$ being the distance modulus. The subsets of sources with different variability types are highlighted in the top (pulsating LPVs, red symbols), middle (LPVs with a LSP, green symbols) and bottom panels (ELL, blue symbols).}
\label{fig:fpr_hrd}
\end{figure}

\begin{figure*}
\centering
\includegraphics[width=\columnwidth]{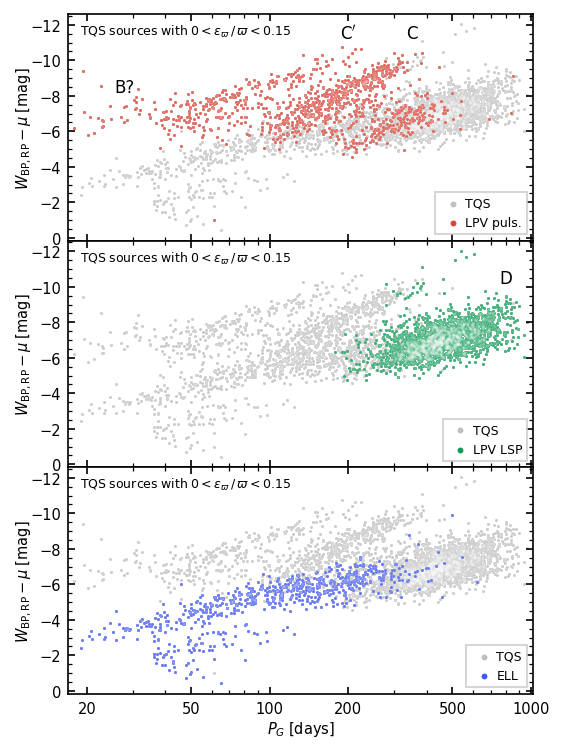}
\includegraphics[width=\columnwidth]{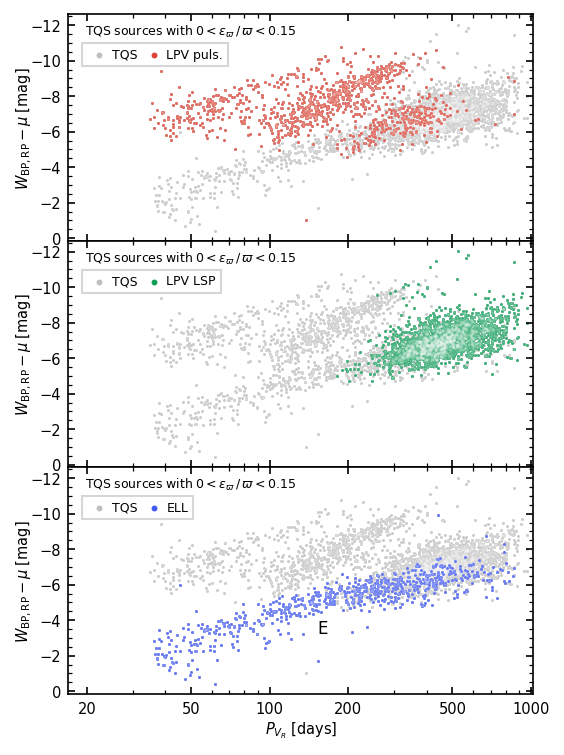}
\caption{Similar to Fig.~\ref{fig:fpr_hrd}, but showing the period~--~luminosity diagram. The panels in the left- and right-hand side columns employ the periods derived from the $G$-band and $\rv$ time series, respectively, and the \Gaia Wesenheit index $\wbprp$ is used as brightness indicator. The extent of the horizontal axis is intentionally kept the same in all panels for the purpose of comparison.}
\label{fig:fpr_pld}
\end{figure*}

\begin{figure}
\centering
\includegraphics[width=\columnwidth]{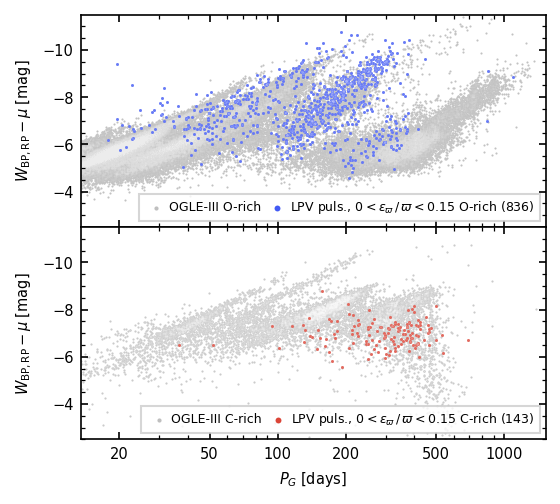}
\caption{Similar to the Fig.~\ref{fig:fpr_pld}, but limited to pulsating LPVs and distinguishing between sources identified as C-stars (red symbols, bottom panel) or not (blue symbols, top panel) according to \Gaia low-resolution spectra \citep{dr3_lpv}. The numbers of these sources are indicated in the legends. Gray points in the background are O-rich (top panel) or C-rich (bottom panel) LPVs in the LMC from the OGLE-III catalog (sources whose OGLE primary period is flagged as LSP are excluded from the bottom panel). For the LMC, we adopt an average distance modulus $\mu_{\rm LMC}=18.49\,\mags$ following \citet{degrijs_etal_2017}.}
\label{fig:fpr_tqs_pld_lpv_chem}
\end{figure}

\begin{figure}
\centering
\includegraphics[width=\columnwidth]{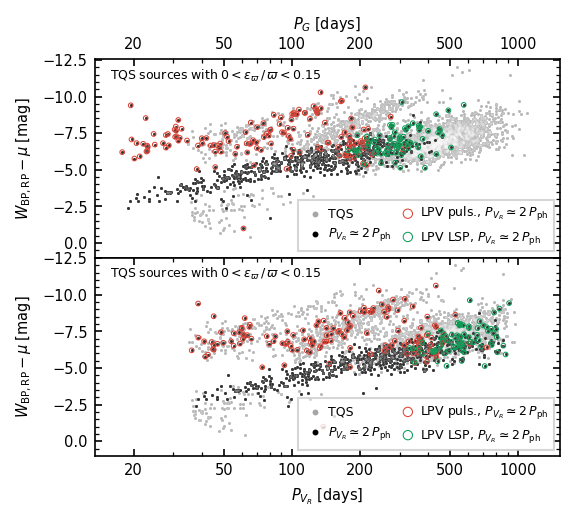}
\caption{Similar to the right-hand side panels in Fig.~\ref{fig:fpr_pld}, but with the sources having $\prv\simeq\,2\pph$ highlighted in black. The PLD constructed with $\pg$ is shown in the top panel and the one constructed with $\prv$ in the bottom panel for the TQS sources with parallax errors better than 15\% (gray symbols), and highlighting the sources with $\prv\simeq\,2\pph$. Red and green circles indicate LPVs showing pulsation and a LSP, respectively. Only five of them are identified as C-rich.}
\label{fig:fpr_pld_Prv2Pg}
\end{figure}

In this section we provide an overview of the FPR catalog, with the purpose of showcasing its content of physical information. We first present the sky distribution of the catalog in Sect.~\ref{sec:catalog_overview:sky_distribution}, then investigate in Sect.~\ref{sec:color_absolute_magnitude_diagram} the distributions in the color~--~absolute magnitude diagram of subsets of sources with good relative parallax precisions, and finally analyze in Sect.~\ref{sec:period_luminosity_diagrams} the period-luminosity relations of various subsets of the top-quality sample.

\subsection{Sky distribution}
\label{sec:catalog_overview:sky_distribution}

Figure~\ref{fig:skymap_rv} illustrates the sky distribution of the sources in this FPR, each being color-coded by the median RV $\medianrv$ resulting from the \Gaia variability processing pipeline. As discussed in Sect.~\ref{sec:catalog_quality:median_radial_velocity}, for the vast majority of the FPR sources the value of $\medianrv$ is compatible with the value of median RV derived from the \Gaia spectroscopic data processing pipeline and published as part of \Gaia DR3, even though the processing details are slightly different. Indeed, the physical picture emerging from Fig.~\ref{fig:skymap_rv} is entirely consistent with the expectations in terms of the Galactic rotation curve.

This also provides support to the expectation that the occurrence of spectral template mismatches has a minor impact on the accuracy of the median RV as an indicator of the center-of-mass velocity of the observed stars, even for the long-period, large-amplitude variables. We note that the RV difference between opposite parts of the Galaxy with respect to the Sun is $\gtrsim200\,\kms$, which is much larger than any difference we encountered between $\medianrv$ and the zero-point RV.

The sky distribution of the FPR shows some clear structures. Some of them are physical, such as the overdensity around the Galactic plane (but not on the plane itself, affected by strong interstellar extinction). The FPR contains a few sources located in the Small Magellanic Cloud (SMC) and Large Magellanic Cloud (LMC), with $\rv\sim150\,\kms$ and $250\lesssim\rv\,/\,\kms\lesssim300$, respectively. Most (possibly all) of them are red supergiant stars, following the period-luminosity relation of fundamental-mode pulsators (see Sect.~\ref{sec:period_luminosity_diagrams}). However, some structures are artificial and result from the \Gaia scanning law and the selection filters applied to construct the catalog. The most evident such structures are the lack of sources in the Galactic center, as well as the largely empty regions on the bottom-left and top-right parts of the map. These areas are aligned on the ecliptic, and are characterized by a relatively small number of \Gaia transits. As a result, they are easily removed by our condition on the number of RV visibility periods (Sect.~\ref{sec:catalog_construction:pre_filtering}). Conversely, the ``stripes'' around the central hole correspond to fields of the sky often observed by \Gaia.

\subsection{Color~--~absolute magnitude diagram}
\label{sec:color_absolute_magnitude_diagram}

We restrict the presentation in this section to the FPR sources having precise parallaxes, so that we can confidently derive their distance moduli $\mu$ and thus their absolute brightnesses. The distributions of the relative parallax errors for the FPR sample and various subsets thereof are displayed in Fig.~\ref{fig:hist_parallax_error_types} (excluding 92 sources with negative parallaxes). The results are largely independent on whether the sources are part of the TQS or not, whereas distinct distributions are observed for the different variability types present in the FPR. Indeed, ellipsoidal variables tend to have better parallax measurements than LPVs, and, within the latter category, LPVs showing an LSP have, on the average, slightly better parallax measurements than pulsating LPVs.

We set the upper limit on the relative parallax error at 15\% (and require the parallax to be positive), thereby obtaining a sample of 5\,977 ``good-parallax'' sources (3\,740 in the TQS), that includes 794 ELL, 2\,136 pulsating LPVs, and 3\,047 LPVs showing a LSP. We examine this sample in the \Gaia color~--~absolute magnitude diagram (CAMD, Fig.~\ref{fig:fpr_hrd}) constructed with the median magnitudes derived from the \Gaia variability pipeline.

As expected, the majority of the LPVs are found on the asymptotic giant branch (AGB) of the CAMD, regardless of whether they show a LSP or a pulsation period. In contrast, ELL candidates are on the average fainter and bluer, and extend to the region of the red giant branch (RGB) in the CAMD. Most of the few stars with absolute $G$ brighter than about 3 mag are identified as red supergiants (RSGs) in the SIMBAD astronomical database \citep{simbad_2000}, while some other ones could include massive AGB stars as well. Their periods tend to be identified as LSPs rather than pulsation by our classification method due to their relatively low variability amplitudes (see Sect.~\ref{sec:catalog_content:lpv}). Their true nature can clearly be revealed once their intrinsic brightness is known.

\subsection{Period~--~luminosity diagrams}
\label{sec:period_luminosity_diagrams}

The sample can be further investigated in the period~--~luminosity diagram (PLD). To reveal the period~--~luminosity (PL) relations we adopt the \Gaia Wesenheit index $\wbprp$, which is an approximately reddening-free luminosity indicator \citep[see][for the definition and details]{lebzelter_etal_2018,lebzelter_etal_2019}. We limit this analysis to the TQS in order to ensure higher period reliability, and note that we can construct two distinct PLDs depending on whether we adopt the period derived from photometric (taking here the $G$ band) or RV time series. The two diagrams, both shown in Fig.~\ref{fig:fpr_pld}, are entirely consistent with each other, except for the lower period limit, which is at 35 days for $\prv$, whereas it reaches $\sim$17.5~days for $\pg$ due to the presence of sources with $\prv \simeq 2\pg$ in the sample.

The main features of the PLDs are the following. Regardless of the adopted period ($\pg$ or $\prv$), pulsating LPVs are found primarily on the period-luminosity sequences C$^{\prime}$ and C associated with pulsation in the first-overtone mode and fundamental mode, respectively (top panels in Fig.~\ref{fig:fpr_pld}). Some of them, that we further examine below, have a period in the area of sequence D, which is associated with LSP variability \citep[see e.g.,][and references therein]{pawlak_2021}. A few of them have $\pg\lesssim40\,\days$, possibly on the bright part of sequence B. These might be identified with the type of variables that, in the context of the Optical Gravitational Lensing Experiment \citep[OGLE,][]{udalski_etal_1992}, are known as OGLE Small-Amplitude Red Giants \citep[OSARGs,][]{wray_etal_2004}.
The LPVs whose variability we identify as LSP, on the other hand, are mostly found on sequence D (middle panels in Fig.~\ref{fig:fpr_pld}), except for the RSGs that are located on the bright end of sequence C. The RV periods of ellipsoidal red giants align on the PL sequence E (bottom panels in the figure), that forms a continuity with sequence D \citep[see fig.~1 of][]{soszynski_etal_2007}.

In order to understand the presence of pulsation periods of LPVs on the PLD sequence D we have to distinguish between stars having O-rich and C-rich surface chemistry. Indeed, the PL relations corresponding to these two chemical types are different when examined through an optical-band Wesenheit index such as $\wbprp$. A similar effect can be seen, for instance, in fig.~1 of \citet{soszynski_etal_2007} \citep[see also][]{lebzelter_etal_2018,lebzelter_etal_2019}. 
To identify potential C-stars among the FPR sources we take advantage of the \texttt{is\_cstar} flag provided with the \Gaia DR3 catalog of LPV candidates \citep{dr3_lpv}. About 7\% of the FPR sources are classified as C-stars by this method, whereas 86\% are identified as O-rich (another 7\% are unclassified). After limiting the sample to pulsating LPV stars from the TQS and having good parallaxes, we find that 836 of them are O-rich, and 143 are C-rich. They are displayed in the PLD in Fig.~\ref{fig:fpr_tqs_pld_lpv_chem}, constructed using the photometric period $\pg$. In order to provide a visual reference for the positions of the PL relations, in the same diagrams we also show the sources from the OGLE-III catalog of LPVs in the LMC published by \citet{soszynski_etal_2009}. These authors also provide a photometry-based spectral-type classification which we use to discriminate between the O-rich and C-rich OGLE-III LPVs in Fig.~\ref{fig:fpr_tqs_pld_lpv_chem}.

As can be seen from the top panel of Fig.~\ref{fig:fpr_tqs_pld_lpv_chem}, there are some O-rich stars that we identified as pulsating LPVs and whose periods lie on the short-period, faint end of sequence D. These are LSPs mistakenly identified as pulsation periods by applying Eq.~\ref{eq:condition_lsp} because they lie close to the dividing line (between about 200 and 400 days, see Fig.~\ref{fig:ampg_Pg_2panels}). Let us now consider the C-rich sources in the bottom panel of Fig.~\ref{fig:fpr_tqs_pld_lpv_chem}. For the purpose of clear visualization, we do not show the C-rich OGLE sources whose primary period is flagged as a LSP. Indeed, the C-rich LPVs pulsating in the fundamental mode often lie below the corresponding PL relation (sequence C), overlapping with sequence D. This is a consequence of self-extinction by circumstellar dust which causes these sources to appear fainter than less dusty stars having similar periods \citep[see e.g.,][and references therein]{lebzelter_etal_2019}. This seems to be the case for the C-stars in our selection. Therefore, the presence of pulsating stars on sequence D can be correct if it results from circumstellar extinction.

Finally, we consider the TQS sources with good parallaxes and with $\prv\simeq2\,\pg$. These are displayed in the PLDs in Fig.~\ref{fig:fpr_pld_Prv2Pg}, constructed with $\pg$ (top panel) and $\prv$ (bottom panel). It is clear that the majority of these sources obey the PL relation E of ellipsoidal red giants, leaving little doubt on the nature of their variability. Nonetheless, some of them populate other regions of the PLD, in particular the pulsation sequences C and C$^{\prime}$ (and possibly B), as well as the LSP sequence D. This contradicts the naive expectation that the occurrence of $\prv$ and $\pg$ in a 2:1 ratio must indicate binary-induced variability. Let us then focus on the sources showing this feature and classified as LPVs (both pulsating and showing a LSP), which are highlighted in color in Fig.~\ref{fig:fpr_pld_Prv2Pg}.

To begin with, we consider the $G$-band periods shorter than about 35 days, that are possibly compatible with the bright end of sequence B and that correspond to RV periods in the middle of sequence C$^{\prime}$. If these periods are correct, it could mean that the variability pipeline identified the signature of first-overtone mode pulsation in the RV time series, and of second-overtone mode pulsation in the $G$-band time series. This would be consistent with the fact that, upon visual inspection, their photometric time series appear poorly regular, which can be taken as an indication of multi-periodicity. However, the first-overtone and second-overtone modes do not occur in a 2:1 ratio in LPVs \citep[e.g.,][and references therein]{wood_2015}. Furthermore, both sequences B and C$^{\prime}$ are actually thought to result from pulsation in the first overtone mode \citep{trabucchi_etal_2017,yu_etal_2020}. Another explanation could be that these short $G$-band periods are associated with spurious frequencies resulting from scan-angle dependent signals \citep[]{holl_etal_2023}. However, the origin of these spurious frequencies as described in that paper should affect only the $G$ band and not the $\gbp$ and $\grp$ measurements. Yet another explanation could be that the periods derived from the RV time series are not correct. We note that this kind of stars often display irregular variations that may result in the detection of spurious periods.

Entirely similar arguments can be put forward for the $G$-band periods along sequence C$^{\prime}$ in the top panel of Fig.~\ref{fig:fpr_pld_Prv2Pg} (between about 40 and 200 days), which correspond to RV periods on sequence C, in the range 80-500 days. However, in this case it is more likely that they actually correspond, respectively, to the first-overtone mode and fundamental mode periods, which can occur in a 2:1 ratio \citet[see e.g., fig.~6 of][]{soszynski_etal_2007}. The photometric and RV periods would then both be correct and consistent with a pulsational origin.

Finally, we consider the LPVs that appear on sequence D in either panel of Fig.~\ref{fig:fpr_pld_Prv2Pg}. The vast majority of them are probably LSPs, even though some are identified as pulsating LPVs (red points in the figure). Indeed, as discussed above, they could be pulsating C-rich stars suffering from circumstellar extinction, but only a few of the sources shown in Fig.~\ref{fig:fpr_pld_Prv2Pg} are classified as probable C-stars. Another explanation for these stars showing $\prv\simeq2\,\pg$ is that their LSPs are indeed caused by binarity, which would be consistent with the scenario outlined by \citet{soszynski_etal_2021}.

\section{Summary and conclusions}
\label{sec:summary_and_conclusions}

The \Gaia DR3 saw the publication of average RVs for over 33 million stars based on 34 months of observations with the \Gaia RVS, whereas epoch radial velocities were published only for a very restricted list of variable sources. Anticipating the publication of the full RV data with the fourth \Gaia data release, we present RV time series for a selected sample of long-period variables as part of this \Gaia FPR.

We describe the construction of the catalog, starting with the set of \Gaia LPV candidates with a median RV published in \Gaia DR3, and applying several filtering steps to ensure the highest quality of the final sample, leading to 9\,614 sources. In addition to the RV time series, for each source we provide the model-derived frequency and amplitude of RV variability, as well as the RV statistics, determined by the \Gaia variability processing pipeline. In addition, we publish a flag allowing for the identification of a subset of 6\,093 sources that show a high degree of compatibility between the periods derived from the RV and photometric time series. We consider them to be of superior quality as all four of their \Gaia time series (three photometric bands and the RV channel) are likely to carry a strong signature from the same physical process, enabling detailed studies of their variability.

We show how the catalog includes three groups of sources exhibiting different types of variability, namely ellipsoidal red giants, pulsating LPVs, and LPVs displaying a long secondary period. Stars from the first group are characterized by comparatively large RV amplitudes (that cannot be attained by pulsating stars) and small photometric amplitudes. They also frequently show RV periods that are twice as long as the periods derived from the photometric time series. They represent between 10 to 15\% of the catalog. The remainder of the FPR consists in roughly equal parts of LPVs showing pulsation and long secondary periods, which we tentatively distinguish by their different positions in the period-amplitude diagram.

We further verify the quality of the FPR by comparison with \Gaia DR3 products as well as other literature data. We show that, despite the use of a different pipeline, the median RV derived by our variability processing is entirely consistent with previous \Gaia data.

When using the median RV of LPVs as a measure of the systemic velocity, one has to consider the following limitations. The first one is connected to the uneven time sampling of \Gaia observations. Combined with the long periods and large amplitudes of the sources we examined, it can lead to a substantial degree of undersampling of specific phases of the variability cycle, skewing the RV distribution away from the true systemic velocity. In this case, the maximum uncertainty is given by half the peak-to-peak (true) RV variability amplitude. The occurrence of multi-periodicity might affect the median in a similar way.

The second limitation is connected with the possibly asymmetric shape of the RV curves of pulsating LPVs, in which case the median does not necessarily trace the systemic velocity, regardless of the sampling. Once again, the maximum deviation cannot be larger than half the peak-to-peak RV amplitude. Finally, a limitation could arise from the possible occurrence of template mismatches, that is the adoption, for the purpose of deriving RVs, of a spectral template whose atmospheric parameters are not suited for the target star. LPVs with C-rich chemistry, whose spectra show very distinctive molecular absorption features, are especially exposed to this risk, as all the adopted templates are for O-rich composition. In this case, it is not easy to assess the maximum deviation. However, our analysis suggests that none of these three aspects significantly affect the RV variability parameters published here.

We then present an overview of the catalog. We first show that the distribution of the RVs on the sky is physically consistent with the Galactic rotation curve. We then analyze, in the color~--~absolute magnitude diagram and in the period~--~luminosity diagram, the distribution of a subsample with good parallaxes. The sources identified as LPVs showing a LSP and as ellipsoidal variables by our classification scheme are seen to follow the period-luminosity relations D and E, respectively, as expected. The periods of pulsating LPVs, on the other hand, are found mainly on sequences C$^{\prime}$ and C, corresponding to the first-overtone mode and the fundamental mode, respectively. Some of these sources show a 2:1 ratio between the RV period and the photometric period, which is consistent with simultaneous pulsation in those two modes. These results indicate the rich content of physical information available in both the RV and photometric time series.

This FPR includes the largest dataset of RV time series of LPVs to date. Moreover, it covers sources over a largely unexplored (in terms of RV time-series analysis) range of distances, intermediate between the extragalactic investigations of variable stars in the Magellanic Clouds and the studies of nearby LPVs. The RV time series, together with the photometric time series published in \Gaia DR3 and spanning the same time baseline, offer an unprecedented opportunity to investigate, from different perspectives, the behavior of three different types of red giant variables, including the LSPs whose nature is still a matter of debate. Finally, the high-quality epoch RV data of this FPR will provide the astrophysical community with a means to prepare for the \Gaia data release 4.

\begin{acknowledgements}
This work presents results from the European Space Agency (ESA) space mission \Gaia.
\Gaia data are being processed by the \Gaia Data Processing and Analysis Consortium (DPAC).
Funding for the DPAC is provided by national institutions, in particular the institutions participating in the \Gaia MultiLateral Agreement (MLA).
The \Gaia mission website is \url{https://www.cosmos.esa.int/gaia}.
The \Gaia archive website is \url{https://archives.esac.esa.int/gaia}.
Full acknowledgements are given in Appendix~\ref{app:acknowledgements}.
\end{acknowledgements}

\bibliographystyle{aa}
\bibliography{fprlpv}

\begin{appendix}

\section{Additional details on the classification of the catalog content}
\label{asec:lsp}

This FPR contains stellar sources displaying photometric and RV variability of different origins. In Sect.~\ref{sec:catalog_content} we separated ellipsoidal variables from LPVs by comparing the amplitudes of their RV and $G$-band time series models. Then, among the LPV candidates, we found periods that result from pulsation as well as LSPs, and we tentatively distinguished the two types of variability in the $G$-band period-amplitude diagram. The separation of these two variability types in this diagram is not as clean as that between LPVs and ELLs in the $G$ versus RV amplitude diagram. Therefore we examined in some more detail the two groups, in order to corroborate our classification.

\begin{figure}
\centering
\includegraphics[width=\columnwidth]{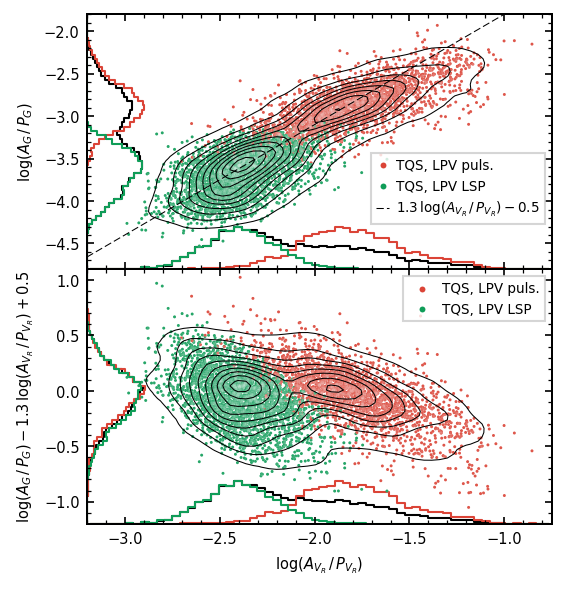}
\caption{$G$-band versus RV period-amplitude ratios for the TQS sources classified as LPVs. Black contour lines represent a smoothed density map of the sample, and correspond to the black histograms on the sides of each panel. Each source is color-coded according to whether its period is identified as resulting from pulsation (red) or as an LSP (green). The same color-code is used for the histograms on the sides (normalized to their maximum). In the bottom panel the quantity $\log(\ampg/\pg)$ is replaced with its distance from the dashed line displayed in the top panel, taken as an arbitrary reference.}
\label{fig:lpv_puls_lsp_P_amp_g_rv_2panels}
\end{figure}

We recall that, in absence of accurate and homogeneous knowledge about absolute brightness (i.e., distance), we have to rely on the variability parameters period and amplitude for this purpose. In addition to the variability parameters emerging from the analysis of the photometric time series, we can inspect the ones resulting from the RV curves. A combination of these four quantities is shown in Fig.~\ref{fig:lpv_puls_lsp_P_amp_g_rv_2panels} (top panel), which displays the ratio between the semi-amplitude and the period of the $G$-band time series model against the same ratio for the RV time series model. For simplicity, we explicitly show the decimal logarithm of both quantities. There exists a clear correlation between the two ratios, indicating that the occurrence of a long period characterized by a small photometric amplitude is often associated with a relatively small RV amplitude as well.

The density contour lines in the top panel of Fig.~\ref{fig:lpv_puls_lsp_P_amp_g_rv_2panels}, as well as the histograms along the axes of the diagram, show that there are two regions of overdensity in the diagram. In order to better visualize them, in the bottom panel of Fig.~\ref{fig:lpv_puls_lsp_P_amp_g_rv_2panels} we replaced the quantity on the vertical axis with its distance from the arbitrary reference line shown in the top panel. Fig.~\ref{fig:lpv_puls_lsp_P_amp_g_rv_2panels} shows in two different colors the sources classified on the basis of Eq.~\ref{eq:condition_lsp}. The latter is consistent with the distributions highlighted here with the inclusion of information related to RV variability, although it appears to lead to some contamination. It is possible that a classification that makes use of the periods and amplitudes resulting from both photometric and RV time series could help reduce such contamination. However, it would be more complex and less intuitive from a physical point of view, which supports the choice made in Sect.~\ref{sec:catalog_content}.

\FloatBarrier

\section{Median RV in unevenly sampled time series}
\label{app:median_RV_in_unevenly_sampled_time_series}

\begin{figure*}
\centering
\includegraphics[width=\textwidth]{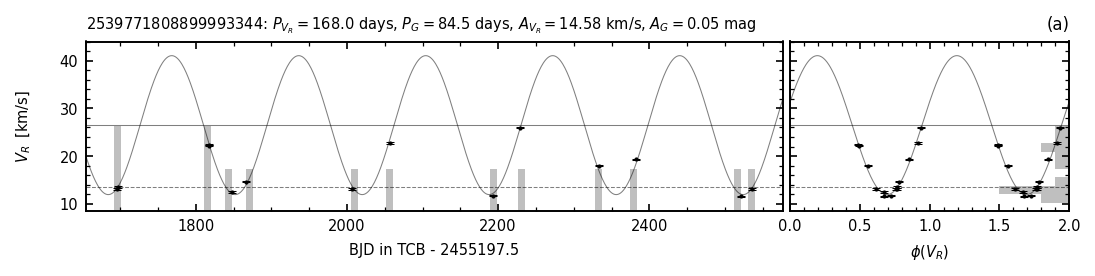}
\includegraphics[width=\textwidth]{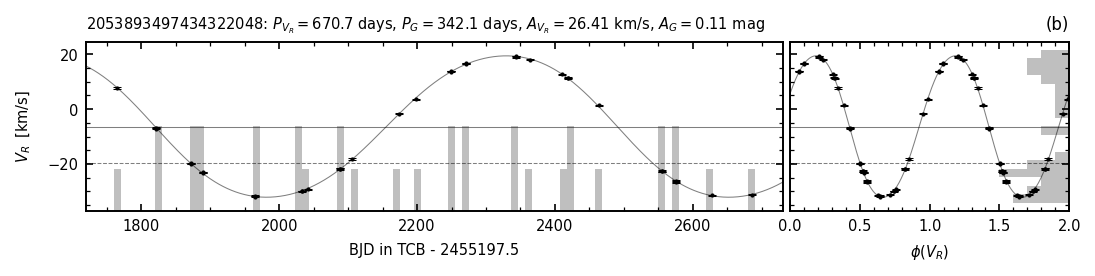}
\includegraphics[width=\textwidth]{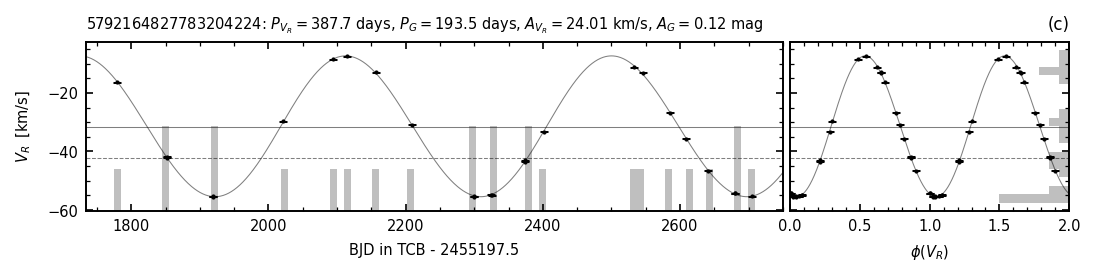}
\caption{Examples of RV time series in which only a small number of cycles is covered, and the minimum phase is oversampled compared to other phases, leading the median RV to underestimate the systemic RV. In each row, the left panel shows the RV time series and best-fit model, whereas their folded counterparts are displayed in the right panel. Histograms in both panels aid to visualize the distribution of measurements both in time (to identify clustered measurements) and in RV. The solid and dashed lines mark the values of the zero-point and median RVs, respectively.}
\label{fig:RVTS_ex_FewCycles}
\end{figure*}

In this section we present a few representative examples of sources showing relatively large differences between the median RV and the zero-point RV, so that the former is not necessarily an accurate indicator of the center-of-mass velocity. The RV curves displayed in Fig.~\ref{fig:RVTS_ex_FewCycles} belong to sources with relatively long periods, for which \Gaia observations cover only a few cycles. They also have large RV amplitudes and highly sinusoidal variations, indicating that their variability is due to binarity, and leaving little doubt concerning the authenticity of its physical origin. In all the examples displayed, the distribution of RV epochs is asymmetric as the phase near minimum RV is oversampled compared with other phases of the variability cycle. As a result, the median RV is underestimated, and shows a large difference with respect to $\zprv$.

\begin{figure*}
\centering
\includegraphics[width=\textwidth]{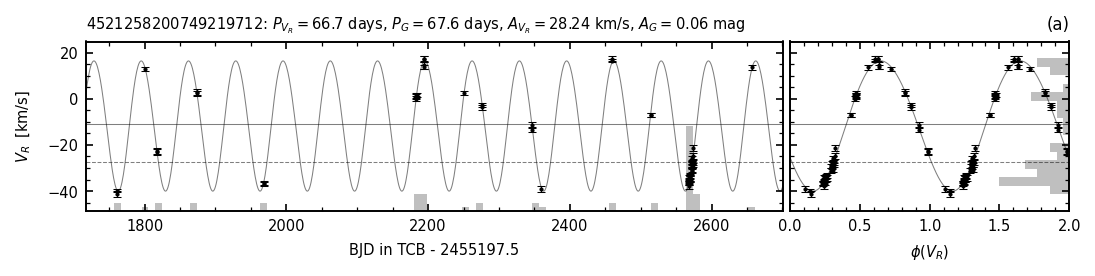}
\includegraphics[width=\textwidth]{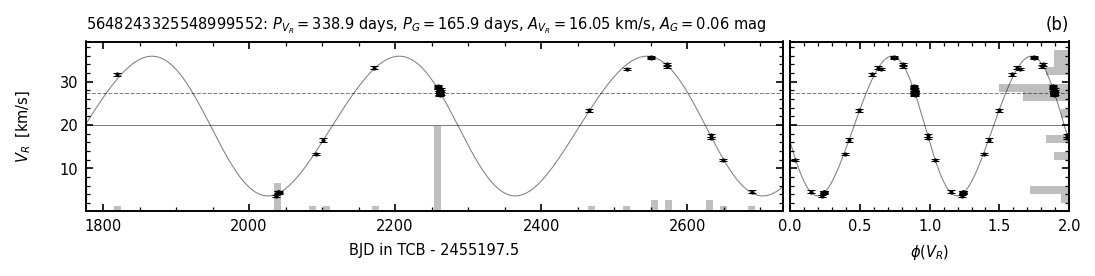}
\includegraphics[width=\textwidth]{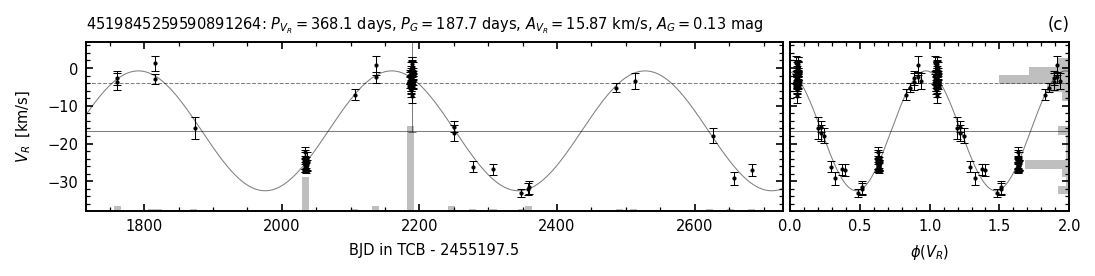}
\includegraphics[width=\textwidth]{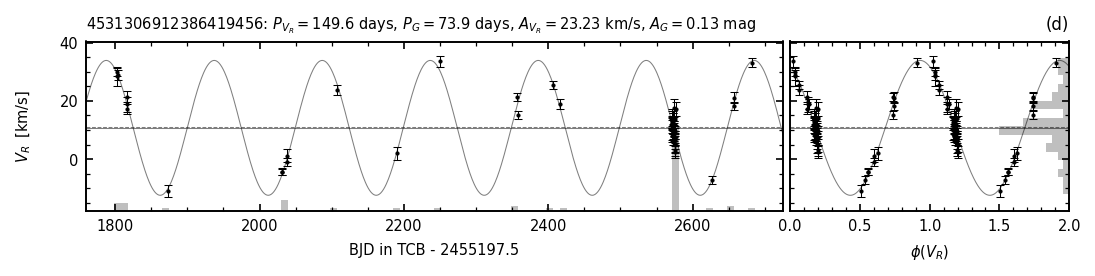}
\includegraphics[width=\textwidth]{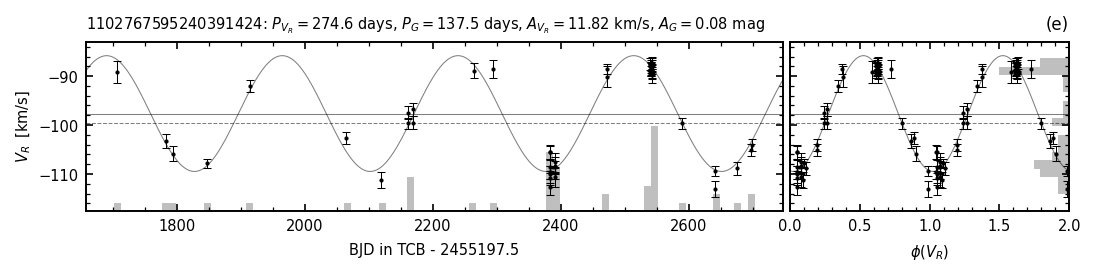}
\caption{Similar to Fig.~\ref{fig:RVTS_ex_FewCycles}, but showing examples of time series dominated by a large group of RV epochs spanning an interval of time much shorter than the typical period of the source, thereby distorting the time series statistics and causing a large difference between $\medianrv$ and $\zprv$. Cases (d) and (e) show examples where the clustered data points are, by chance, either  located near the mid-point of the RV time series (case d), or cancel out with each other (case e).}
\label{fig:RVTS_ex_Clusters}
\end{figure*}

Further examples are provided in Fig.~\ref{fig:RVTS_ex_Clusters}, illustrating how the median of the RV time series ceases to be representative of the systemic RV due to the occurrence of clusters of RV epochs spanning an interval of time much shorter than the variability period of a given source. The RV distributions of the example time series are dominated by the effect of these groups causing large discrepancies between $\medianrv$ and $\zprv$, except in the fortuitous circumstances in which the clusters end up near the mid-point of the RV curve or cancel out with each other, as illustrated in panels (d) and (e) of Fig.~\ref{fig:RVTS_ex_Clusters}, respectively.

Overall, there exist rather specific circumstances in which the median $\medianrv$ cannot be considered a good indicator of the systemic RV. These arise primarily from the uneven and irregular time sampling of \Gaia observations, as well as the relatively small number of epochs, and affect mainly the sources with large RV amplitudes and long periods. In particular, when only a few variability cycles are covered, even a relatively homogeneous phase coverage can be insufficient to obtain a well-centered median value.

\FloatBarrier

\section{Numerical differences with respect to \Gaia DR3}
\label{asec:numerical_differences_with_respect_to_gaia_dr3}

The production of this FPR required running the LPV variability pipeline on a subset of the sample of LPV candidates published as part of \Gaia DR3. Despite having run the same operations on the photometric time series, we detected numerical differences with respect to the results previously obtained and published in \Gaia DR3. We traced the origin of these numerical differences to the execution of the Apache Math Commons \texttt{LevenbergMarquardtOptimizer}, which produces different results depending on the adopted Java Development Kit (JDK) version. The differences arised after updating from JDK-8 to JDK-17.

The differences are systematically reproducible\footnote{
    See the github repository \url{https://github.com/gjevardat/ReproduceFloatingPointBug} for an executable proof of the numerical differences.
}, and a bug report was submitted to the Oracle bug submission system late in 2022, without receiving any feedback. The most plausible explanation we found for the numerical differences is that JDK-17 enforces the floating point standard IEEE 754, whereas in JDK-8 the runtime could decide to deviate from this standard in order to optimize the generated code\footnote{
    See \url{https://openjdk.org/jeps/306} for more details.
}.

In order to assess the impact of this bug we consider the periods of the $G$-band time series resulting from the variability pipeline before and after upgrading to JDK-17, and examine the differences between them. We use for this purpose the sample adopted for constructing the FPR catalog after pre-filtering that consists of 110\,654 sources. From this set we exclude 1\,413 sources for which either of the two periods is longer than the duration of the $G$-band time series. Within this reference set, 50\% of the time series result in the exact same period regardless of the JDK version employed in the pipeline.

\begin{figure}
\centering
\includegraphics[width=\columnwidth]{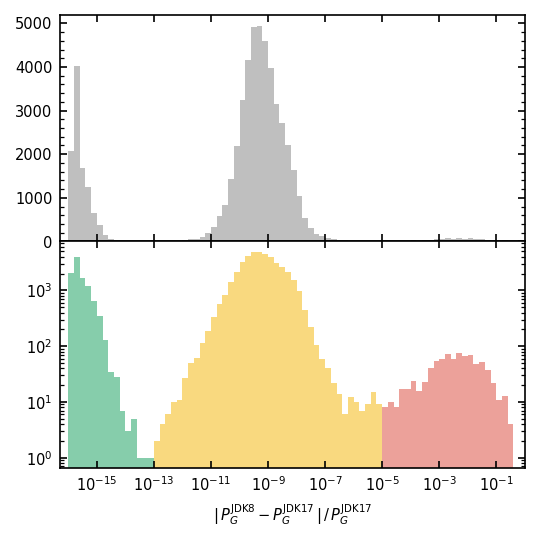}
\caption{Distribution of relative differences between the period derived for $G$-band time series before and after the upgrade of the variability pipeline from JDK-8 to JDK-17. A linear scale is used in the top panel for the vertical axis, while a log-scale is used in the bottom panel. Three distinct groups are identified, colored in the bottom panel in green, orange and red, corresponding to various levels of difference. We note that only in the latter group the differences are non-negligible (being typically of order 1\%). This group is so small that it is barely visible in the top panel.}
\label{fig:hist_reldif_P_jdk}
\end{figure}

In Fig.~\ref{fig:hist_reldif_P_jdk} we display the distribution of the relative difference between the JDK-8 and JDK-17 periods for the sources that have a non-null difference. It shows three well-distinct groups. About 60\% of the time series display relative period differences that are at the level of machine precision, and thus entirely negligible ($<10^{-13}$, and possibly zero). A second group contains sources that display relative differences between $10^{-13}$ and $10^{-5}$, and consists of about 40\% of the sample. In this case the absolute differences are on the order of a few minutes for periods of 1000 days, or a few seconds for periods of 10 days, hence they are negligible as well. Finally, there exists a third group containing fewer than 1\% of the examined dataset, whose sources display relative period differences above $10^{-5}$, but only 28 of them show a discrepancy above the 10\% level (at most as large as 35\%).

\begin{figure}
\centering
\includegraphics[width=\columnwidth]{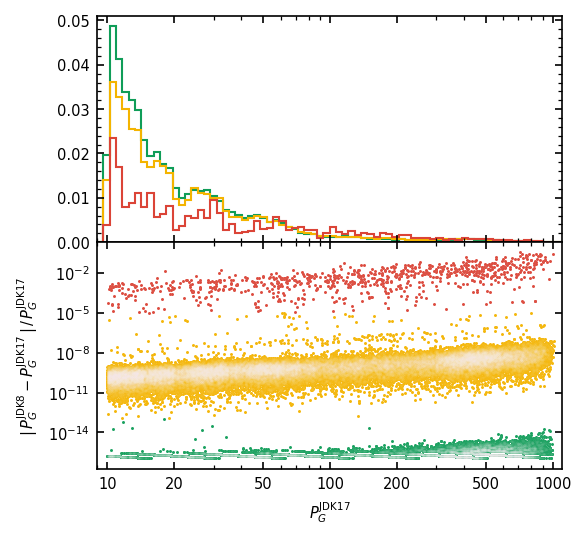}
\caption{Period distribution for the three groups shown in Fig.~\ref{fig:hist_reldif_P_jdk} (top panel), and relation of the period with the relative period difference. The period shown is the one obtained after the upgrade to JDK-17. The same color scheme as in Fig.~\ref{fig:hist_reldif_P_jdk} is adopted. The histograms are normalized by area.}
\label{fig:hist_reldif_P_jdk_P}
\end{figure}

\begin{figure}
\centering
\includegraphics[width=\columnwidth]{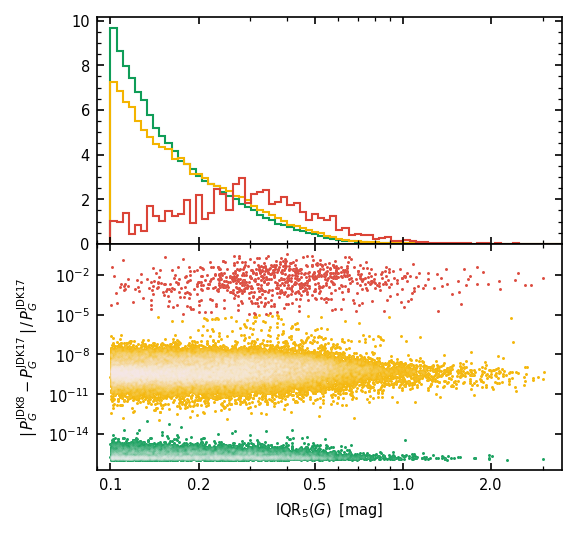}
\caption{Similar to Fig.~\ref{fig:hist_reldif_P_jdk_P}, but for the 5-95\% interquantile range of the $G$-band time series.}
\label{fig:hist_reldif_P_jdk_trange}
\end{figure}

We examined the distribution of $G$-band time series statistics separately for these three groups, looking for systematic effects and features that might be triggering a strong effect of the bug. However, given the nature of this bug, one could expect it to impact different sources in a random fashion, and to be independent of the specific properties of the time series being processed. Our analysis tends to confirm this expectation. Overall, all three groups of sources show similar distributions of number of observations, time series duration, and average brightness. Only the parameters related with the amplitude and the period itself display a connection with the strength of the effect of the bug. In particular, there is a slight tendency for the third group (shown in red in Figs.~\ref{fig:hist_reldif_P_jdk_P} and~\ref{fig:hist_reldif_P_jdk_trange}) to lack sources with relatively small amplitudes and short periods, whereas the opposite is true for the first (green) group. Moreover, there is a clear correlation between the relative difference and the period within each group. The most likely explanation is that these sources are more exposed to numerical instabilities in the period determination as their time series cover a small number of cycles, which makes the period difficult to constrain in any case.

\FloatBarrier

\section{Catalog retrieval}
\label{app:catalog_retrieval}

Here are two examples on how to retrieve LPV-related data from the \Gaia archive.
A query like the following allows for retrieving selected quantities from the source catalog and LPV FPR catalog for all LPV FPR sources.

\begin{footnotesize}
\begin{verbatim}
SELECT gs.source_id,
    gs.ra, gs.dec,
    fprlpv.frequency_rv, fprlpv.frequency_rv_error,
    fprlpv.amplitude_rv, fprlpv.flag_rv,
    ...
FROM gaiadr3.gaia_source AS gs
LEFT JOIN gaiafpr.vari_long_period_variable AS fprlpv
  ON gs.source_id = fprlpv.source_id
WHERE fprlpv.source_id IS NOT NULL
\end{verbatim}
\end{footnotesize}

In order to retrieve all data from the source table, the LPV FPR table, the RV statistics table, and the tables of data on LPVs from DR3 for all sources published in the LPV FPR, a query like the following can be used.

\begin{footnotesize}
\begin{verbatim}
SELECT src.*,
    fprlpv.*,
    rvstat.*,
    lpv.*,
    varistat.*
FROM gaiadr3.gaia_source AS src
LEFT JOIN gaiafpr.vari_long_period_variable AS fprlpv
    ON src.source_id = fprlpv.source_id
LEFT JOIN gaiafpr.vari_rad_vel_statistics AS rvstat
    ON src.source_id = rvstat.source_id
LEFT JOIN gaiadr3.vari_long_period_variable AS lpv
    ON src.source_id = lpv.source_id
LEFT JOIN gaiadr3.vari_summary AS varistat
    ON src.source_id = varistat.source_id
WHERE fprlpv.source_id IS NOT NULL
\end{verbatim}
\end{footnotesize}

Finally, one can download all the RV time series published as part of this FPR using a source like the following.

\begin{footnotesize}
\begin{verbatim}
SELECT fprlpv.source_id,
    varirv.transit_id,
    varirv.rv_obs_time,
    varirv.radial_velocity,
    varirv.radial_velocity_error
    varirv.rejected_by_variability
FROM gaiafpr.vari_long_period_variable AS fprlpv
LEFT JOIN gaiafpr.vari_epoch_radial_velocity AS varirv
    ON fprlpv.source_id = varirv.source_id
WHERE fprlpv.source_id IS NOT NULL
\end{verbatim}
\end{footnotesize}

We also recall that it is possible to download ancillary data such as the RV and photometric time series using the \Gaia datalink service\footnote{\url{https://www.cosmos.esa.int/web/gaia-users/archive/datalink-products}}.

\section{Acknowledgements}
\label{app:acknowledgements}
This work presents results from the European Space Agency (ESA) space mission \Gaia. \Gaia data are being processed by the \Gaia Data Processing and Analysis Consortium (DPAC). Funding for the DPAC is provided by national institutions, in particular the institutions participating in the \Gaia MultiLateral Agreement (MLA). The \Gaia mission website is \url{https://www.cosmos.esa.int/gaia}. The \Gaia archive website is \url{https://archives.esac.esa.int/gaia}.

The \Gaia mission and data processing have financially been supported by, in alphabetical order by country:
\begin{itemize}
\item the Algerian Centre de Recherche en Astronomie, Astrophysique et G\'{e}ophysique of Bouzareah Observatory;
\item the Austrian Fonds zur F\"{o}rderung der wissenschaftlichen Forschung (FWF) Hertha Firnberg Programme through grants T359, P20046, and P23737;
\item the BELgian federal Science Policy Office (BELSPO) through various PROgramme de D\'{e}veloppement d'Exp\'{e}riences scientifiques (PRODEX) grants and the Polish Academy of Sciences - Fonds Wetenschappelijk Onderzoek through grant VS.091.16N, and the Fonds de la Recherche Scientifique (FNRS), and the Research Council of Katholieke Universiteit (KU) Leuven through grant C16/18/005 (Pushing AsteRoseismology to the next level with TESS, GaiA, and the Sloan DIgital Sky SurvEy -- PARADISE);  
\item the Brazil-France exchange programmes Funda\c{c}\~{a}o de Amparo \`{a} Pesquisa do Estado de S\~{a}o Paulo (FAPESP) and Coordena\c{c}\~{a}o de Aperfeicoamento de Pessoal de N\'{\i}vel Superior (CAPES) - Comit\'{e} Fran\c{c}ais d'Evaluation de la Coop\'{e}ration Universitaire et Scientifique avec le Br\'{e}sil (COFECUB);
\item the Chilean Agencia Nacional de Investigaci\'{o}n y Desarrollo (ANID) through Fondo Nacional de Desarrollo Cient\'{\i}fico y Tecnol\'{o}gico (FONDECYT) Regular Project 1210992 (L.~Chemin);
\item the National Natural Science Foundation of China (NSFC) through grants 11573054, 11703065, and 12173069, the China Scholarship Council through grant 201806040200, and the Natural Science Foundation of Shanghai through grant 21ZR1474100;  
\item the Tenure Track Pilot Programme of the Croatian Science Foundation and the \'{E}cole Polytechnique F\'{e}d\'{e}rale de Lausanne and the project TTP-2018-07-1171 ``Mining the Variable Sky,'' with the funds of the Croatian-Swiss Research Programme;
\item the Czech-Republic Ministry of Education, Youth, and Sports through grant LG 15010 and INTER-EXCELLENCE grant LTAUSA18093, and the Czech Space Office through ESA PECS contract 98058;
\item the Danish Ministry of Science;
\item the Estonian Ministry of Education and Research through grant IUT40-1;
\item the European Commission’s Sixth Framework Programme through the European Leadership in Space Astrometry (\href{https://www.cosmos.esa.int/web/gaia/elsa-rtn-programme}{ELSA}) Marie Curie Research Training Network (MRTN-CT-2006-033481), through Marie Curie project PIOF-GA-2009-255267 (Space AsteroSeismology \& RR Lyrae stars, SAS-RRL), and through a Marie Curie Transfer-of-Knowledge (ToK) fellowship (MTKD-CT-2004-014188); the European Commission's Seventh Framework Programme through grant FP7-606740 (FP7-SPACE-2013-1) for the \Gaia European Network for Improved data User Services (\href{https://gaia.ub.edu/twiki/do/view/GENIUS/}{GENIUS}) and through grant 264895 for the \Gaia Research for European Astronomy Training (\href{https://www.cosmos.esa.int/web/gaia/great-programme}{GREAT-ITN}) network;
\item the European Cooperation in Science and Technology (COST) through COST Action CA18104 ``Revealing the Milky Way with \Gaia (MW-Gaia)'';
\item the European Research Council (ERC) through grants 320360, 647208, and 834148 and through the European Union’s Horizon 2020 research and innovation and excellent science programmes through Marie Sk{\l}odowska-Curie grant 745617 (Our Galaxy at full HD -- Gal-HD) and 895174 (The build-up and fate of self-gravitating systems in the Universe) as well as grants 687378 (Small Bodies: Near and Far), 682115 (Using the Magellanic Clouds to Understand the Interaction of Galaxies), 695099 (A sub-percent distance scale from binaries and Cepheids -- CepBin), 716155 (Structured ACCREtion Disks -- SACCRED), 951549 (Sub-percent calibration of the extragalactic distance scale in the era of big surveys -- UniverScale), and 101004214 (Innovative Scientific Data Exploration and Exploitation Applications for Space Sciences -- EXPLORE);
\item the European Science Foundation (ESF), in the framework of the \Gaia Research for European Astronomy Training Research Network Programme (\href{https://www.cosmos.esa.int/web/gaia/great-programme}{GREAT-ESF});
\item the European Space Agency (ESA) in the framework of the \Gaia project, through the Plan for European Cooperating States (PECS) programme through contracts C98090 and 4000106398/12/NL/KML for Hungary, through contract 4000115263/15/NL/IB for Germany, and through PROgramme de D\'{e}veloppement d'Exp\'{e}riences scientifiques (PRODEX) grant 4000127986 for Slovenia;  
\item the Academy of Finland through grants 299543, 307157, 325805, 328654, 336546, and 345115 and the Magnus Ehrnrooth Foundation;
\item the French Centre National d’\'{E}tudes Spatiales (CNES), the Agence Nationale de la Recherche (ANR) through grant ANR-10-IDEX-0001-02 for the ``Investissements d'avenir'' programme, through grant ANR-15-CE31-0007 for project ``Modelling the Milky Way in the \Gaia era'' (MOD4Gaia), through grant ANR-14-CE33-0014-01 for project ``The Milky Way disc formation in the \Gaia era'' (ARCHEOGAL), through grant ANR-15-CE31-0012-01 for project ``Unlocking the potential of Cepheids as primary distance calibrators'' (UnlockCepheids), through grant ANR-19-CE31-0017 for project ``Secular evolution of galaxies'' (SEGAL), and through grant ANR-18-CE31-0006 for project ``Galactic Dark Matter'' (GaDaMa), the Centre National de la Recherche Scientifique (CNRS) and its SNO \Gaia of the Institut des Sciences de l’Univers (INSU), its Programmes Nationaux: Cosmologie et Galaxies (PNCG), Gravitation R\'{e}f\'{e}rences Astronomie M\'{e}trologie (PNGRAM), Plan\'{e}tologie (PNP), Physique et Chimie du Milieu Interstellaire (PCMI), and Physique Stellaire (PNPS), the ``Action F\'{e}d\'{e}ratrice \Gaia'' of the Observatoire de Paris, the R\'{e}gion de Franche-Comt\'{e}, the Institut National Polytechnique (INP) and the Institut National de Physique nucl\'{e}aire et de Physique des Particules (IN2P3) co-funded by CNES;
\item the German Aerospace Agency (Deutsches Zentrum f\"{u}r Luft- und Raumfahrt e.V., DLR) through grants 50QG0501, 50QG0601, 50QG0602, 50QG0701, 50QG0901, 50QG1001, 50QG1101, 50\-QG1401, 50QG1402, 50QG1403, 50QG1404, 50QG1904, 50QG2101, 50QG2102, and 50QG2202, and the Centre for Information Services and High Performance Computing (ZIH) at the Technische Universit\"{a}t Dresden for generous allocations of computer time;
\item the Hungarian Academy of Sciences through the Lend\"{u}let Programme grants LP2014-17 and LP2018-7 and the Hungarian National Research, Development, and Innovation Office (NKFIH) through grant KKP-137523 (``SeismoLab'');
\item the Science Foundation Ireland (SFI) through a Royal Society - SFI University Research Fellowship (M.~Fraser);
\item the Israel Ministry of Science and Technology through grant 3-18143 and the Tel Aviv University Center for Artificial Intelligence and Data Science (TAD) through a grant;
\item the Agenzia Spaziale Italiana (ASI) through contracts I/037/08/0, I/058/10/0, 2014-025-R.0, 2014-025-R.1.2015, and 2018-24-HH.0 to the Italian Istituto Nazionale di Astrofisica (INAF), contract 2014-049-R.0/1/2 to INAF for the Space Science Data Centre (SSDC, formerly known as the ASI Science Data Center, ASDC), contracts I/008/10/0, 2013/030/I.0, 2013-030-I.0.1-2015, and 2016-17-I.0 to the Aerospace Logistics Technology Engineering Company (ALTEC S.p.A.), INAF, and the Italian Ministry of Education, University, and Research (Ministero dell'Istruzione, dell'Universit\`{a} e della Ricerca) through the Premiale project ``MIning The Cosmos Big Data and Innovative Italian Technology for Frontier Astrophysics and Cosmology'' (MITiC);
\item the Netherlands Organisation for Scientific Research (NWO) through grant NWO-M-614.061.414, through a VICI grant (A.~Helmi), and through a Spinoza prize (A.~Helmi), and the Netherlands Research School for Astronomy (NOVA);
\item the Polish National Science Centre through HARMONIA grant 2018/30/M/ST9/00311 and DAINA grant 2017/27/L/ST9/03221 and the Ministry of Science and Higher Education (MNiSW) through grant DIR/WK/2018/12;
\item the Portuguese Funda\c{c}\~{a}o para a Ci\^{e}ncia e a Tecnologia (FCT) through national funds, grants SFRH/\-BD/128840/2017 and PTDC/FIS-AST/30389/2017, and work contract DL 57/2016/CP1364/CT0006, the Fundo Europeu de Desenvolvimento Regional (FEDER) through grant POCI-01-0145-FEDER-030389 and its Programa Operacional Competitividade e Internacionaliza\c{c}\~{a}o (COMPETE2020) through grants UIDB/04434/2020 and UIDP/04434/2020, and the Strategic Programme UIDB/\-00099/2020 for the Centro de Astrof\'{\i}sica e Gravita\c{c}\~{a}o (CENTRA);  
\item the Slovenian Research Agency through grant P1-0188;
\item the Spanish Ministry of Economy (MINECO/FEDER, UE), the Spanish Ministry of Science and Innovation (MICIN), the Spanish Ministry of Education, Culture, and Sports, and the Spanish Government through grants BES-2016-078499, BES-2017-083126, BES-C-2017-0085, ESP2016-80079-C2-1-R, ESP2016-80079-C2-2-R, FPU16/03827, PDC2021-121059-C22, RTI2018-095076-B-C22, and TIN2015-65316-P (``Computaci\'{o}n de Altas Prestaciones VII''), the Juan de la Cierva Incorporaci\'{o}n Programme (FJCI-2015-2671 and IJC2019-04862-I for F.~Anders), the Severo Ochoa Centre of Excellence Programme (SEV2015-0493), and MICIN/AEI/10.13039/501100011033 (and the European Union through European Regional Development Fund ``A way of making Europe'') through grant RTI2018-095076-B-C21, the Institute of Cosmos Sciences University of Barcelona (ICCUB, Unidad de Excelencia ``Mar\'{\i}a de Maeztu'') through grant CEX2019-000918-M, the University of Barcelona's official doctoral programme for the development of an R+D+i project through an Ajuts de Personal Investigador en Formaci\'{o} (APIF) grant, the Spanish Virtual Observatory through project AyA2017-84089, the Galician Regional Government, Xunta de Galicia, through grants ED431B-2021/36, ED481A-2019/155, and ED481A-2021/296, the Centro de Investigaci\'{o}n en Tecnolog\'{\i}as de la Informaci\'{o}n y las Comunicaciones (CITIC), funded by the Xunta de Galicia and the European Union (European Regional Development Fund -- Galicia 2014-2020 Programme), through grant ED431G-2019/01, the Red Espa\~{n}ola de Supercomputaci\'{o}n (RES) computer resources at MareNostrum, the Barcelona Supercomputing Centre - Centro Nacional de Supercomputaci\'{o}n (BSC-CNS) through activities AECT-2017-2-0002, AECT-2017-3-0006, AECT-2018-1-0017, AECT-2018-2-0013, AECT-2018-3-0011, AECT-2019-1-0010, AECT-2019-2-0014, AECT-2019-3-0003, AECT-2020-1-0004, and DATA-2020-1-0010, the Departament d'Innovaci\'{o}, Universitats i Empresa de la Generalitat de Catalunya through grant 2014-SGR-1051 for project ``Models de Programaci\'{o} i Entorns d'Execuci\'{o} Parallels'' (MPEXPAR), and Ramon y Cajal Fellowship RYC2018-025968-I funded by MICIN/AEI/10.13039/501100011033 and the European Science Foundation (``Investing in your future'');
\item the Swedish National Space Agency (SNSA/Rymdstyrelsen);
\item the Swiss State Secretariat for Education, Research, and Innovation through the Swiss Activit\'{e}s Nationales Compl\'{e}mentaires and the Swiss National Science Foundation through an Eccellenza Professorial Fellowship (award PCEFP2$\_$194638 for R.~Anderson);
\item the United Kingdom Particle Physics and Astronomy Research Council (PPARC), the United Kingdom Science and Technology Facilities Council (STFC), and the United Kingdom Space Agency (UKSA) through the following grants to the University of Bristol, the University of Cambridge, the University of Edinburgh, the University of Leicester, the Mullard Space Sciences Laboratory of University College London, and the United Kingdom Rutherford Appleton Laboratory (RAL): PP/D006511/1, PP/D006546/1, PP/D006570/1, ST/I000852/1, ST/J005045/1, ST/K00056X/1, ST/\-K000209/1, ST/K000756/1, ST/L006561/1, ST/N000595/1, ST/N000641/1, ST/N000978/1, ST/\-N001117/1, ST/S000089/1, ST/S000976/1, ST/S000984/1, ST/S001123/1, ST/S001948/1, ST/\-S001980/1, ST/S002103/1, ST/V000969/1, ST/W002469/1, ST/W002493/1, ST/W002671/1, ST/W002809/1, and EP/V520342/1.
\end{itemize}

The GBOT programme  uses observations collected at (i) the European Organisation for Astronomical Research in the Southern Hemisphere (ESO) with the VLT Survey Telescope (VST), under ESO programmes
092.B-0165,
093.B-0236,
094.B-0181,
095.B-0046,
096.B-0162,
097.B-0304,
098.B-0030,
099.B-0034,
0100.B-0131,
0101.B-0156,
0102.B-0174, and
0103.B-0165;
and (ii) the Liverpool Telescope, which is operated on the island of La Palma by Liverpool John Moores University in the Spanish Observatorio del Roque de los Muchachos of the Instituto de Astrof\'{\i}sica de Canarias with financial support from the United Kingdom Science and Technology Facilities Council, and (iii) telescopes of the Las Cumbres Observatory Global Telescope Network.
\end{appendix}

\end{document}